\documentclass[11pt]{article}
\usepackage[utf8]{inputenc}
\usepackage[T1]{fontenc}
\usepackage{natbib}
\usepackage{booktabs}
\usepackage{rotating}
\usepackage{subfigure}
\usepackage{multirow}
\usepackage[official]{eurosym}
\usepackage{accessibility}
\usepackage{amsmath}
\usepackage{amssymb}
\usepackage{url}
\usepackage{breakurl}

\usepackage[a4paper, top=1in, bottom=1in, left=0.55in, right=0.55in]{geometry}

\title{The Impact of Device Type, Data Practices, and Use Case Scenarios on Privacy Concerns about Eye-tracked Augmented Reality in the United States and Germany}
\author{
  Efe Bozkir$^{1}$ \and
  Babette Bühler$^{1}$ \and
  Xiaoyuan Wu$^{2}$ \and
  Enkelejda Kasneci$^{1}$ \and
  Lujo Bauer$^{2}$ \and
  Lorrie Faith Cranor$^{2}$
}
\date{}

\begin{document}
\maketitle

\begin{center}
$^{1}$Technical University of Munich, Munich, Germany\\
$^{2}$Carnegie Mellon University, Pittsburgh, PA, USA\\
\end{center}

\begin{abstract}
Augmented reality technology will likely be prevalent with more affordable head-mounted displays. Integrating novel interaction modalities, such as eye trackers into head-mounted displays could lead to collecting vast amounts of biometric data, which may allow inference of sensitive user attributes like health status or sexual preference, posing privacy issues. While previous works broadly examined privacy concerns about augmented reality, ours is the first to extensively explore privacy concerns on behavioral data, particularly eye tracking in augmented reality. We crowdsourced four survey studies in the United States (n1 = 48, n2 = 525) and Germany (n3 = 48, n4 = 525) to understand the impact of user attributes, augmented reality devices, use cases, data practices, and country on privacy concerns. Our findings indicate that participants are generally concerned about privacy when they know what inferences can be made based on the collected data. Despite the more prominent use of smartphones in daily life than augmented reality glasses, we found no indications of differing privacy concerns depending on the device type. In addition, our participants are more comfortable when a particular use case benefits them and less comfortable when other humans can consume their data. Furthermore, participants in the United States are less concerned about their privacy than those in Germany. Based on our findings, we provide several recommendations to practitioners and policymakers for privacy-aware augmented reality.
\end{abstract}

\section{Introduction}
Advancements in imaging, computing, and artificial intelligence hold the potential to profoundly influence daily human life through augmented and virtual reality (AR/VR). Users often experience AR and VR, collectively called extended reality (XR), through head-mounted displays (HMDs), and these devices may become next-generation personal devices in everyday life, comparable to smartphones. While the purpose of the HMDs is mostly for entertainment, the HMD market will grow significantly beyond the entertainment domain, such as education and healthcare~\cite{hmd_market_est_2023}. For instance, prior research has already identified AR and VR HMDs as effective and engaging tools, such as for learning about human anatomy~\cite{DuncanVaidya_and_Stevenson_2021} and building expertise for teaching effectiveness~\cite{gao_etal_2023_ismar}. In healthcare, these devices have been helpful for incision planning in surgeries~\cite{ivan_etal_2021} and for preoperative planning~\cite{Kenngott_etal_2022}. Such examples highlight potential real-world applications in which behavioral data, such as eye-tracking, can be collected and analyzed. One can utilize behavioral data to understand users, and this can be particularly helpful for optimizing the workflows of real-world applications. 

AR and VR HMDs are similar in terms of form factor and types of integrated sensors, such as eye and hand trackers. However, AR is conceptually different and more privacy-invasive than VR because AR utilizes videos of the real-world environment and overlays visualizations on real-world content to augment and enhance users' perception and experience in real-world scenarios instead of working entirely in virtual settings. Integrating eye trackers into AR HMDs to understand users' visual attention and cognition~\citep{HAREZLAK2018176, Strohmaier2020, ojstervsek2019eye} may enable facilitating hands-free interaction~\cite{10342045}, optimizing image quality~\cite{kim_etal_2019}, or utilizing adaptive user interfaces~\cite{PFEUFFER20211}, which increases the utility of HMDs. HMD manufacturers have recently enabled eye tracking in their devices, such as in Apple Vision Pro~\cite{apple_visionpro_specs_2023}, Microsoft HoloLens 2~\cite{hololens2_specs_2023}, Meta Quest Pro~\cite{meta_questpro_et_2023}, and Varjo XR-4~\cite{varjo_xr4_et_2023}. Despite these, using eye-gaze and pupillometry information, it is possible to accurately identify users and infer their sensitive attributes, including health status~\cite{shic_etal_2008}, activity~\cite{bozkir_etal_2021}, gender~\cite{steil_etal_2019}, and sexual preference~\cite{wenzlaff_etal_2016}, which also raises privacy concerns. Recent work has also indicated the importance of privacy protection when these devices record gaze data in AR and VR~\cite{liebling_preibusch_2014, silva_etal_2019, Kroeger_etal_2020, bozkir2023eyetracked_VR}, with methods to preserve privacy~\cite{davidjohn_etal_2021, bozkir_etal_2020_etra, bozkir_etal_2021, li_etal_usenix_2021, ozdel2024privacy}. 

As AR becomes pervasive with modern HMDs and smartphone-based AR~\cite{schoeps_etal_2024, jin_etal_2024_chi, valentin_etal_2018}, the amount of collected eye-tracking data will likely increase, especially with more devices enabling eye-tracking capability~\cite{apple_ios18_eyetracking_may2024}. With increased behavioral data from AR and novel machine learning (ML) algorithms, understanding users' privacy concerns in eye-tracked AR in detail is essential. Such understanding may help design human-centered privacy protection methods and facilitate more conscious informed consent procedures for AR. However, little work has focused on understanding user privacy concerns in eye-tracked AR when recorded data reveals sensitive user attributes. 

To address this research gap, we examined users' privacy concerns, considering sensitive attributes derived from eye-tracked AR and how external factors influence these concerns. We first identified several user attributes that eye movements can reveal in AR~\cite{liebling_preibusch_2014, Kroeger_etal_2020, bozkir2023eyetracked_VR}. We then studied the factors that affect user privacy concerns across various use case scenarios, in which we varied data retention durations, entities with whom data is shared, types of AR devices, and valence priming. We crowdsourced four surveys on Prolific, collecting data from participants in two countries: the United States (US) and Germany. In the initial two surveys (n = 48 for each country), we focused on understanding participants' perceptions of the priming texts we introduced before each scenario and made sure that participants observed the beneficialness levels of priming in the same way as we designed them. In the latter two surveys (n = 525 for each country), we aimed to understand privacy concerns in vignette studies. To do this, we ran regression analysis for each user attribute we presented, controlling for factors such as priming, age, and gender. Our work is the first study that addresses user privacy concerns for AR when eye movements and inference possibilities are the focus. 

In summary, we answer two research questions (RQs). \textbf{RQ1:} ``What factors affect users' privacy concerns about AR technology when their attributes are inferred?'' \textbf{RQ2:} ``Whether and how do privacy concerns differ based on countries in eye-tracked AR?'' In answering our RQ1, our findings indicate that participants were less comfortable with data collection that enables inference of attributes like sexual preference and personal identity than data collection that can infer attributes such as alertness and stress. Participants felt more comfortable when their data, which led to sensitive attribute inferences, was not consumed by other humans. Some of our other findings overlap with the findings of previous research in similar contexts, such as discomfort with the human consumption of data~\cite{lee2016information} or comfort with beneficial use cases~\cite{harborth_and_frik_2021}. We did not find a consistent trend between the privacy concerns and data retention times, though participants often preferred their data to be retained for a short period. Finally, we found similar levels of concern and excitement in scenarios that involved AR glasses and smartphone-based AR, and discovered no evidence in participants' answers of any fear or excitement about AR glasses. Answering our RQ2, we found that participants in Germany were more concerned about privacy overall than those in the US, suggesting that community-specific privacy policies may be appropriate for AR. 

\section{Related work}
We introduce the inference possibilities eye movements offer and privacy-aware eye-tracking methods in Section~\ref{lbl_subseceye_rw_mov_privacy}. Then, we review research on user privacy concerns in smart devices, focusing on AR in Section~\ref{lbl_subsec_rw_privacy_concerns}. Finally, we point out the research gaps and emphasize why our research is essential in Section~\ref{subsec_researchgaps}.

\subsection{Eye movements in XR and privacy}
\label{lbl_subseceye_rw_mov_privacy}
Eye movements provide insights into how humans explore and perceive presented stimuli and can help facilitate hands-free interaction. However, due to differences in visual scanning patterns, the combination of stimuli and eye movements reveals sensitive user attributes~\cite{liebling_preibusch_2014}. For instance, when gaze and head directions are combined, it is possible to authenticate users implicitly with ML beyond passwords~\cite{duzgun_etal_2022, katsini_etal_2020_chi}. Pfeuffer et al.\ also showed that gaze information complements body motion and relations in the context of behavioral authentication in VR~\cite{Pfeuffer_etal_2019}. Furthermore, Steil et al.\ and Bozkir et al.\ demonstrated accurate user identification and gender detection solely by eye movement features~\cite{steil_etal_2019, bozkir_etal_2021}. In addition, based on the stimulus, eye movements are representative of various personal attributes such as sexual preference~\cite{wenzlaff_etal_2016, hall_etal_2011}, body mass index (BMI)~\cite{GRAHAM2011577, VELAZQUEZ2014578}, health status~\cite{lian2023evaluating, WILLIAMS2010617}, and cognitive load~\cite{appel_etal_2018, kosch_etal_2018}. The always-on possibility of AR devices with scene recordings increases the likelihood of encountering scenes or stimuli that reveal such sensitive personal characteristics during everyday use. 

To this end, several works emphasized the importance of privacy protection for eye-tracking data~\cite{liebling_preibusch_2014, Kroeger_etal_2020, bozkir2023eyetracked_VR, Adhanom_Etal_2023, Gressel_etal_2023, future_of_privacy_forum_2022}. Steil et al.\ and Liu et al.\ applied standard differential privacy mechanisms on aggregated eye movement features from VR~\cite{steil_etal_2019} and heatmaps~\cite{liu_etal_2019}, respectively, to privatize aggregated eye movement features. Although differential privacy provides a strong privacy guarantee, its standard mechanisms are vulnerable to correlated data. To address this, Bozkir et al.\ and Li et al.\ took temporal correlations~\cite{bozkir_etal_2021} and the possibility of spatio-temporal attacks on gaze streams~\cite{li_etal_usenix_2021} into account. Bozkir et al.\ indicated that despite the need for significant noise, an optimal privacy-utility trade-off is possible for differentially private eye movements~\cite{bozkir_etal_2021}. Li et al.\ showed that users enjoyed the virtual experience even with real-time privacy protection~\cite{li_etal_usenix_2021}. Others have also emphasized the importance of the real-time working principle in a privacy-preserving gaze estimation approach for XR, where a server instance handles the ML task~\cite{bozkir_etal_2020_etra}. However, while these works have mostly proposed technical solutions to protect privacy, they have not considered users' perspectives on privacy for XR, which is an essential research direction for designing human-centered privacy solutions. 

Researchers also focused on probabilistic solutions to preserve eye-tracking data privacy without strong privacy guarantees, which may contradict privacy regulations and policies. For instance, Fuhl et al.\ utilized a reinforcement learning-based ML approach successfully by taking the performance of expertise and task classification as a utility while hiding personal and gender identifiers in the gaze data~\cite{fuhl_etal_2021}. David-John et al.\ hindered user identification using eye-tracking data by spatial and temporal downsampling and indicated that such an approach is a promising way to protect privacy when authentication is not needed~\cite{davidjohn_etal_2021}. Combining scene and eye movement information, Steil et al.\ created a mechanical shutter to preserve the privacy of the scene, which is relevant to AR. The authors showed that when a sensitive scene is identified using sensor data, the shutter turns off the camera stream~\cite{steil_etal_2019b_shutter}. However, what users consider private is not always clear. Therefore, without knowing the users' privacy preferences and concerns for AR, it remains a challenge to design human-centered privacy solutions. 

\subsection{User privacy concerns}
\label{lbl_subsec_rw_privacy_concerns}
User privacy concerns on smart devices, such as life-logging cameras~\cite{hoyle_etal_2014, hoyle_etal_2015, price_etal_2017}, Internet of things devices~\cite{naeini_privacy_iot_soups2017, emami_naeini_iot_sp2021}, smartphones~\cite{lee2016information, harborth_and_frik_2021}, and AR glasses~\cite{koelle_etal_2017, rixen_Etal_2022, gallardo2023speculative, lammerding_etal_2021}, have been well-researched. AR technology is not limited to glasses; smartphones can provide AR functionality using their cameras. However, many of the previous works focused on concerns about AR glasses. In an interview study, Denning et al.\ investigated privacy perspectives of individuals when there were bystanders around AR glasses and found that participants expected to be asked before being recorded~\cite{denning_etal_2014}. Lebeck et al.\ researched multi-user AR's security and privacy aspects and found that bystander privacy is a significant concern~\cite{lebeck_etal_2018}. The authors noted the importance of AR users' privacy, indicating concerns toward physical surroundings and users' behavioral information, such as eye tracking. Furthermore, O'Hagan et al.\ suggested that privacy-enhancing technologies for AR glasses should consider activity type and its relationship to bystanders, adaptively facilitating awareness and consent~\cite{ohagan_etal_2023}. Gallardo et al.\ argued for customizable and multidimensional privacy solutions for AR glasses due to context-dependent reservations about data collection and use~\cite{gallardo2023speculative}. 

Koelle et al.\ studied user attitudes and concerns about HMDs and smartphones and identified situations that raised concerns, such as teaching situations, work environments, driving, and recording of videos~\cite{koelle_etal_2015}. In another work, Koelle et al.\ showed in a multi-year case study that social acceptability is more critical for the short term, and usability is more valuable for the long-term adoption of HMDs~\cite{koelle_etal_2017}. While both studies provide important insights into smartphones and early-generation HMDs, they do not examine potential inferences from the behavioral data, and the provided insights are based on relatively small sample sizes. Similarly, Lee et al.\ evaluated users' information disclosure concerns through a hypothetical head-worn device, understanding different activities of users, and sharing this information with recipients such as the general public or the app server~\cite{lee2016information}. The authors found that when other humans, such as users' friends, work contacts, or the public, consume data, users become significantly more upset than when an application consumes it. A similar study evaluating smartphone users' privacy concerns indicated that data privacy concerns depend on with whom the data is shared~\cite{felt_etal_2012}. For smartphone-based AR, Harborth and Frik argued that justifying data collection can mitigate privacy concerns if the collection benefits the users~\cite{harborth_and_frik_2021}. Furthermore, Harborth and Pape stated that in mobile AR, whether sensitive resources on the smartphone are accessed determines users' privacy concerns~\cite{HARBORTH2021106833}. While the previous research findings have provided insights into users' privacy concerns based on data recipients, perceived benefits, and data types, integrating artificial intelligence into AR systems can now enable the inference of various user characteristics from their data, which may introduce novel privacy issues, and this research area remains underexplored. 

Despite the prior works' focus on privacy concerns related to smart devices, there has also been little research on specific data types, such as eye tracking~\cite{steil_etal_2019, alsakar2023investigating} and user backgrounds and cultural differences~\cite{Wilkowska_etal_2021}. To this end, Steil et al.\ investigated user privacy concerns about their eye-tracking data in AR and VR by technical explanations of eye tracking. The authors found users' comfort in sharing eye-tracking data for early detection of diseases and hands-free interaction~\cite{steil_etal_2019}. In addition, users only agreed to share their data when it was co-owned by a governmental health agency or for research purposes. However, collected data from AR and VR vary significantly due to the nature of the environments, and a distinction is essential to evaluate user privacy concerns precisely. Further, technical explanations for eye tracking are unlikely to be relevant for an everyday user. More recently, Alsakar et al.\ similarly suggested the need for privacy-aware eye tracking on handheld mobile devices after participants watched a video showing the amount of personal information eye movements include~\cite{alsakar2023investigating}. 

Prior work also considered different user backgrounds for smart device acceptance. For instance, Wilkowska et al.\ focused on privacy perceptions of video-based lifelogging between Germans and Turks and found that culture influences the technology acceptance~\cite{Wilkowska_etal_2021}. Other researchers also focused on similar directions for different technologies, such as smartphones~\cite{Harbach_etal_2016} and radio-frequency identification technology~\cite{Hossain_etal_2014}. In both works, cultural differences led to distinct privacy attitudes. Despite these works, no research has considered the effects of different cultures and communities on users' privacy concerns for AR.

\subsection{Research gaps and novelty of this work}
\label{subsec_researchgaps}
Despite the recent broad focus on privacy concerns for AR glasses and smartphones~\cite{gallardo2023speculative, Hadan_Etal_2024, felt_etal_2012, harborth_and_frik_2021, HARBORTH2021106833}, prior research has not focused on users' privacy concerns on behavioral data, such as eye tracking, and inferences from this data. As aforementioned, the association of eye-tracking data with stimuli information can lead to inferences of various user attributes~\cite{bozkir2023eyetracked_VR, liebling_preibusch_2014, steil_etal_2019, Kroeger_etal_2020}, including activity~\cite{bozkir_etal_2021}, alertness~\cite{6518125}, BMI~\cite{VELAZQUEZ2014578}, cognitive load~\cite{8797758}, gender~\cite{steil_etal_2019}, location~\cite{kiefer_etal_2014}, personal identity~\cite{9865991}, sexual preference~\cite{wenzlaff_etal_2016}, stress~\cite{7574455}, and health information~\cite{bradshaw_etal_2019} including mental states~\cite{10318110}. With such inference possibilities, it is essential to ask users about their concerns in order to build the AR technology responsibly and avoid any harm to users. 

Solove conceptualized privacy harms and violations into four groups: information collection, information processing, information dissemination, and invasions~\cite{solove2006taxonomy}. In his taxonomy, information processing refers to the ``use, storage, and manipulation of data that has been collected.'' One form of information processing is identification, which refers to connecting information to individuals. Identification is particularly interesting for behavioral data from AR since it can reveal various individual characteristics, especially with modern ML algorithms. In Solove's taxonomy, information dissemination is another potential harm and deals with concepts around spreading individuals' data. In AR, information dissemination may be coupled with identification, as the identification can occur in different locations (e.g., computers) with possible human data consumption. As such, identification is contextual with AR data, and privacy concerns should be associated with context. To this end, Nissenbaum's ``Contextual Integrity'' can be useful as it argues for understanding privacy concepts based on appropriateness with contextual informational norms, including data subject, sender, recipient, type of information, and transmission principle~\cite{nissenbaum2004privacy}. 

In light of these privacy concepts, prior work focused on the effects of data types and uses~\cite{gallardo2023speculative}, users' benefits~\cite{harborth_and_frik_2021}, retention times~\cite{naeini_privacy_iot_soups2017}, and types of users or systems with access~\cite{lee2016information}, on users' privacy concerns for smart devices, including AR glasses~\cite{gallardo2023speculative, ohagan_etal_2023}. However, no focused research has studied user privacy concerns on AR devices, considering the inference of sensitive user attributes with behavioral data, particularly eye tracking. To address this research gap, we conducted four user studies on users' privacy concerns about eye-tracked AR, considering what happens with the data, with whom it is shared, how long it is retained, and the context of information processing, for two types of AR devices to answer our RQ1. To address RQ2, we performed this research in two countries with substantial AR market share and significant advancements in AR and eye-tracking technologies, namely the US and Germany~\cite{ar_vr_market_2024, eyetracking_market_2022}. In addition to their contributions to technological advancements, these countries also historically differ in their cultural~\cite{whitman2003two} and regulatory approaches~\cite{schwartz2019global} to privacy, and these differences may lead to potential cross-cultural privacy concern variations about AR, which are essential to understand. 

\section{Methods}
To understand user privacy concerns in eye-tracked AR, we designed a two-step study with several factors based on the inferences using eye movements. The institutional review boards of Carnegie Mellon University and the Technical University of Munich approved our study protocols in the US and Germany, respectively. In this section, we discuss the details of the experimental design, participants and their recruitment, and data analysis methods. 

\subsection{Experimental design}
We aim to understand the effects of AR device type, use case scenario, priming types (describing uses designed to appear \textit{beneficial} or \textit{not-beneficial} to the device user, or \textit{no priming}), data retention time, data-receiving entity, the country in which users live, and inferred user attributes on privacy concerns toward eye-tracked AR. To this end, we designed two surveys: privacy concerns and calibration. To address our RQ1 and RQ2, we asked participants to respond to vignettes that incorporated our interest variables in the privacy concerns survey. We created use case scenarios involving 11 user attributes we identified from the eye-tracking research literature that eye movements and stimulus data can reveal in AR~\cite{liebling_preibusch_2014, Kroeger_etal_2020, bozkir2023eyetracked_VR}. Table~\ref{lbl_attribute_tbl} lists those attributes. We designed our surveys by implying that both AR glasses and smartphones are personal devices and that no multi-user device use is intended. We conducted separate calibration surveys before the privacy concerns surveys to validate the priming type's effectiveness and achieve statistical control for different priming perceptions in privacy concern survey analysis. To this end, in the calibration surveys, we asked participants to rate each of our use cases for beneficialness. While the notions of beneficialness may vary for each attribute based on participants' perception and each user attribute's sensitivity, statistically controlling for these differences by incorporating the mean beneficialness scores from our calibration survey allowed us to account for the variation in the framing of each attribute in the best possible way. We first designed our surveys in English and then translated them into German in two iterations by two bilinguals, including a native German speaker. We conducted our user studies in English in the US and German in Germany. We explain the experimental design details of our surveys in the following.

\begin{table*}[ht]
    \centering
    \caption{User attributes and English versions of the priming texts for beneficial and not-beneficial priming.}
    \alt{Table that includes evaluated user attributes, beneficial and not-beneficial priming texts for the US study. The first column includes eleven user attributes, whereas the second and third columns include beneficial and not-beneficial priming texts for corresponding attributes.}
    \footnotesize
    \begin{tabular}{|p{1.5cm}|p{7.5cm}|p{7.5cm}|}
        \hline
        \textbf{Attribute} & \textbf{Beneficial priming text} & \textbf{Not-beneficial priming text} \\
        \hline
        activity & Suppose that your personal assistant app on your [AR glasses/smartphone] provides cues while working or studying to help you optimize your work. & Suppose that your personal assistant app on your [AR glasses/smartphone] observes your activity while working or studying and reports your productivity level to your employer or school. \\
        \hline
        alertness & Suppose that your personal assistant app on your [AR glasses/smartphone] warns you about potential hazards that you might not otherwise notice while driving or cycling. & Suppose that your personal assistant app on your [AR glasses/smartphone] warns you constantly that you are tired while driving. \\
        \hline
        body mass index & Suppose that your personal assistant app on your [AR glasses/smartphone] suggests healthy food and exercises based on your body mass index. & Suppose that your personal assistant app on your [AR glasses/smartphone] suggests snacks you might like based on your unhealthy dietary habits. \\
        \hline
        cognitive load & Suppose that your personal assistant app on your [AR glasses/smartphone] recognizes that you are cognitively overloaded during your task and suggests you cool down for a while. & Suppose that your personal assistant app on your [AR glasses/smartphone] tries to maximize your cognitive load constantly during your daily routines. \\
        \hline
        depression & Suppose that your personal assistant app on your [AR glasses/smartphone] notices that you are depressed and provides you with solutions to improve your well-being. & Suppose that your personal assistant app on your [AR glasses/smartphone] keeps suggesting ways to improve your mental health based on content from the internet. \\
        \hline
        gender & Suppose that your personal assistant app on your [AR glasses/smartphone] suggests products you might be interested in based on your gender. & Suppose that your personal assistant app on your [AR glasses/smartphone] provides product suggestions while working based on gender stereotypes. \\
        \hline
        heart \mbox{condition} & Suppose that your personal assistant app on your [AR glasses/smartphone] informs you that you might have a heart condition and suggests it would be good to see a doctor. & Suppose that your personal assistant app on your [AR glasses/smartphone] informs your insurance provider and employer about your pre-existing heart condition. \\
        \hline
        location & Suppose that your personal assistant app on your [AR glasses/smartphone] offers you hands-free help for navigation purposes. & Suppose that your personal assistant app on your [AR glasses/smartphone] tracks you based on your location in indoor places, such as a coffee shop, office, or home to target advertising to you based on your location. \\
        \hline
        personal identity & Suppose that your personal assistant app on your [AR glasses/smartphone] allows you to authenticate without a password and personalizes the device for you. & Suppose that your personal assistant app on your [AR glasses/smartphone] allows you to authenticate without a password and personalizes the device for you but this increases risk of identity theft.\\
        \hline
        sexual \mbox{preference} & Suppose that your personal assistant app on your [AR glasses/smartphone] offers an opinion about whether the person you are interested in finds you attractive. & Suppose that your personal assistant app on your [AR glasses/smartphone] keeps suggesting events based on your sexual preference while working. \\
        \hline
        stress & Suppose that your personal assistant app on your [AR glasses/smartphone] suggests taking a break from a stressful activity and doing activities you enjoy. & Suppose that your personal assistant app on your [AR glasses/smartphone] constantly reminds you of items on your to-do list. \\
        \hline
    \end{tabular}
    \label{lbl_attribute_tbl}
\end{table*}

\subsubsection{Calibration survey}
\label{subsec_exp_priming}
In the privacy concerns survey, one factor we studied was priming, which included beneficial, not-beneficial, and no priming. However, as our perceptions of the beneficialness of priming texts might differ from the participants' perceptions, we designed a separate calibration survey to ask participants whether their perceptions of the priming texts overlapped with ours. This survey had one factor (i.e., device type), and the condition was either AR glasses or smartphones in a between-subjects design. The dependent variable was a 5-point Likert scale with a range of ``very harmful,'' ``somewhat harmful,'' ``neither beneficial nor harmful,'' ``somewhat beneficial,'' and ``very beneficial.'' 

After participants provided online consent to participate in the calibration survey, we randomly assigned them to a condition and informed them that they would answer questions about AR glasses or smartphones. We provided them with beneficial and not-beneficial priming texts, depicted in Table~\ref{lbl_attribute_tbl}, in a randomized order for each user attribute, followed by the question \emph{How beneficial or harmful would you find a personal assistant app on your [AR glasses/smartphone] that does this?}. We have 11 attributes and two priming types (i.e., beneficial and not-beneficial priming), leading to a total of 22 questions about priming. Then, we asked nine demographic questions, totaling 31 questions in this survey. 

We used the calibration survey to confirm that the differences we introduced in beneficial and not-beneficial priming texts resulted in beneficialness scores for beneficial texts being greater than those for non-beneficial texts. When analyzing the subsequent privacy concerns survey, we used the mean values of the beneficialness distributions to statistically control for the priming. 

\subsubsection{Privacy concerns survey}
\label{subsec_exp_main}
We designed our privacy concerns survey using three factors: priming strategy, data retention time, and data-receiving entity. With these factors, we had 11 use case scenarios to represent the 11 user attributes. The priming strategy consisted of three conditions: beneficial, not-beneficial, and no priming. We asked our questions directly without any priming text for the no-priming condition. The data retention time factor also included three conditions: one day, one month, and indefinitely. Considering that statistical models and ML algorithms that utilize temporally dependent data are often used to improve the performance of inferring user attributes in user modeling, it is essential to assess users' privacy concerns with different data retention times. For the data-receiving entity, we identified six conditions, namely, data remaining in the device, personal assistant app cloud, a service that is accessible by friends, a service that is accessible by work contacts, a service that is accessible by employees of the personal assistant app company, and a service that is accessible by the public. For the data remaining in the device condition, there was neither explicit mention of data sharing nor data remaining in the device. We applied randomization to decide the conditions of the factors. Our study used a between-subjects design, and the dependent variable was a 5-point Likert scale ranging from ``very uncomfortable'' to ``very comfortable,'' with the middle option ``neutral.'' 

We used the following question template (and the German equivalent) in our privacy concerns survey for each attribute with minor grammatical adaptations. The use case scenarios were directly associated with priming types and attributes provided in Table~\ref{lbl_attribute_tbl}. 

\begin{quote}
\emph{
[use case scenario text unless no-priming condition]\\ 
How would you feel if your personal assistant app on your [device type] can determine your [user attribute] using data collected by your [device type], share the collected data with [data-receiving entity], and store the collected data for [data retention time]? 
}
\end{quote}

After participants consented to be study participants, they received an introductory text (provided in the Appendix) that our study was about AR glasses and smartphone practices. We randomly assigned our participants to receive the AR glasses or the smartphone questions block first. We also randomized the question order within the blocks to prevent the ordering effect. At the end of each question block, we asked the question \emph{What other capabilities of your personal assistant app on your [AR glasses/smartphone] would make you uncomfortable?}, with a free text answer option. After the first question block, participants answered nine demographic questions, the same as the demographic questions in the calibration survey. After the demographics, participants continued to the other question block. Question blocks for AR glasses and smartphones also included briefing texts in the page headers describing AR glasses or smartphone capabilities to the participants in a summarized way, and these are available in the Appendix. We asked all participants the same multiple-choice questions for AR glasses and smartphones, totaling 22 questions about privacy concerns, with 11 questions per AR device. In addition, we had two free-text answer questions (one per AR device) and nine demographics questions, totaling 33 questions in our privacy concerns survey.

\subsection{Recruitment and participants}
We conducted two separate surveys (i.e., calibration and privacy concerns) in the US and Germany, leading to four user studies. We used the crowdsourcing platform Prolific to recruit participants, as it provides high-quality data~\cite{Peer_etal_2022}. After participants expressed interest, we directed them from Prolific to our surveys on Qualtrics. Upon completion of the surveys on Qualtrics, we redirected the participants to Prolific for compensation. We collected data from adults, and our surveys did not include any sensitive information that could directly identify participants, such as IP addresses or names. All participants in the US and Germany reported fluency in English and German, respectively, and none of the participants participated in multiple surveys. We conducted the surveys in the US before those in Germany. In addition, for both countries, we conducted calibration surveys prior to privacy concerns surveys. Furthermore, we ran pilot tests for each survey to assess the functionality, flow, and timing. We present our participant demographics in Table~\ref{table_demographics}. 

\subsubsection{Calibration survey}
For the calibration surveys, we sampled 48 participants in each of the two countries. We used the equal sex distribution option on Prolific and compensated the participants \$2.5 and \euro{2.5} for a survey predicted to take 10 minutes. As we obtained demographics both from Prolific and our survey questions, we excluded participants when an age mismatch of more than one year occurred. We also excluded the data of the participants who revoked their consent or who indicated that they did not use the Internet despite filling out our questionnaires online, leading to final sample sizes of $n^{US}_{c}=45$ and $n^{GER}_{c}=48$ for the US and Germany, respectively, totaling $n_{c} = 93$. 

\subsubsection{Privacy concerns survey}
For the privacy concerns surveys, we sampled 525 participants in each country. We used the representative sample distribution on Prolific for the US study according to sex, age, and ethnicity (i.e., simplified US Census), as it was available for large sample sizes. However, this option was unavailable in Germany; therefore, we used the equal sex distribution option for Germany. We compensated participants \$2.75 and \euro{2.75} for a survey predicted to take 11 minutes. We adapted the predicted completion time of this survey from 20 minutes to 11 minutes by adhering to our hourly compensation rate after the pilot test in the US ($n=20$), and the sample pool of the US also includes the participants from the pilot test. Similar to the calibration surveys, we excluded participants when there was an age mismatch of more than one year, a revocation of consent, or an inconsistency in Internet use. We also checked the contextual relationship of free-text responses with participants' survey responses for each device to avoid automated and bot-like responses. We finally ended up with $n^{US}_{p}=504$ and $n^{GER}_{p}=511$ for the US and Germany, respectively, totaling $n_{p} = 1015$. 

\begin{table*}
    \centering
    \caption{Demographics in our surveys. ``h/d'' indicates hours per day.}
    \alt{Table that provides the demographics in our surveys. The first column corresponds to the demographics types. The second and third columns correspond to the actual demographics of the calibration and privacy concerns surveys from the US, respectively. Similarly, the fourth and fifth columns correspond to the actual demographics of the calibration and privacy concerns surveys from Germany, respectively.}
    \begin{tabular}{c c|r|r|r|r}
        \toprule
        \multirow{2}{*}{} & \multirow{2}{*}{\textbf{Demographics}} & \multicolumn{2}{c|}{$n^{US}_{c} = 45, n^{US}_{p} = 504$} & \multicolumn{2}{c}{$n^{GER}_{c} = 48, n^{GER}_{p} = 511$} \\
        \cline{3-6}
         & & \multicolumn{1}{c|}{Calibration} & \multicolumn{1}{c|}{Privacy} & \multicolumn{1}{c|}{Calibration} & \multicolumn{1}{c}{Privacy} \\
        \hline
        \multirow{5}{*}{\rotatebox{90}{\textbf{Gender}}} 
         & Woman & $24~(53\%)$ & $252~(50\%)$ & $25~(52\%)$ & $242~(47\%)$ \\
        \cline{2-6}
         & Man & $19~(42\%)$ & $240~(48\%)$ & $23~(48\%)$ & $252~(49\%)$ \\
        \cline{2-6}
         & Non-binary & $2~(4\%)$ & $8~(2\%)$ & $0~(0\%)$ & $11~(2\%)$ \\
        \cline{2-6}
         & Self-describe & $0~(0\%)$ & $2~(<1\%)$ & $0~(0\%)$ & $4~(1\%)$ \\
        \cline{2-6}
         & Do not say & $0~(0\%)$ & $2~(<1\%)$ & $0~(0\%)$ & $2~(<1\%)$ \\
        \hline
        \multicolumn{2}{c|}{\textbf{Age}} & $32.8 \pm 11.54$ & $44.4 \pm 15.8$ & $29.4 \pm 9.7$ & $29.9 \pm 8.9$ \\
        \hline
        \multicolumn{2}{c|}{\textbf{Internet use (h/d)}} & $6.5 \pm 3.7$ & $6.4 \pm 3.5$ & $7.5 \pm 4.3$ & $6.2 \pm 3.3$ \\
        \hline
        \multicolumn{2}{c|}{\textbf{Smartphone use (h/d)}} & $5.2 \pm 3.1$ & $4.4 \pm 3.3$ & $4.8 \pm 3.7$ & $4.1 \pm 2.4$ \\
        \hline
        \multicolumn{2}{c|}{\textbf{Own AR glasses}} & $1~(2\%)$ & $31~(6\%)$ & $8~(17\%)$ & $65~(13\%)$ \\
        \bottomrule
    \end{tabular}
    \label{table_demographics}
\end{table*}

\subsection{Analysis}
Despite our exploratory study, we applied statistical analyses to better understand privacy concerns. For calibration surveys, we compared user perceptions of beneficial and not-beneficial priming texts separately for AR glasses and smartphones for each attribute listed in Table~\ref{lbl_attribute_tbl} by using Wilcoxon signed-rank test with $\alpha = .05$. For the privacy concerns surveys, we used ordinal regression with a logit model for every user attribute separately, controlling for the factors of priming, age, gender, daily internet and smartphone use, and ownership of AR glasses with $\alpha = .05$. In the regression analyses, we used the mean values of the beneficialness distributions to statistically control for the priming effect. In addition, we used a dummy coding scheme by treating conditions and factors for data remaining in the device, one day of data retention, the AR glasses, the US, owning AR glasses, and women as references, and compared the remaining conditions and factors with these references. Before we ran regression analyses, we validated that none of the factors led to multicollinearity by analyzing correlation coefficients and variance inflation factors. In addition, we inspected proportional odds using the Brant test~\cite{brant_1990} and validated that our dataset meets the ordinal regression assumptions. 

To understand additional privacy concerns about AR glasses and smartphones, we analyzed the free-text answers to the question \emph{What other capabilities of your personal assistant app on your [AR glasses/smartphone] would make you uncomfortable?} separately for each AR device and country. We focused on concerns about the data types similar to our user attributes rather than other details. To identify those, two researchers, fluent in English and German, conducted inductive content analysis. They first separately identified potential data types in the free-text answers data and later finalized these categories for the qualitative coding by mutual agreement. The data types we identified for the qualitative coding are financial information, video/audio recordings, environment/bystander information, health information, sexuality, emotions, routines, thoughts, social relationships/contacts, online tracking/messaging, political views, socio-economic information, psychological characteristics, physiological characteristics, and gaze. For the free-text coding, we omitted mentions of heart condition and depression in the health-related concerns category and sexual preference in the sexuality category, as these already exist in our surveys, and we explicitly sought additional attributes. We also coded the situations when participants explicitly mentioned overlapping concerns for devices, such as ``Same as for AR glasses'' in response to the smartphone question. We applied the coding information from one device to another in these cases. The two researchers separately coded the answers, obtaining Krippendorff's alpha inter-rater reliability scores of 0.74 and 0.77 for the US and Germany, respectively. When a disagreement occurred, the raters discussed and reached a consensus. We provide our codebook in the Appendix. 

\section{Results}
\label{sec_Results}
This section provides the results of our survey analyses. Calibration survey analyses demonstrated that priming texts are perceived by participants as intended, as beneficial priming scores are always greater than not-beneficial priming scores. Answering our \textbf{RQ1}, findings of our privacy concerns survey indicate that participants are particularly concerned about data collection, which enables the inference of sensitive attributes. However, participants are less concerned when other humans do not consume their data or when they benefit from the use case. Despite mixed trends, participants tend to be more comfortable when their data is retained for one day compared to longer durations. We found similar privacy concerns for each scenario involving AR glasses and smartphones. Focusing on our \textbf{RQ2}, country-wise analyses indicate that the participants in Germany are more concerned about their privacy than those in the US. Plots in the following sections indicate statistical significances of ($p < .0001$), ($p < .001$), ($p < .01$), and ($p < .05$), with ****, ***, **, and *, respectively and orange triangles indicate mean values. In addition to the results in this section, we report detailed supporting statistics and regression tables in the Appendix.

\subsection{Calibration survey}
\label{subsec_results_priming}
\begin{figure*}[ht!]
\centering
\subfigure[Activity.]{
    {\includegraphics[width=0.23\linewidth]{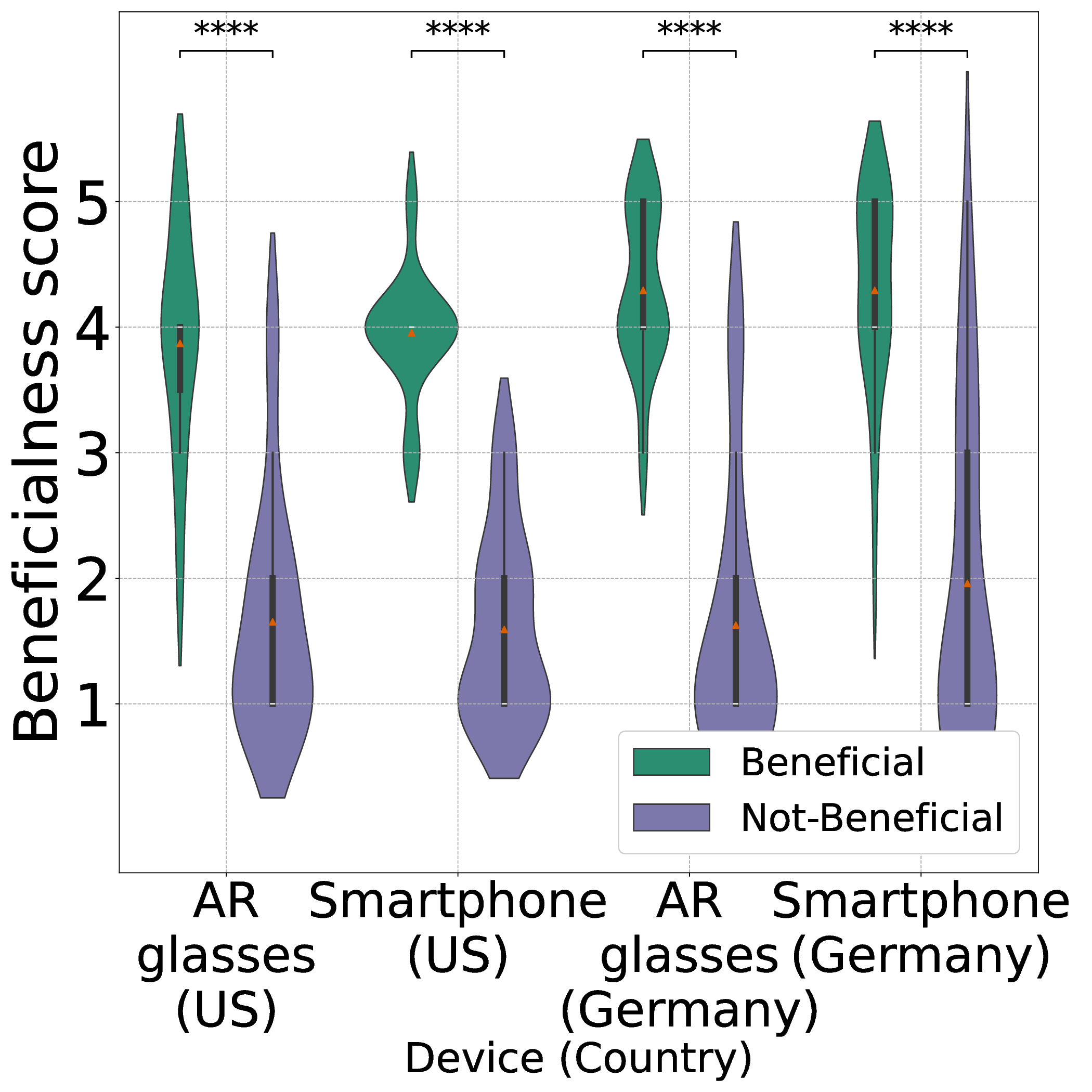}}
}
\subfigure[Alertness.]{
    {\includegraphics[width=0.23\linewidth,keepaspectratio]{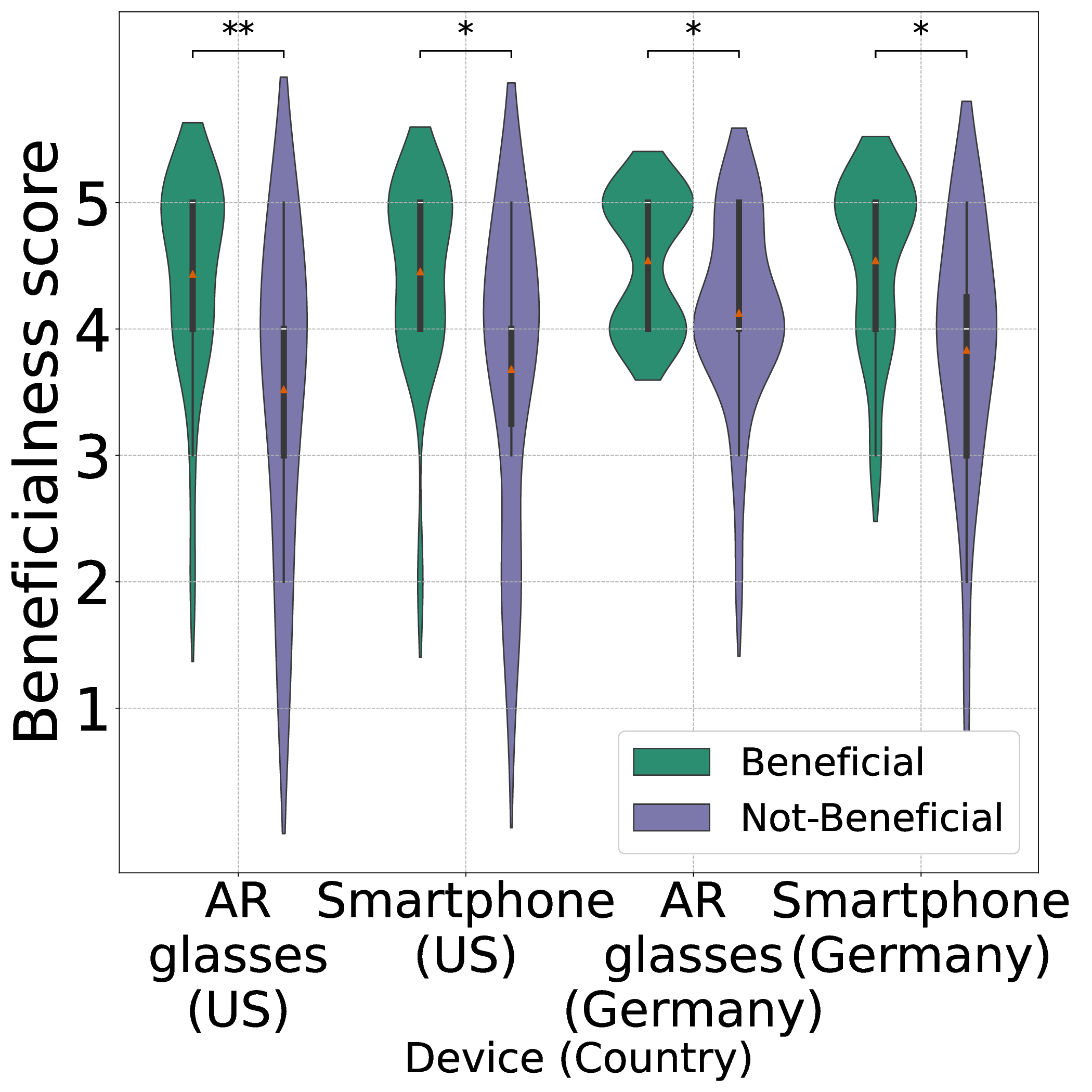}}
}
\subfigure[BMI.]{
    {\includegraphics[width=0.23\linewidth,keepaspectratio]{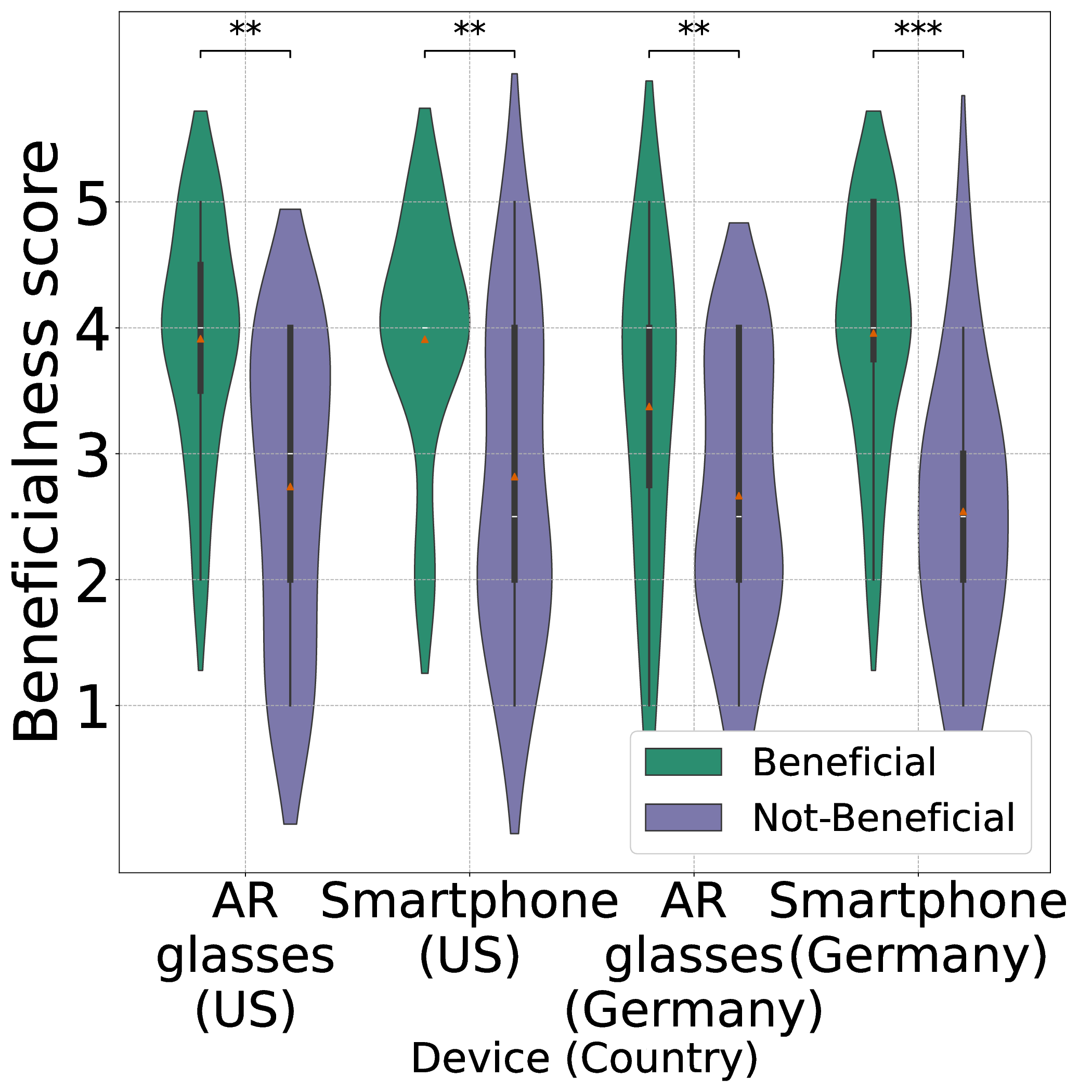}}
}
\subfigure[Cognitive load.]{
    {\includegraphics[width=0.23\linewidth,keepaspectratio]{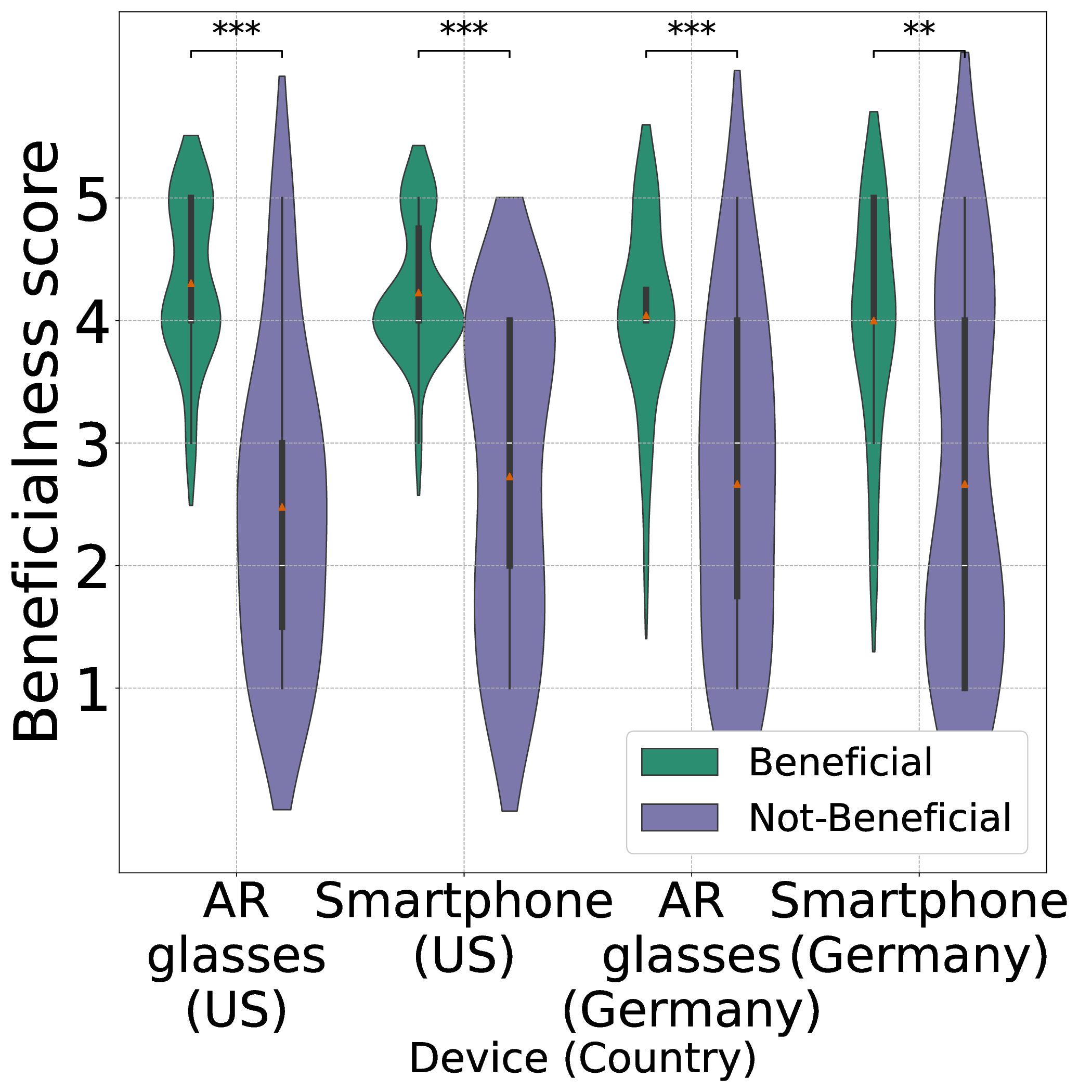}}
}
\subfigure[Depression.]{
    {\includegraphics[width=0.23\linewidth,keepaspectratio]{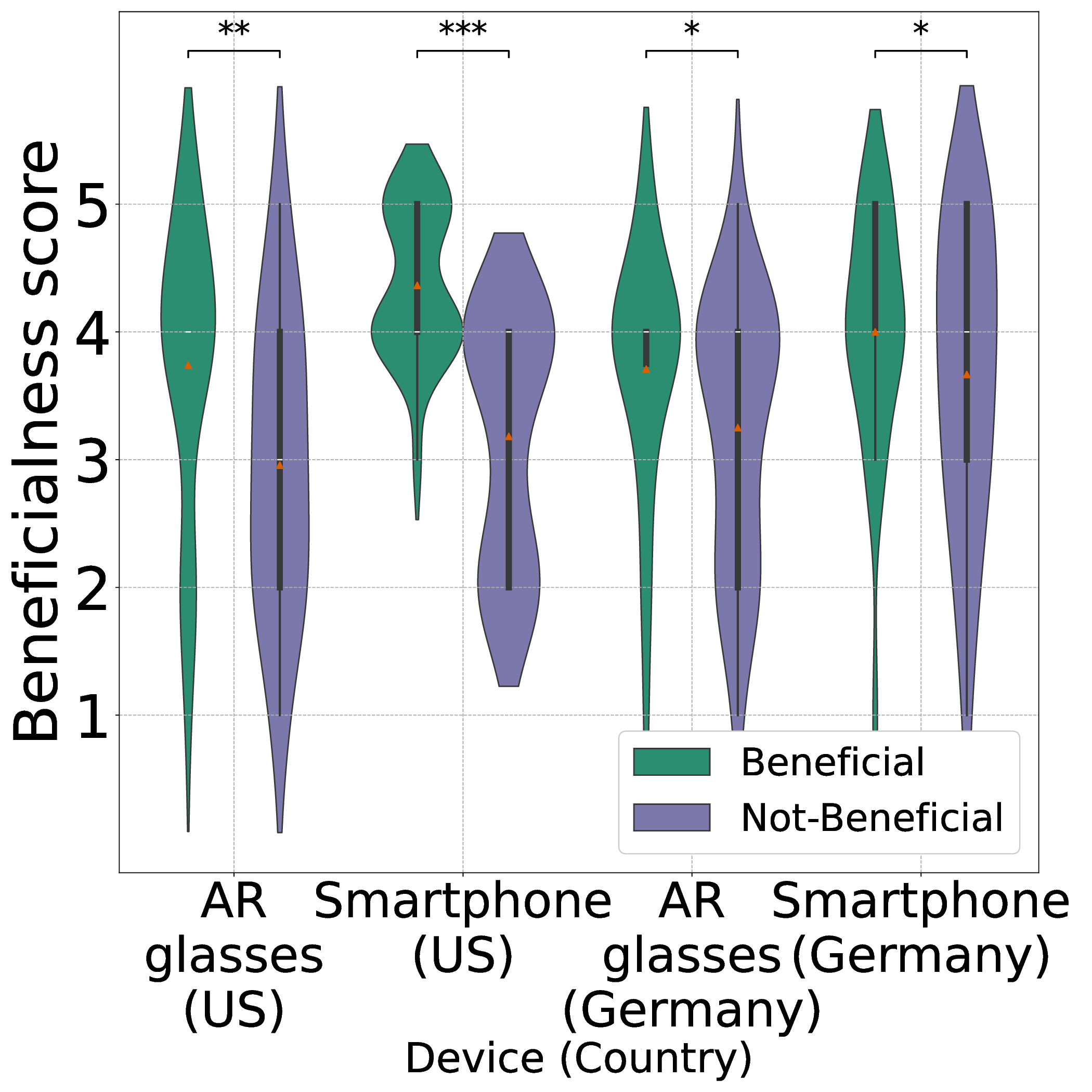}}
}
\subfigure[Gender.]{
    {\includegraphics[width=0.23\linewidth,keepaspectratio]{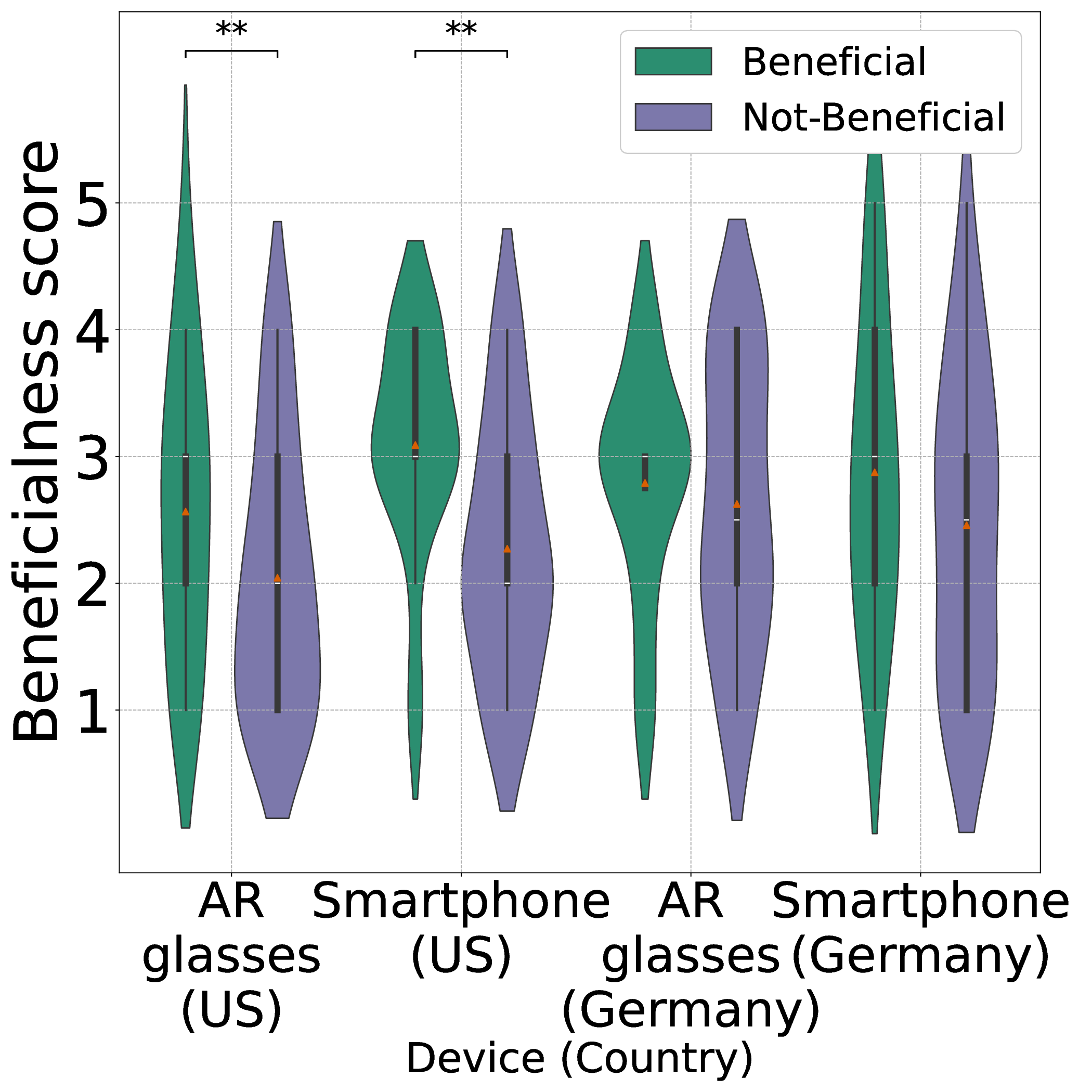}}
}
\subfigure[Heart condition.]{
    {\includegraphics[width=0.23\linewidth,keepaspectratio]{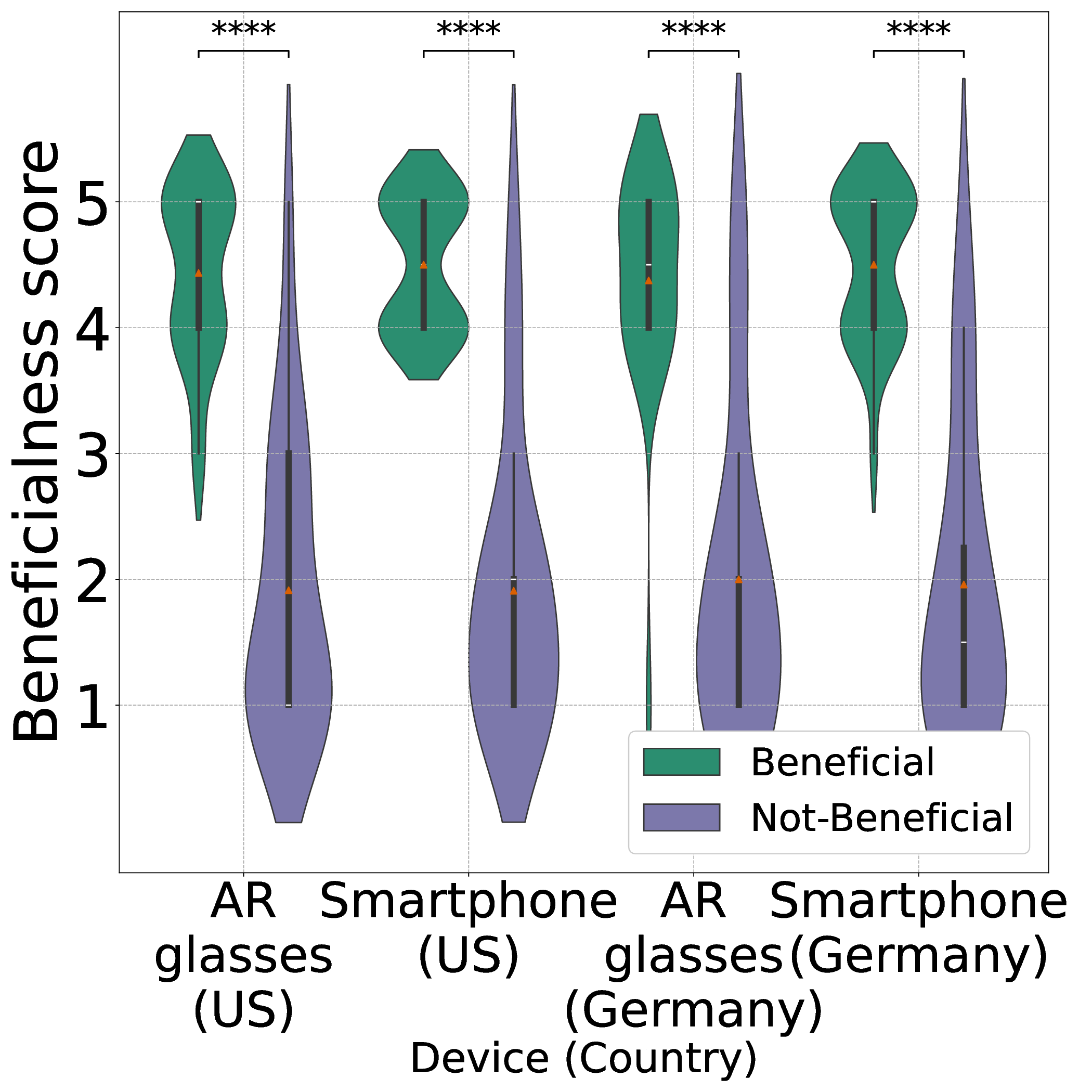}}
}
\subfigure[Location.]{
    {\includegraphics[width=0.23\linewidth,keepaspectratio]{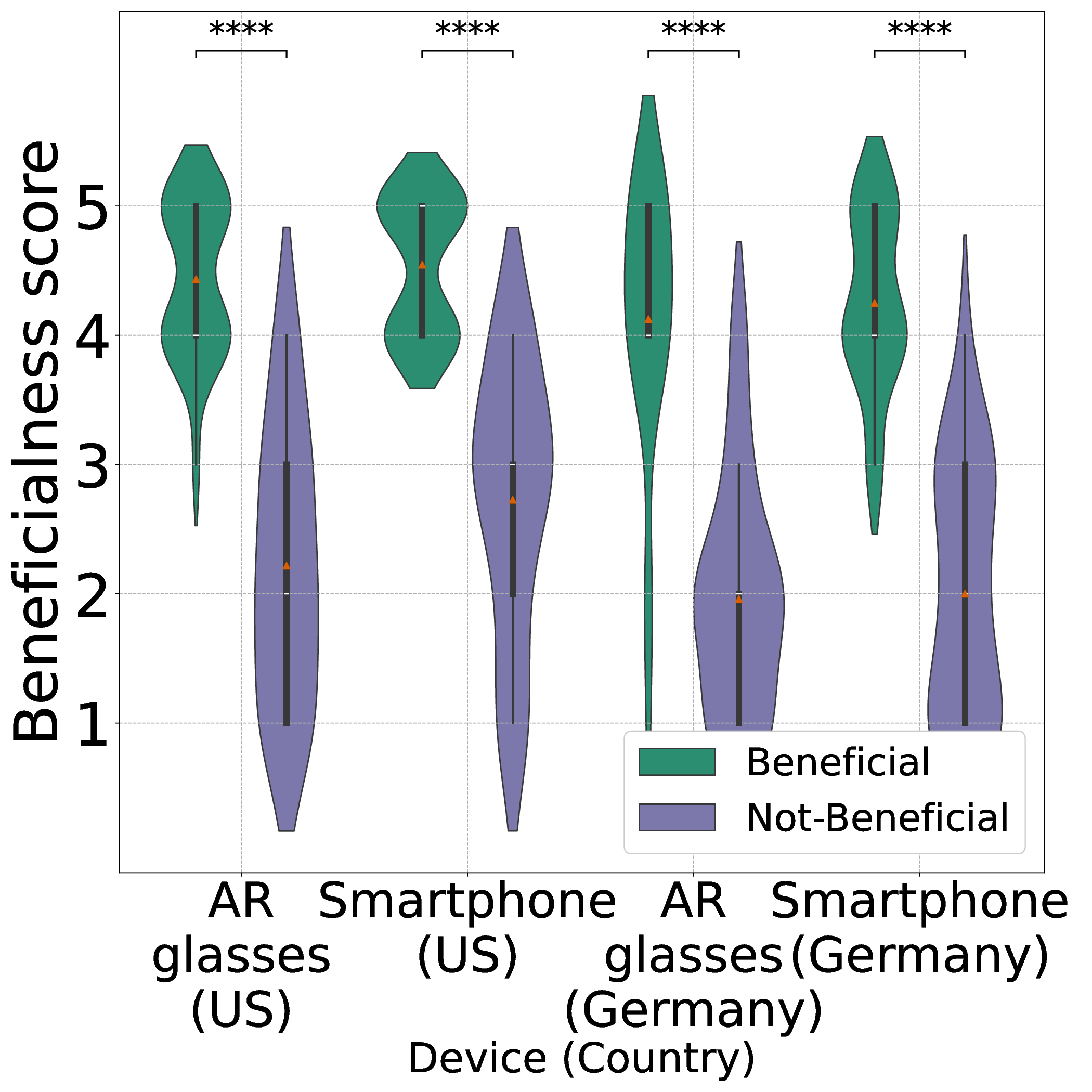}}
}
\subfigure[Personal identity.]{
    {\includegraphics[width=0.23\linewidth,keepaspectratio]{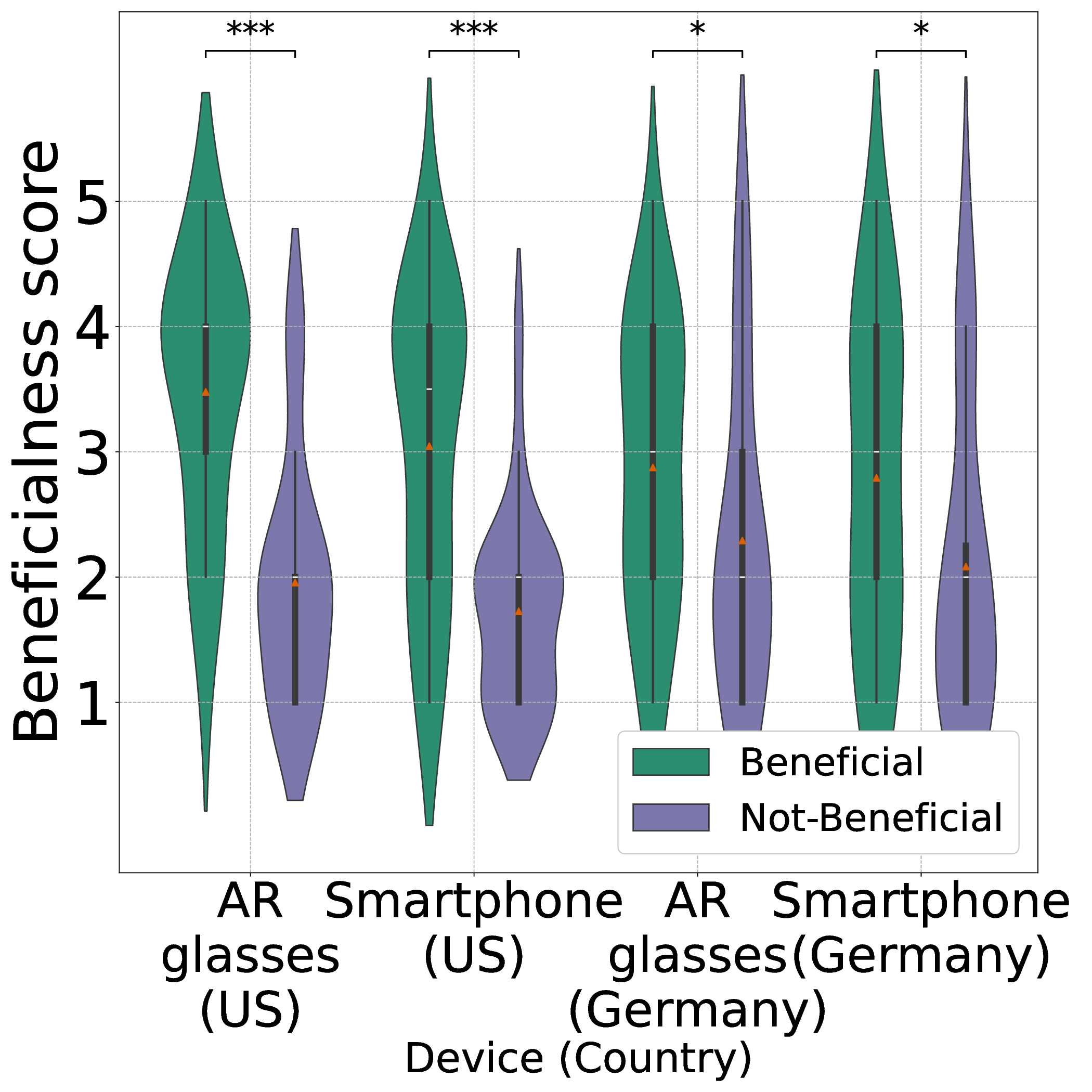}}
}
\subfigure[Sexual preference.]{
    {\includegraphics[width=0.23\linewidth,keepaspectratio]{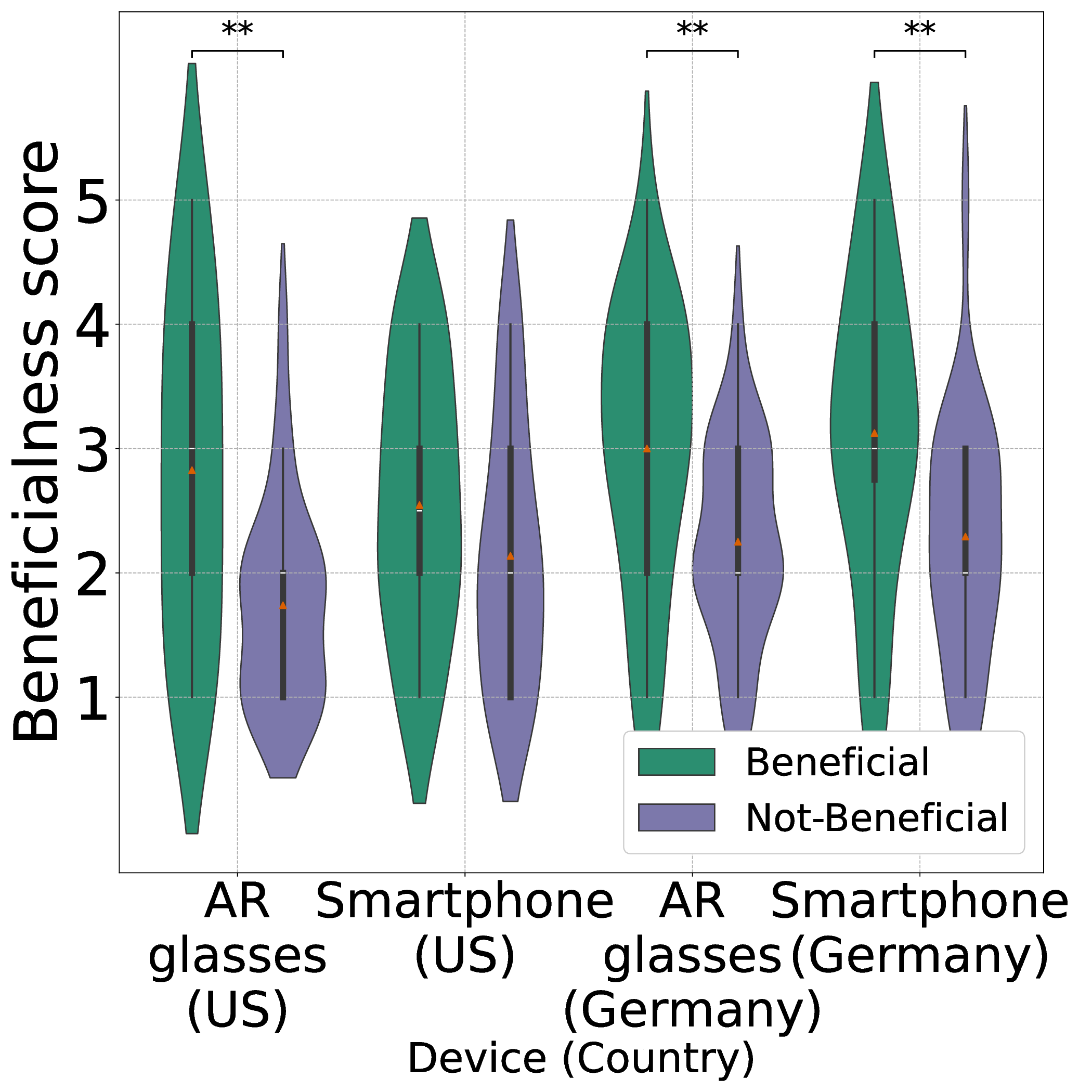}}
}
\subfigure[Stress.]{
    {\includegraphics[width=0.23\linewidth,keepaspectratio]{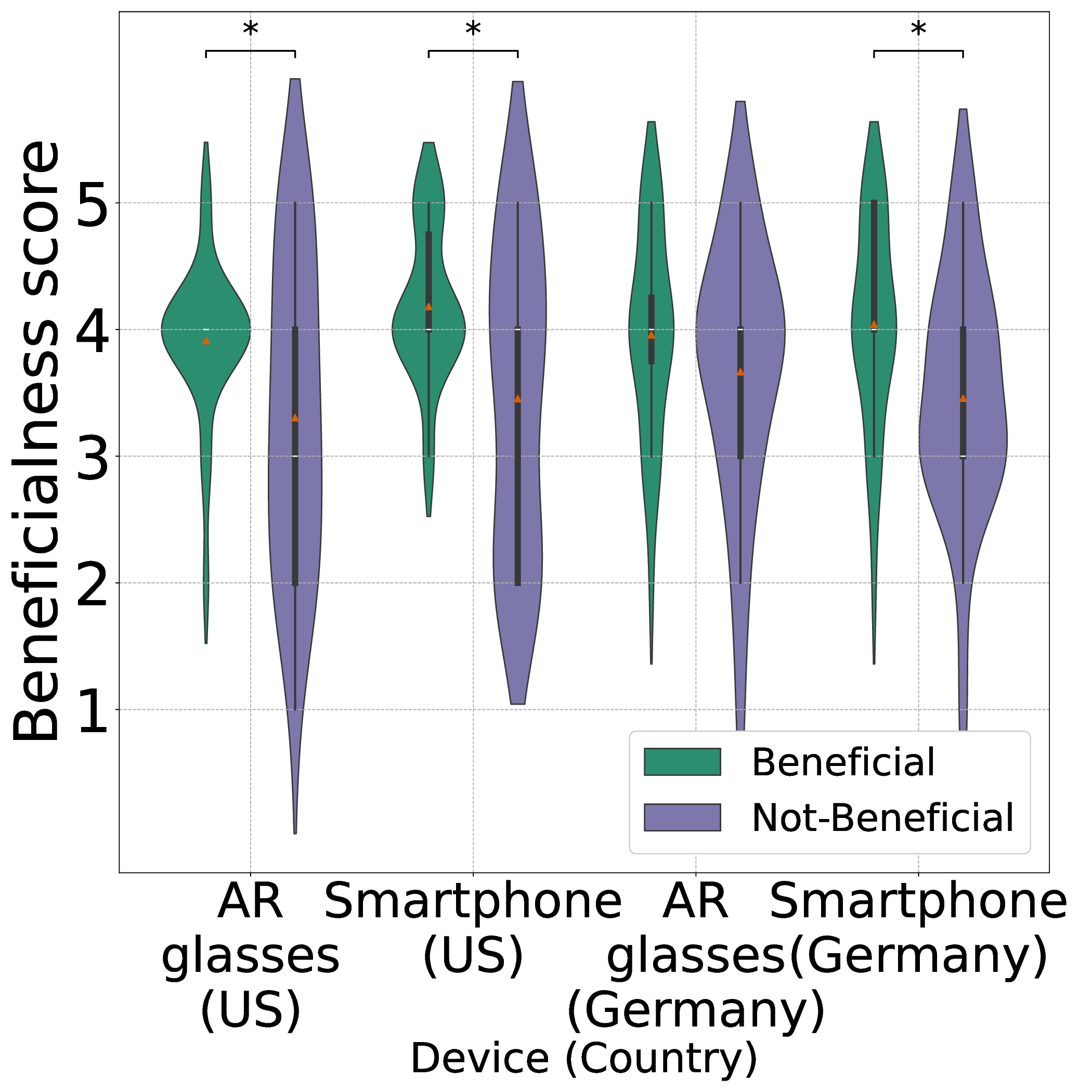}}
}
\caption{\label{fig_priming_results} Violin plots representing the calibration survey analyses for each attribute, separated by device type and country.}
\alt{Violin plots representing the relationship between beneficialness scores and device types for each country in our calibration survey. Each subplot corresponds to an evaluated user attribute. The Y-axes in the plots represent the beneficialness scores ranging from 1 to 5, with 1 being very harmful and 5 being very beneficial. The X-axes correspond to the AR glasses (US), Smartphone (US), AR glasses (Germany), and Smartphone (Germany).}
\end{figure*}

We applied the Wilcoxon signed-rank test to the data of beneficial and not-beneficial priming conditions separately for each device and country. Our analyses show that participants perceived the priming texts as intended. Beneficialness scores for the beneficial priming condition are always greater than for the not-beneficial priming condition, regardless of device type and country. Therefore, we did not iterate our priming texts before conducting privacy concerns surveys. We depict the results for each attribute in Figure~\ref{fig_priming_results}, where beneficialness scores range between 1 and 5, from ``very harmful'' to ``very beneficial.''

For the US, the beneficialness scores of our designed beneficial and not-beneficial text differed significantly for all attributes in varying levels from ($p < .05$) to ($p < .0001$), except the sexual preference in the smartphone condition ($p = .13$). Even though there are no significant trends for this attribute, the beneficial priming score outscored the not-beneficial priming score, like the other attributes. We obtained optimal results for both AR glasses and smartphones for activity, BMI, cognitive load, heart condition, location, and personal identity attributes, as their beneficialness scores of the beneficial priming condition are greater than neutral (i.e., greater than 3) on average. In contrast, beneficialness scores of not-beneficial priming conditions are less than neutral. For alertness, depression, and stress, the distributions are slightly skewed toward being seen as beneficial, whereas, for gender and sexual preference attributes, this trend is in the harmful direction. 

The data for Germany shows similar trends. Statistical analysis did not reveal significant differences for stress ($p = .24$) and gender ($p = .41$) in the AR glasses condition. Analysis of the gender attribute also did not yield significant differences for smartphone condition ($p = .11$). Despite these, the beneficialness scores of beneficial priming texts are greater than those of not-beneficial priming texts for these attributes. We found significant differences in varying levels for the remaining attributes from ($p < .05$) to ($p < .0001$). Similar to the US data, for activity, BMI, cognitive load, heart condition, and location attributes, beneficialness scores of the beneficial priming condition are greater than the neutral level on average. The scores of not-beneficial priming conditions for the same attributes are lower than the neutral level. For alertness, depression, and stress attributes, there is a shift toward the beneficial direction, whereas, for gender, personal identity, and sexual preference attributes, scores slightly shift toward the harmful direction. Hence, similar to our insights on the US data, it is reasonable to employ our priming texts in the privacy concerns survey and treat the beneficial, not-beneficial, and no priming differently, for Germany. 

\subsection{Privacy concerns survey}
\label{sec_main_survey_results}
\begin{figure*}[t!]
\centering
\subfigure[Activity.]{
    {\includegraphics[width=0.23\linewidth,keepaspectratio]{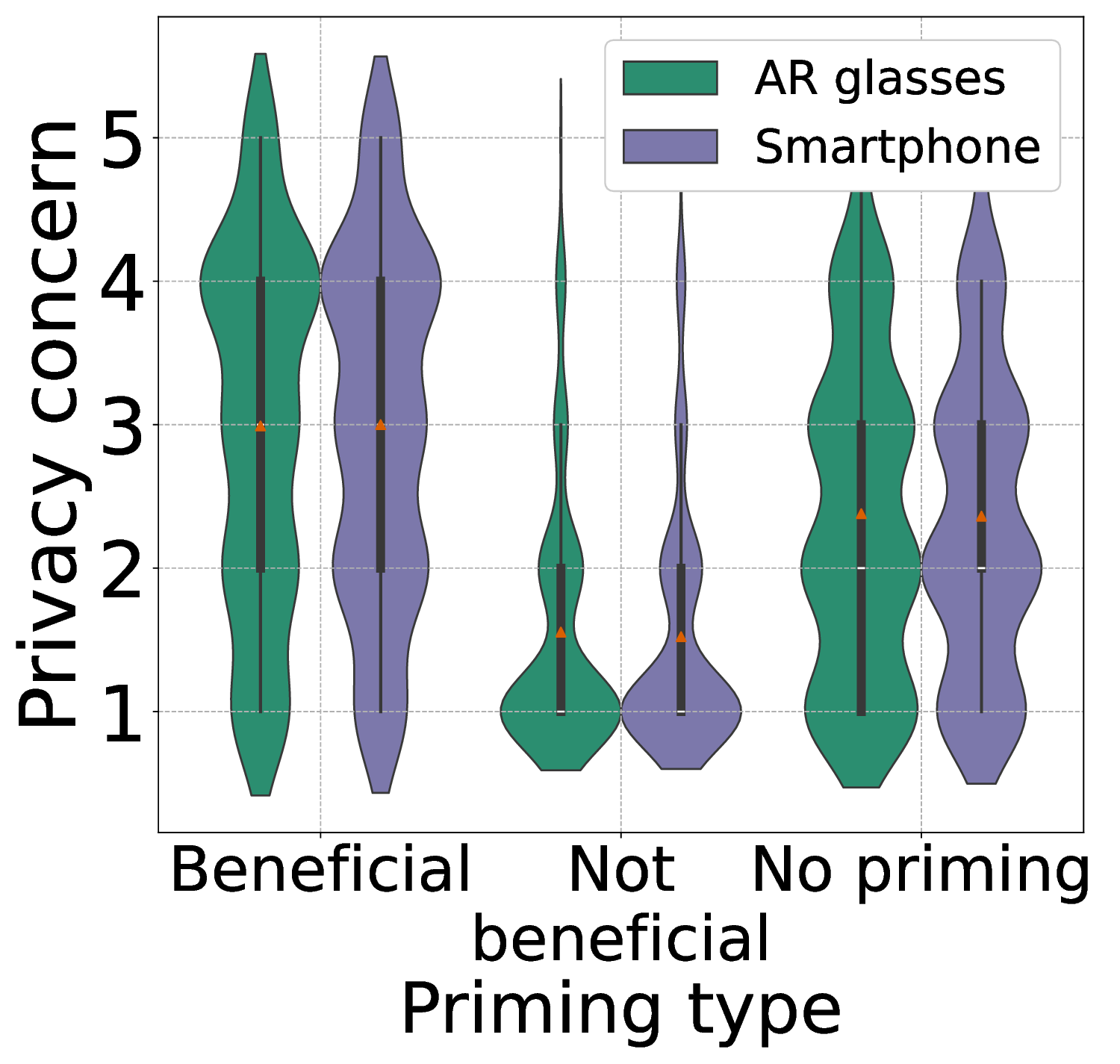}}
}
\subfigure[Alertness.]{
    {\includegraphics[width=0.23\linewidth,keepaspectratio]{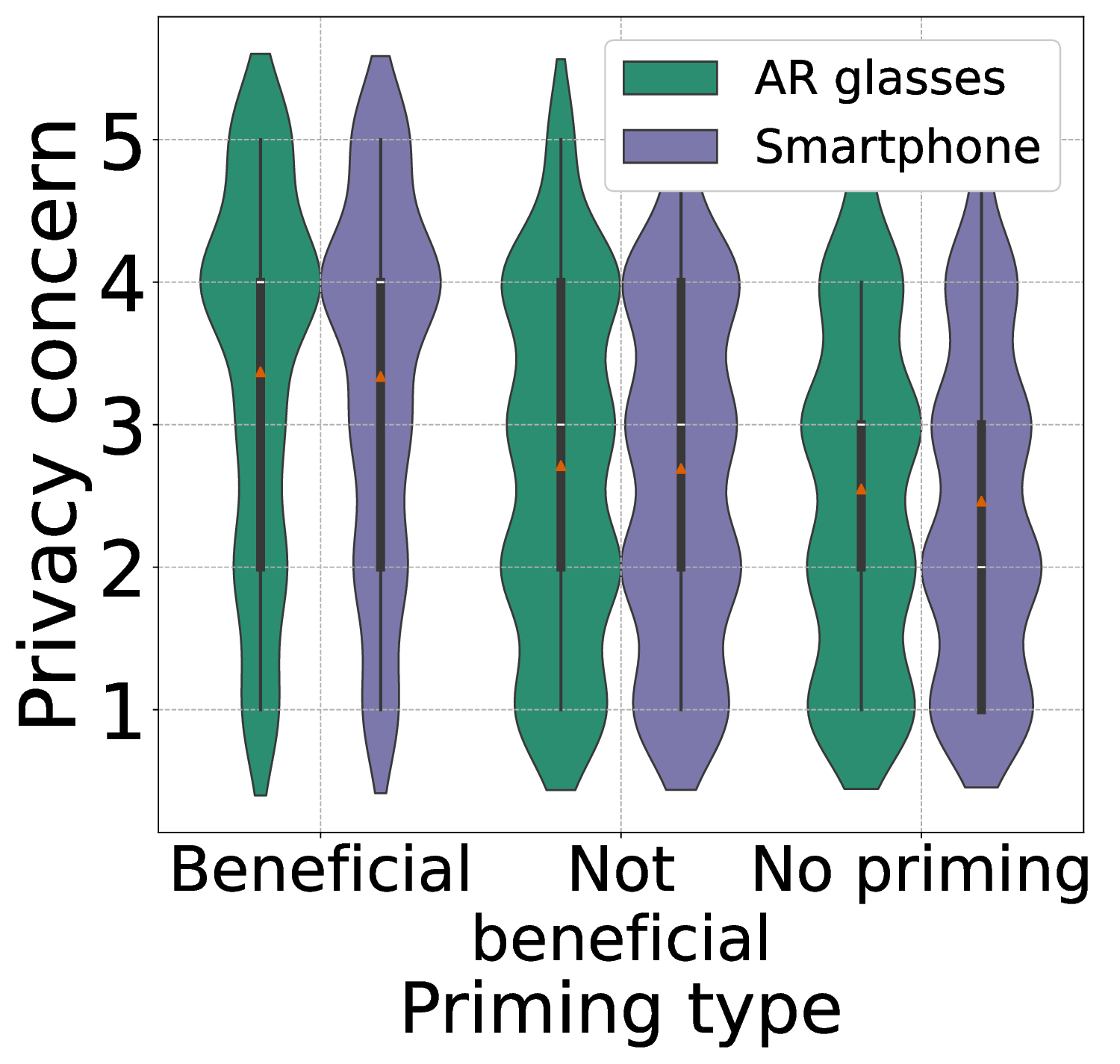}}
}
\subfigure[BMI.]{
    {\includegraphics[width=0.23\linewidth,keepaspectratio]{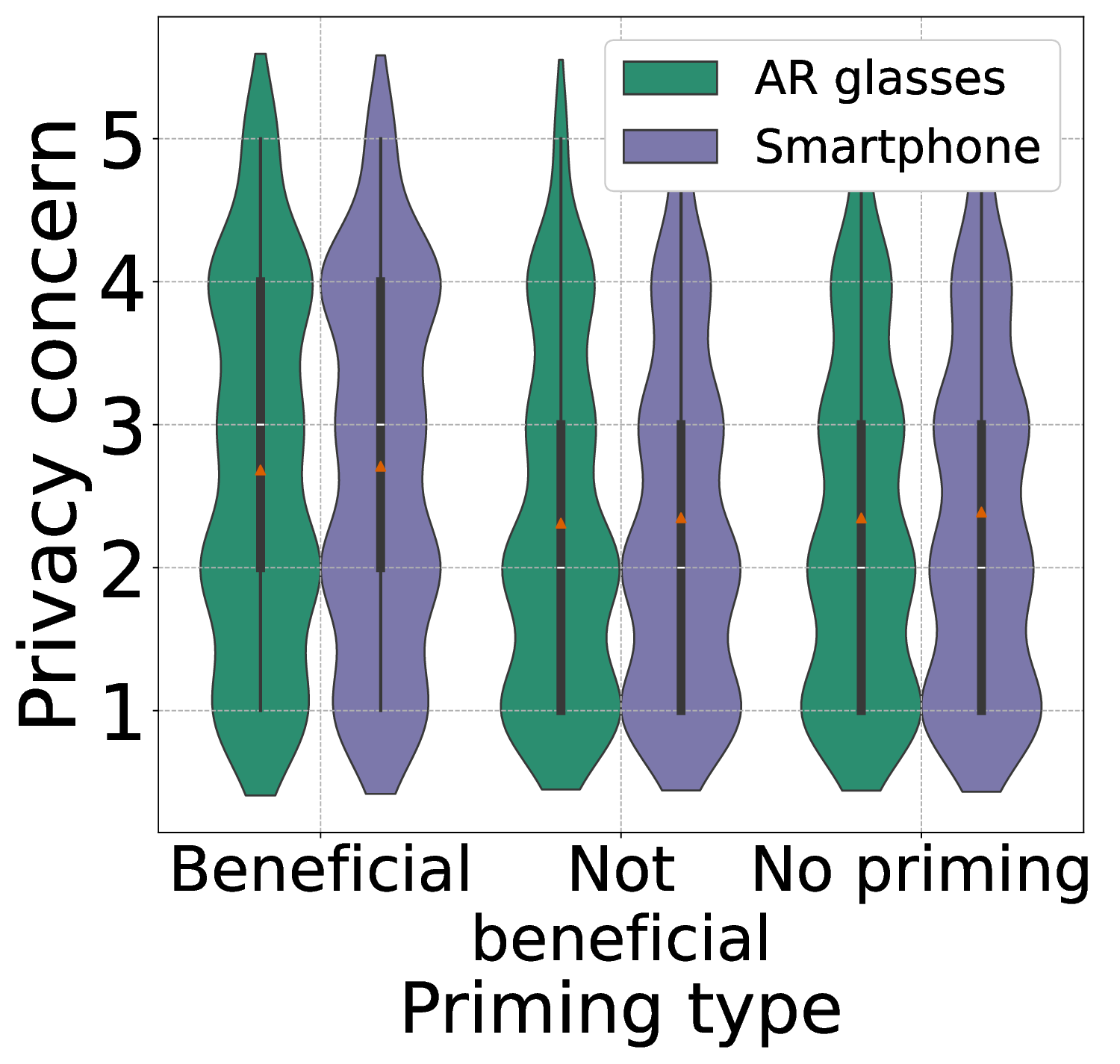}}
}
\subfigure[Cognitive load.]{
    {\includegraphics[width=0.23\linewidth,keepaspectratio]{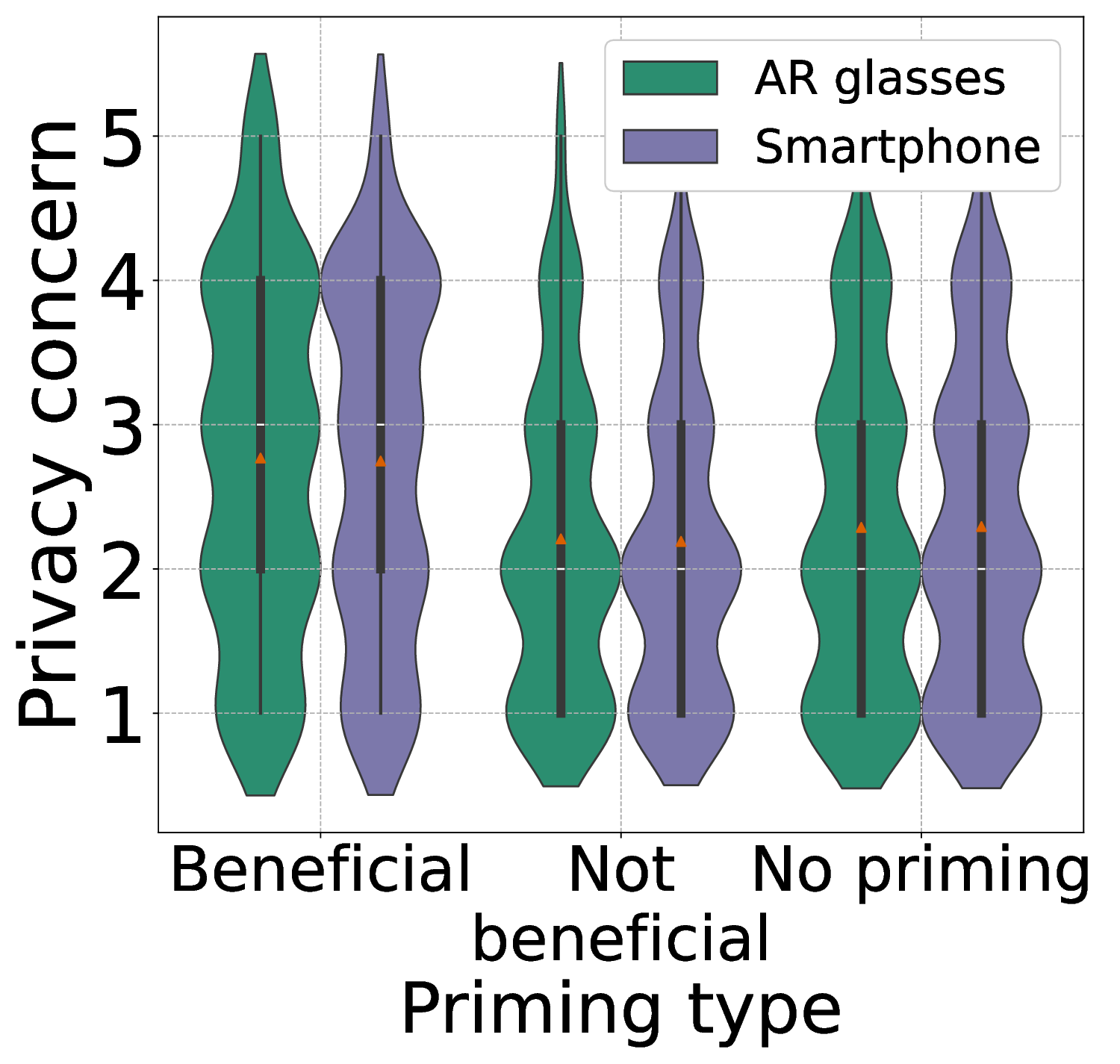}}
}
\subfigure[Depression.]{
    {\includegraphics[width=0.23\linewidth,keepaspectratio]{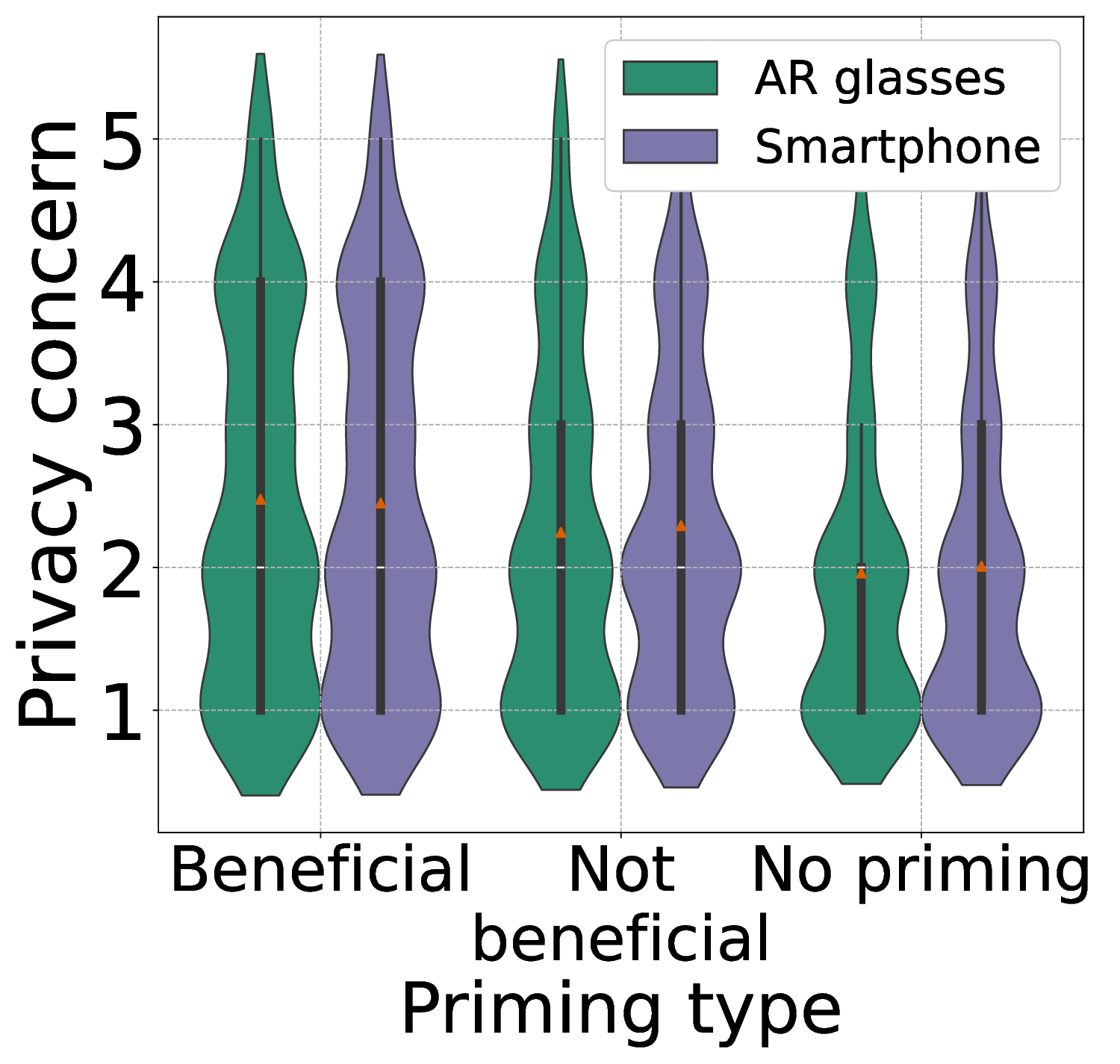}}
}
\subfigure[Gender.]{
    {\includegraphics[width=0.23\linewidth,keepaspectratio]{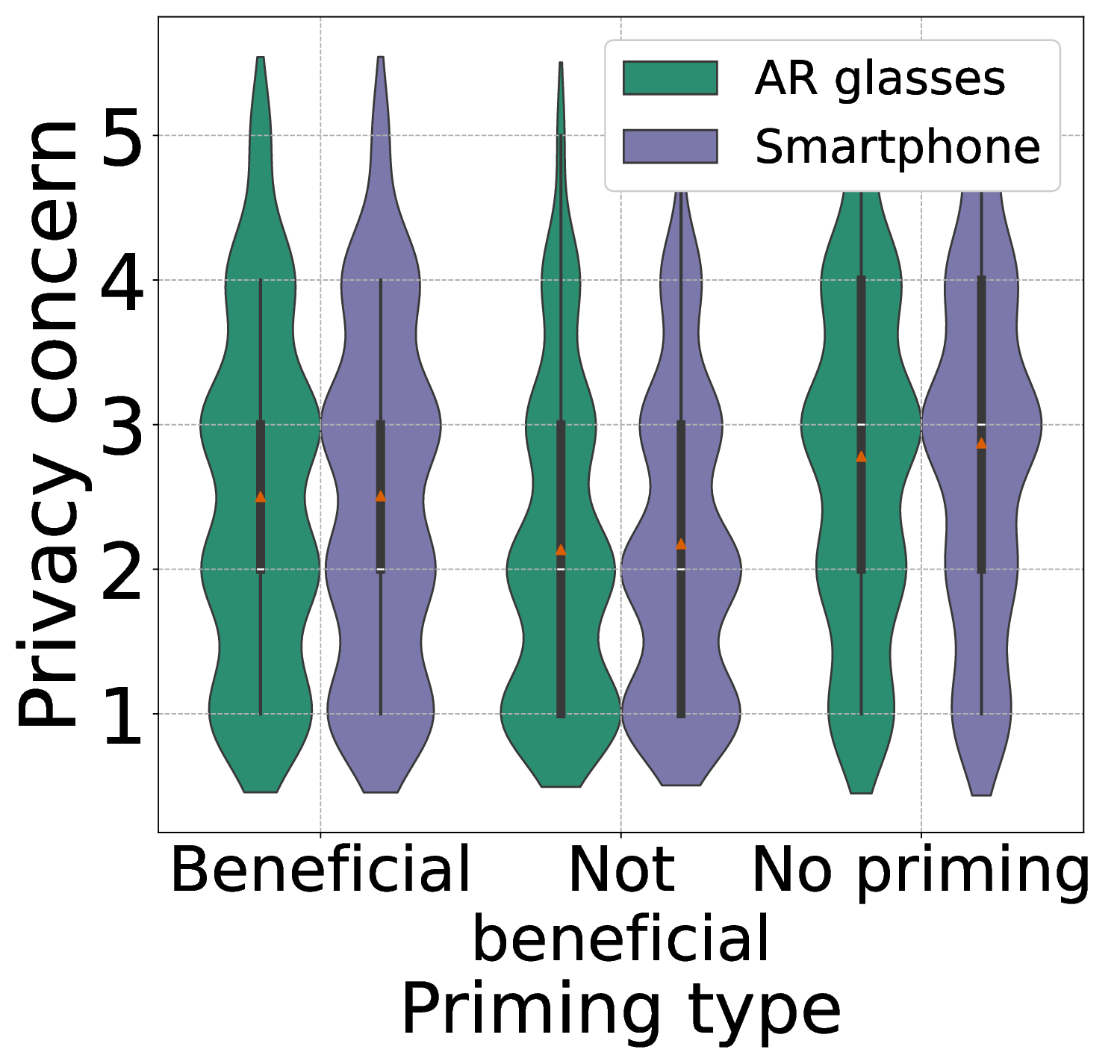}}
}
\subfigure[Heart condition.]{
    {\includegraphics[width=0.23\linewidth,keepaspectratio]{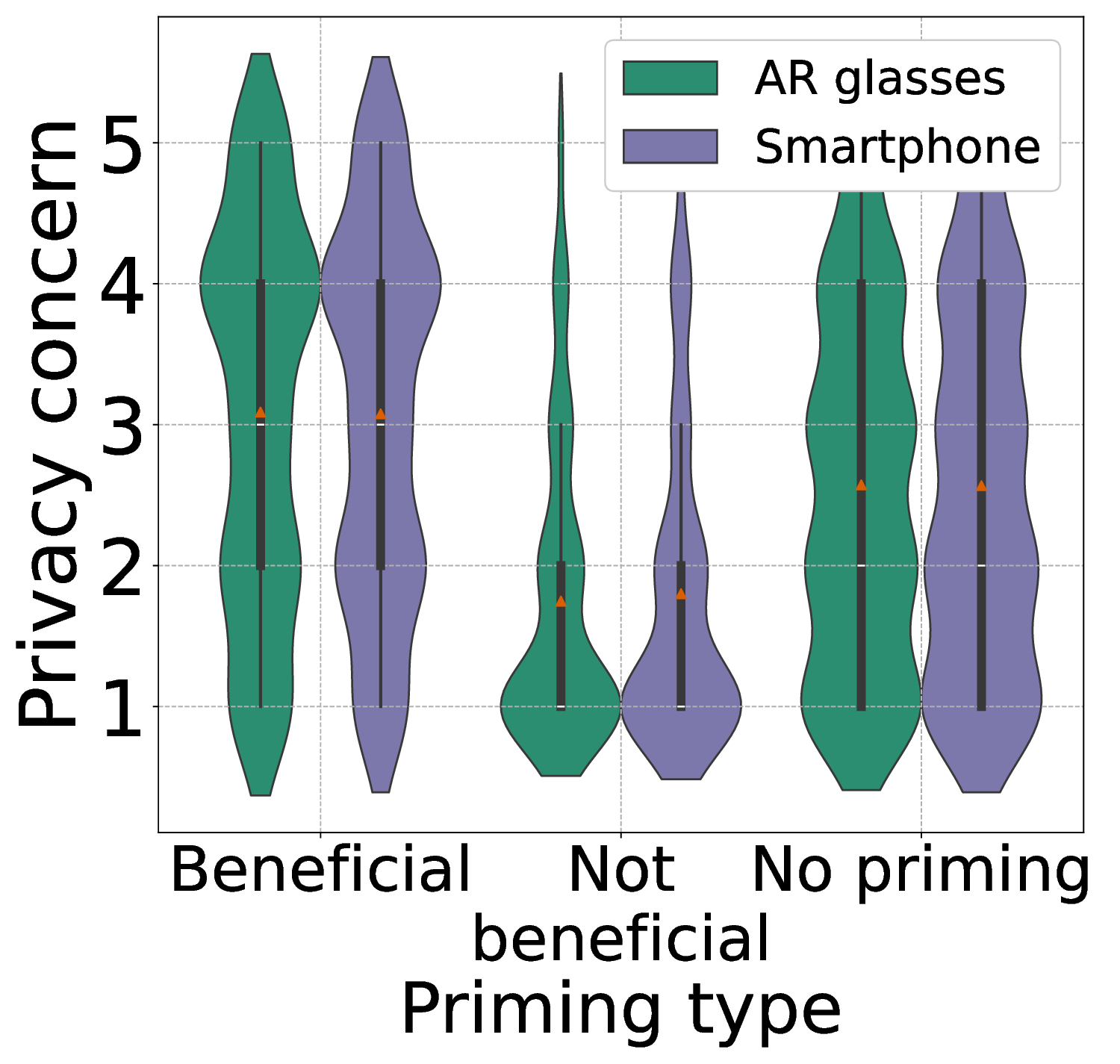}}
}
\subfigure[Location.]{
    {\includegraphics[width=0.23\linewidth,keepaspectratio]{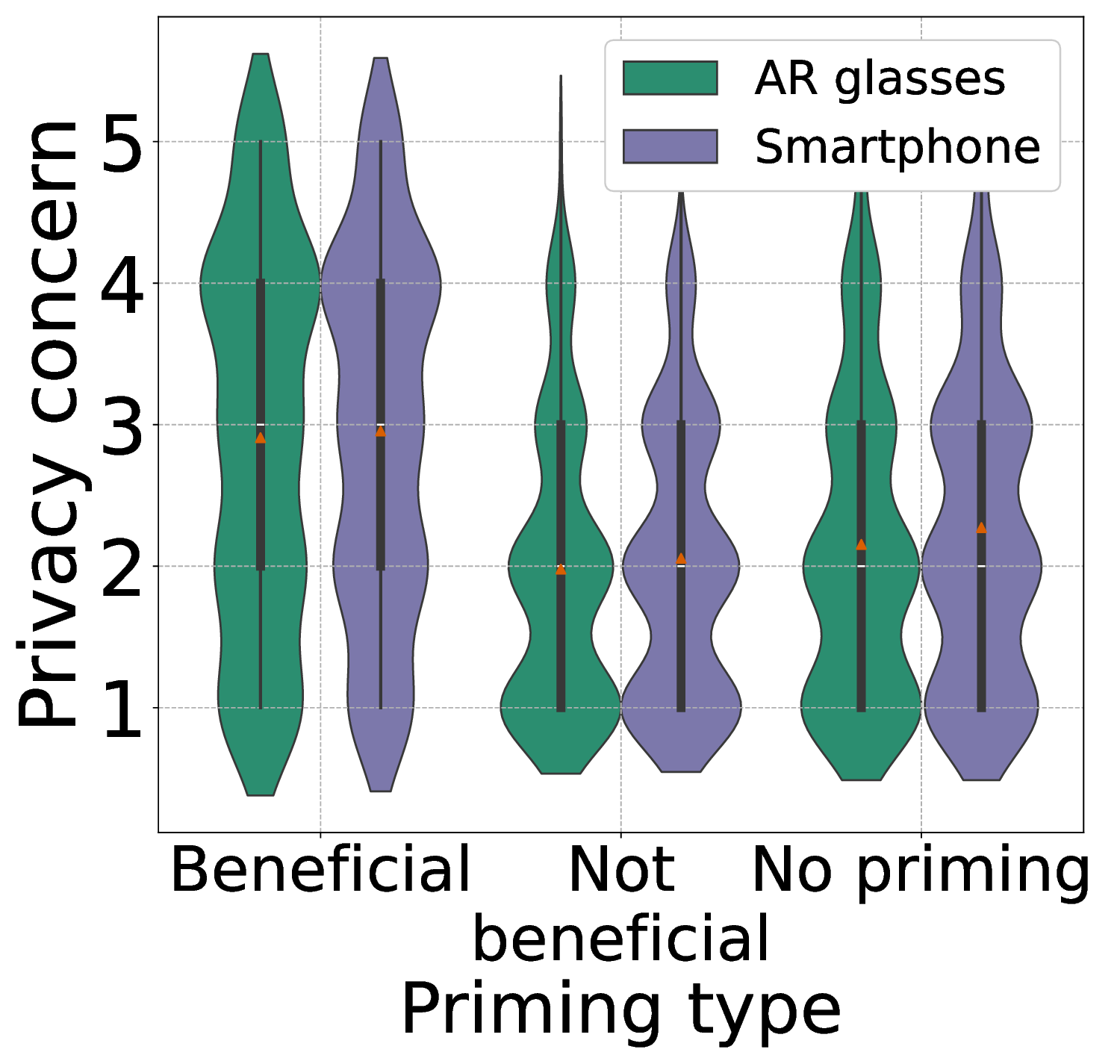}}
}
\subfigure[Personal identity.]{
    {\includegraphics[width=0.23\linewidth,keepaspectratio]{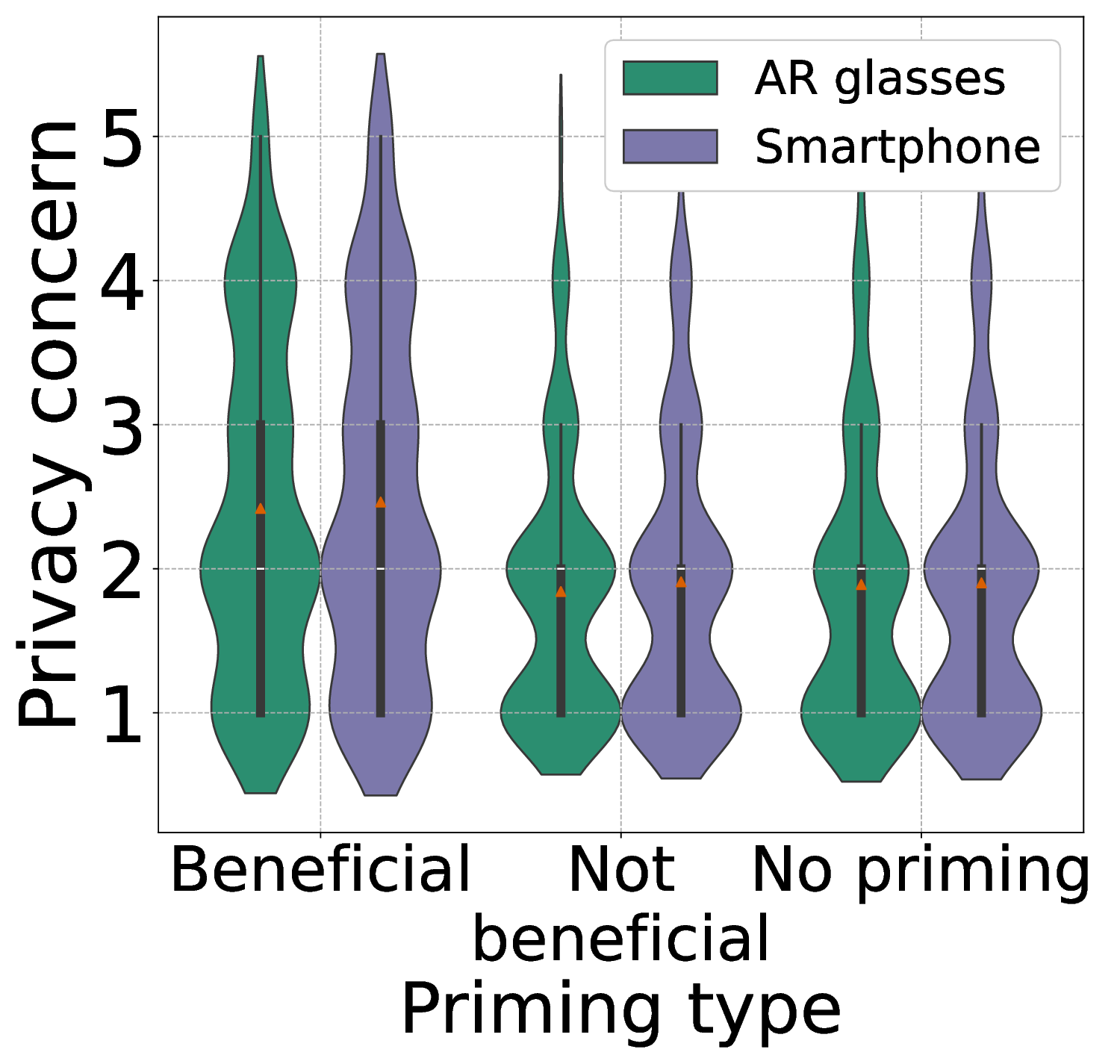}}
}
\subfigure[Sexual preference.]{
    {\includegraphics[width=0.23\linewidth,keepaspectratio]{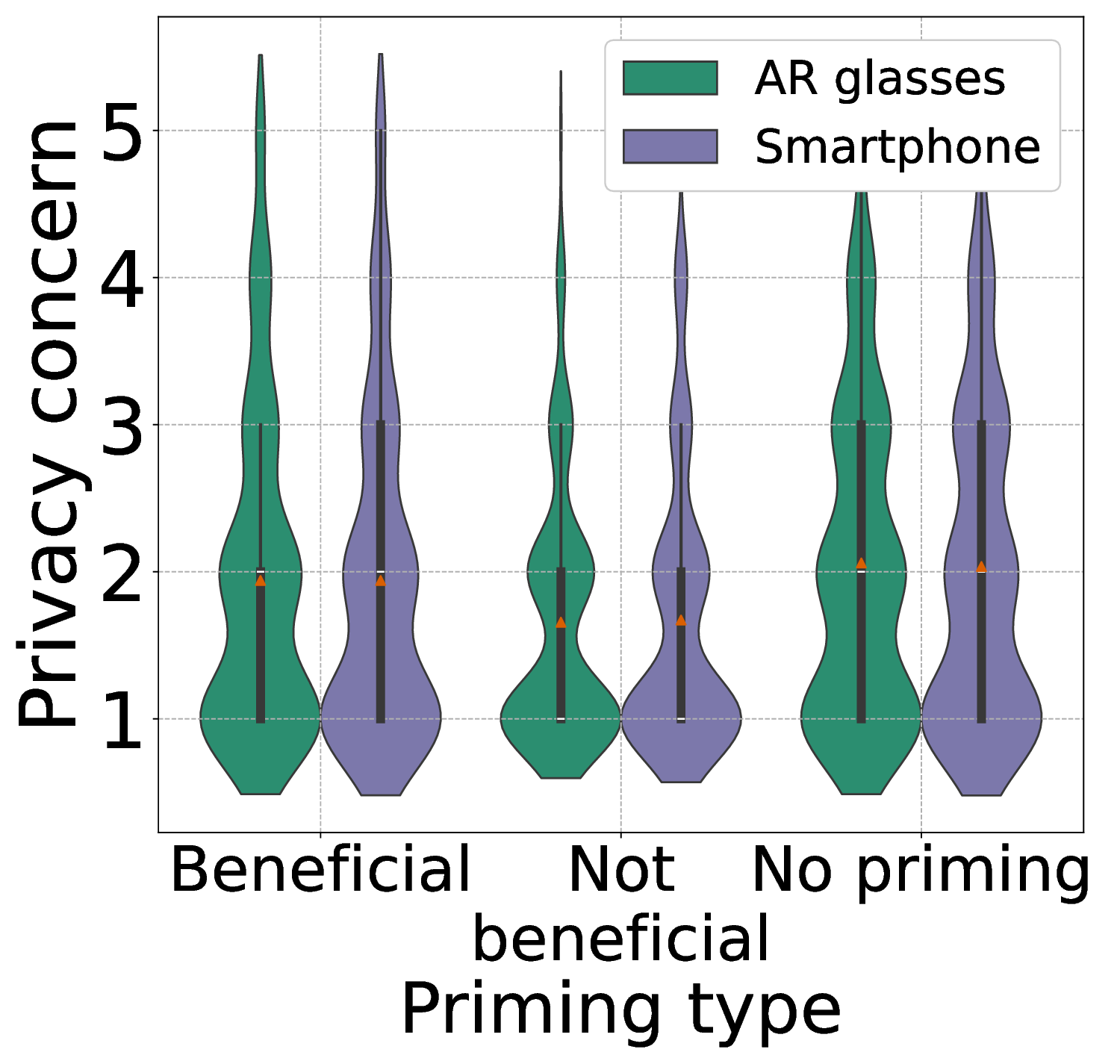}}
}
\subfigure[Stress.]{
    {\includegraphics[width=0.23\linewidth,keepaspectratio]{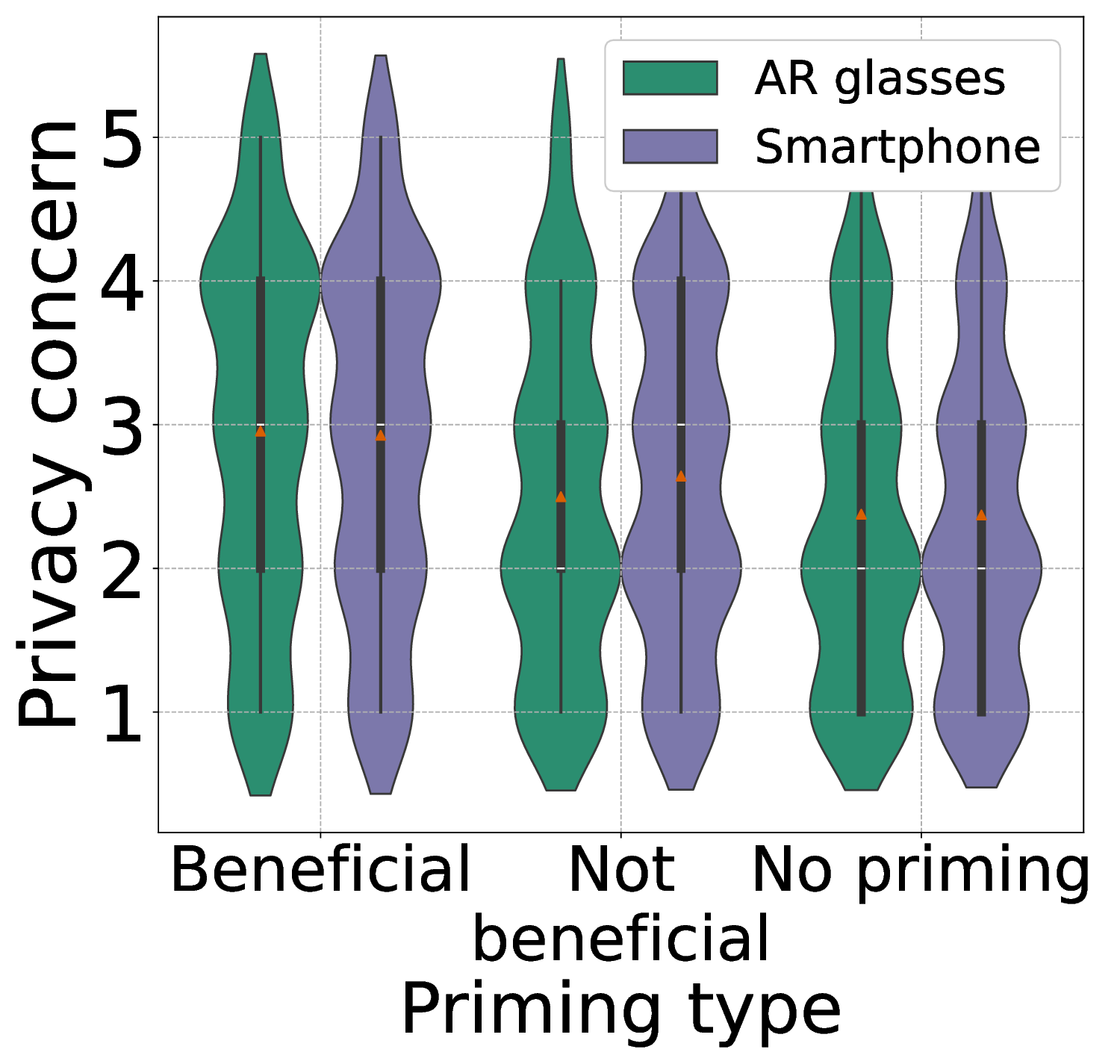}}
}
\caption{\label{fig_main_priming_results} Violin plots representing the relationship between privacy concerns and priming types for each user attribute.}
\alt{Violin plots representing the relationship between privacy concerns and priming types for our privacy concerns survey. Each subplot corresponds to an evaluated user attribute. The Y-axes in the plots represent privacy concerns ranging from 1 to 5, with 1 being very uncomfortable and 5 being very comfortable. The X-axes correspond to beneficial, not-beneficial, and no priming levels, each with data for AR glasses and smartphones.}
\end{figure*}

\begin{figure*}[ht!]
\centering
\subfigure[Activity.]{
    {\includegraphics[width=0.31\linewidth,keepaspectratio]{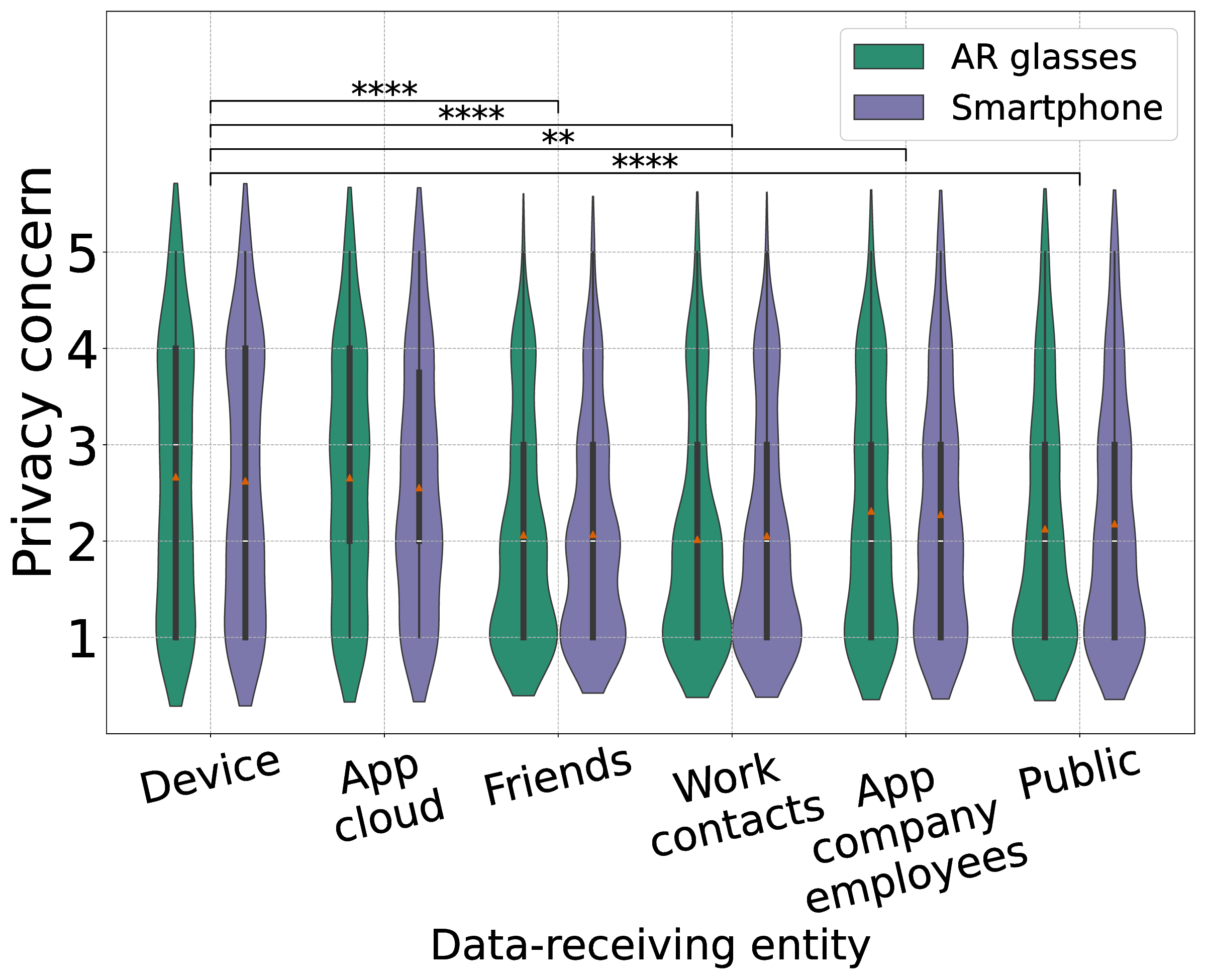}}
}
\subfigure[Alertness.]{
    {\includegraphics[width=0.31\linewidth,keepaspectratio]{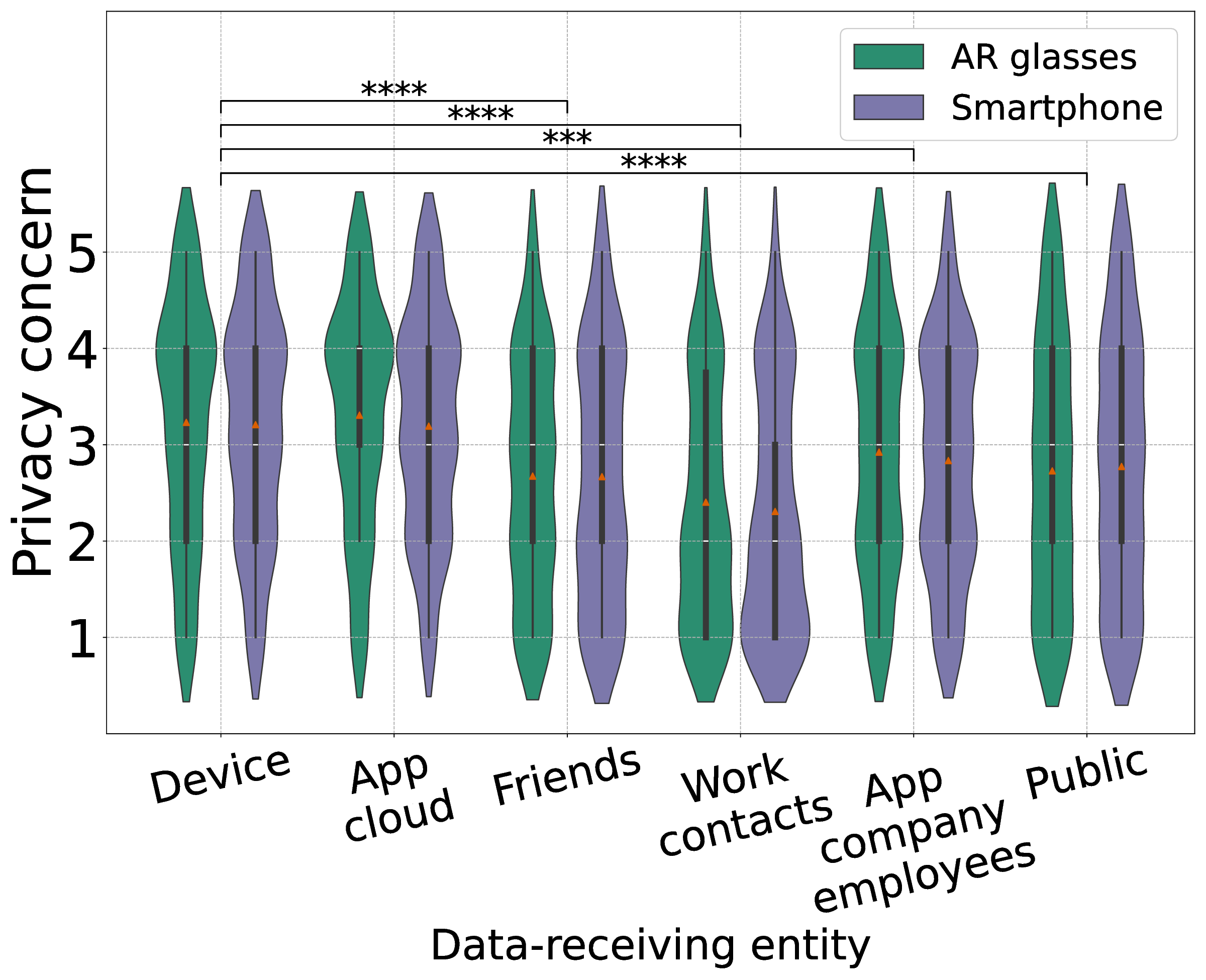}}
}
\subfigure[BMI.]{
    {\includegraphics[width=0.31\linewidth,keepaspectratio]{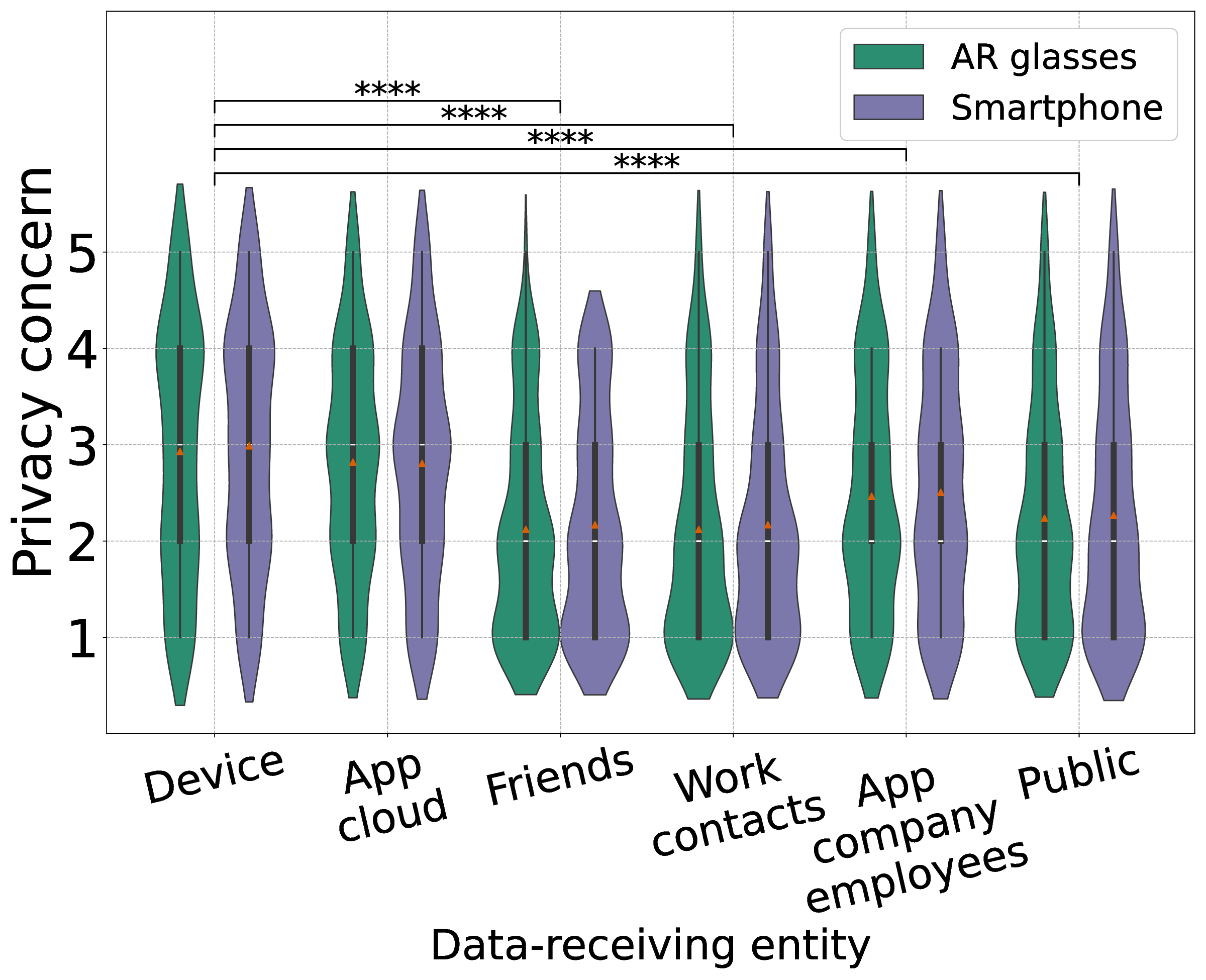}}
}
\subfigure[Cognitive load.]{
    {\includegraphics[width=0.31\linewidth,keepaspectratio]{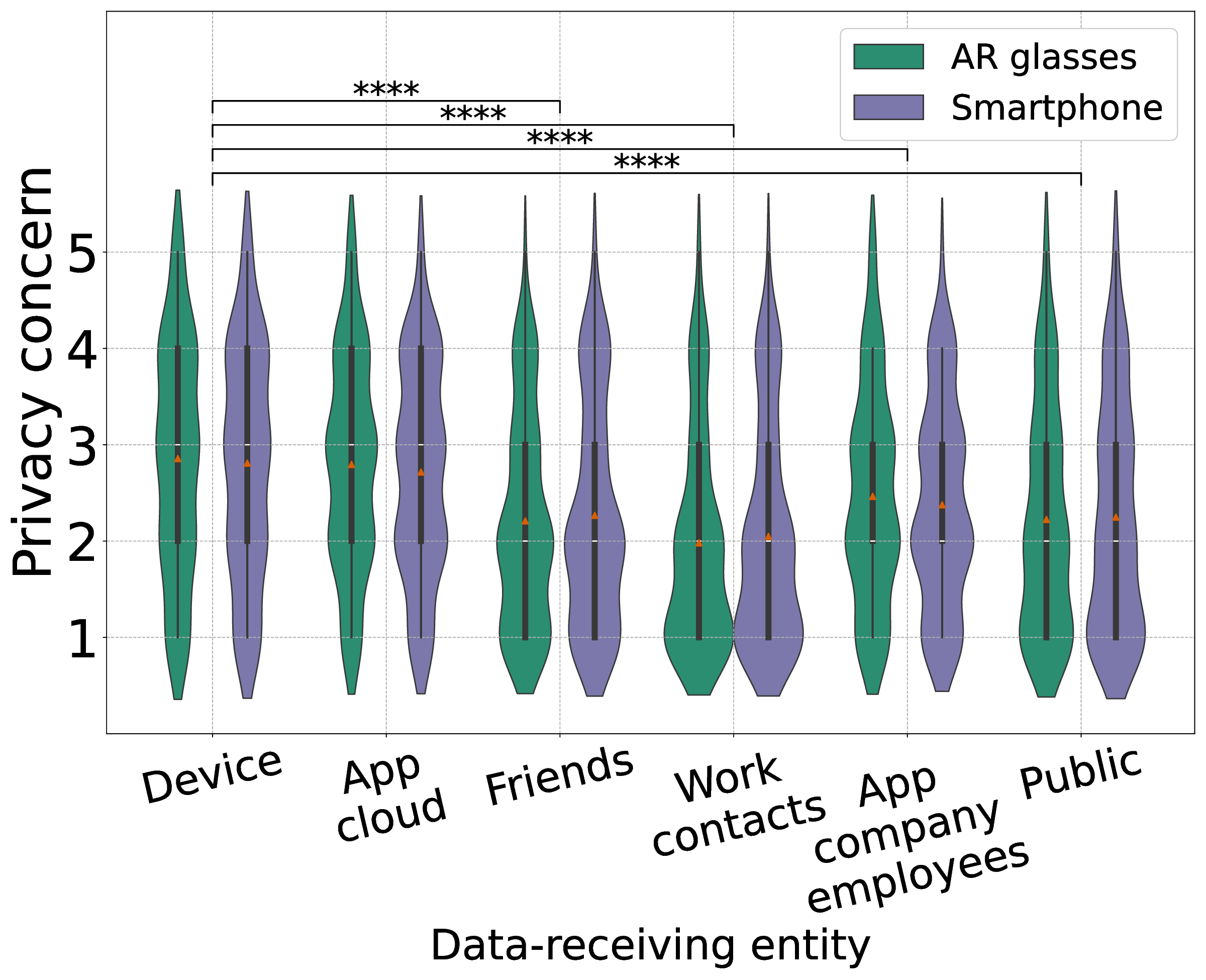}}
}
\subfigure[Depression.]{
    {\includegraphics[width=0.31\linewidth,keepaspectratio]{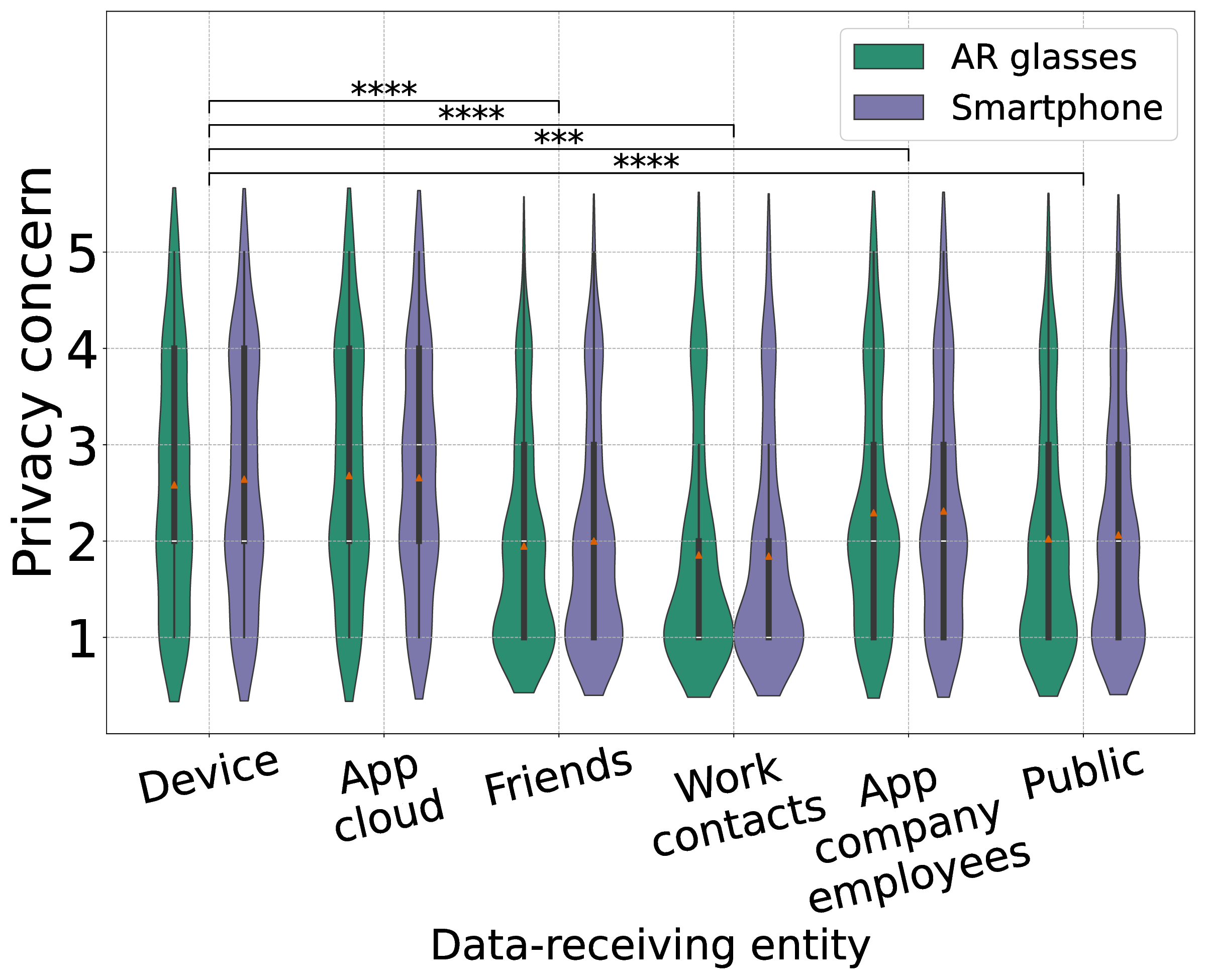}}
}
\subfigure[Gender.]{
    {\includegraphics[width=0.31\linewidth,keepaspectratio]{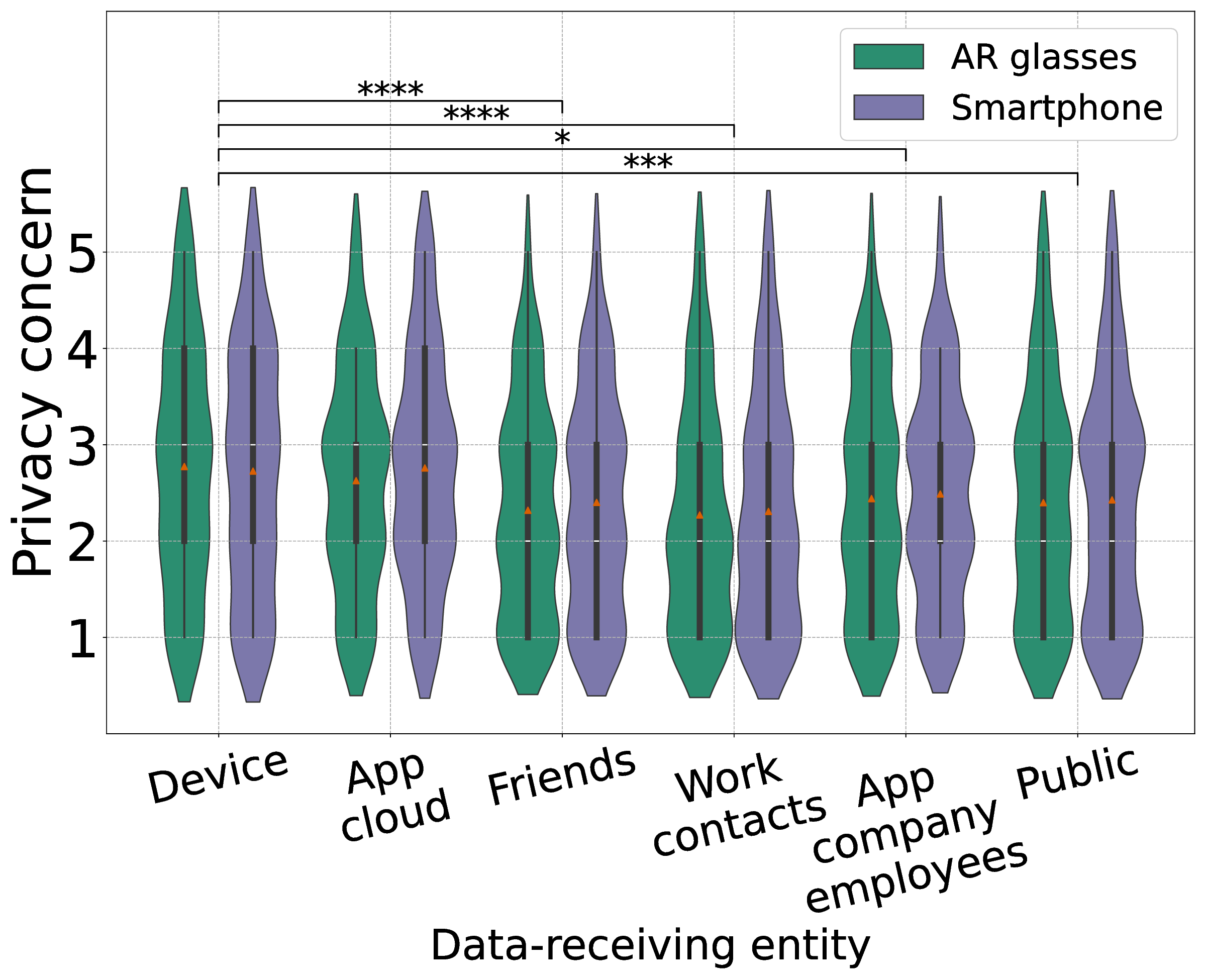}}
}
\subfigure[Heart condition.]{
    {\includegraphics[width=0.31\linewidth,keepaspectratio]{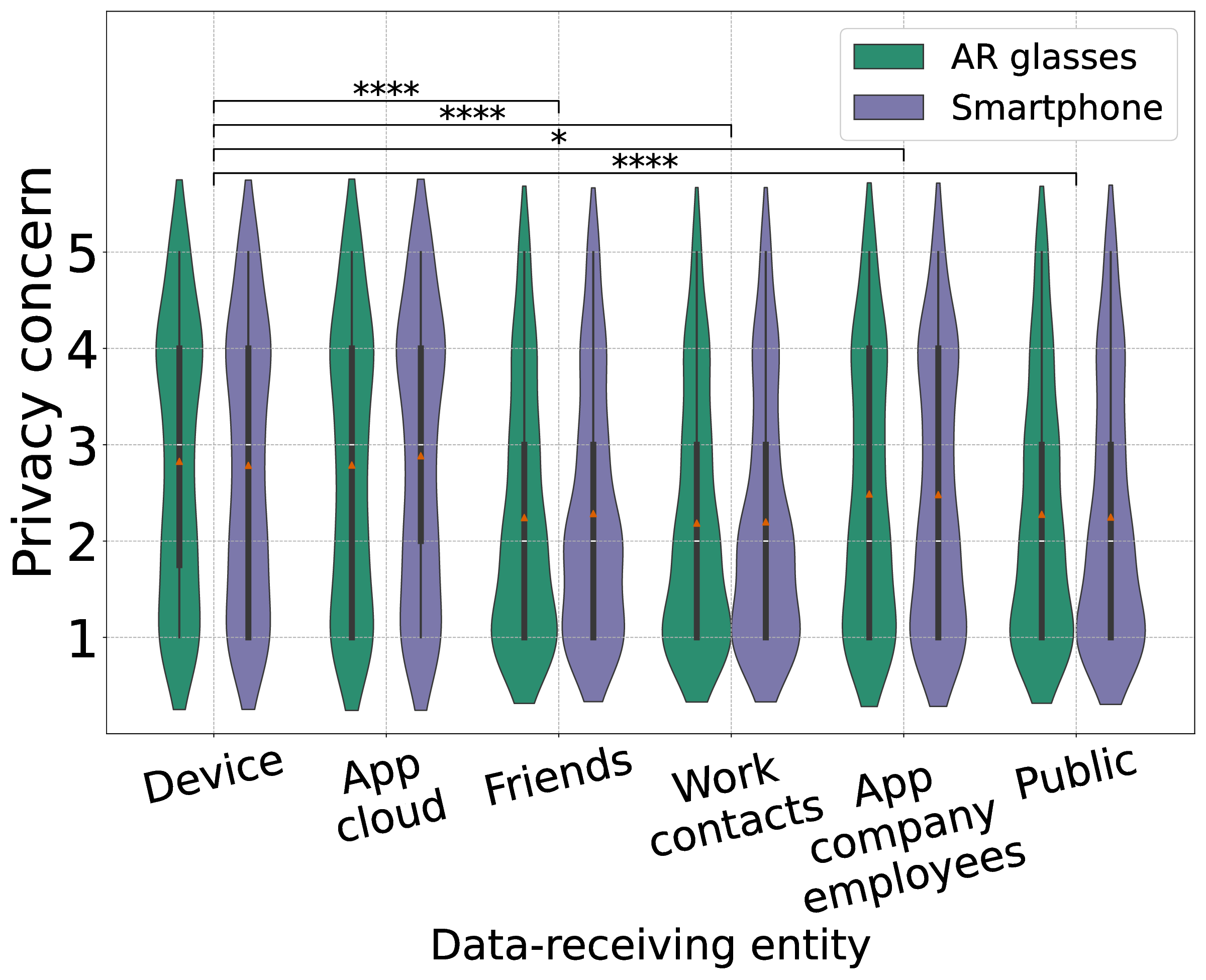}}
}
\subfigure[Location.]{
    {\includegraphics[width=0.31\linewidth,keepaspectratio]{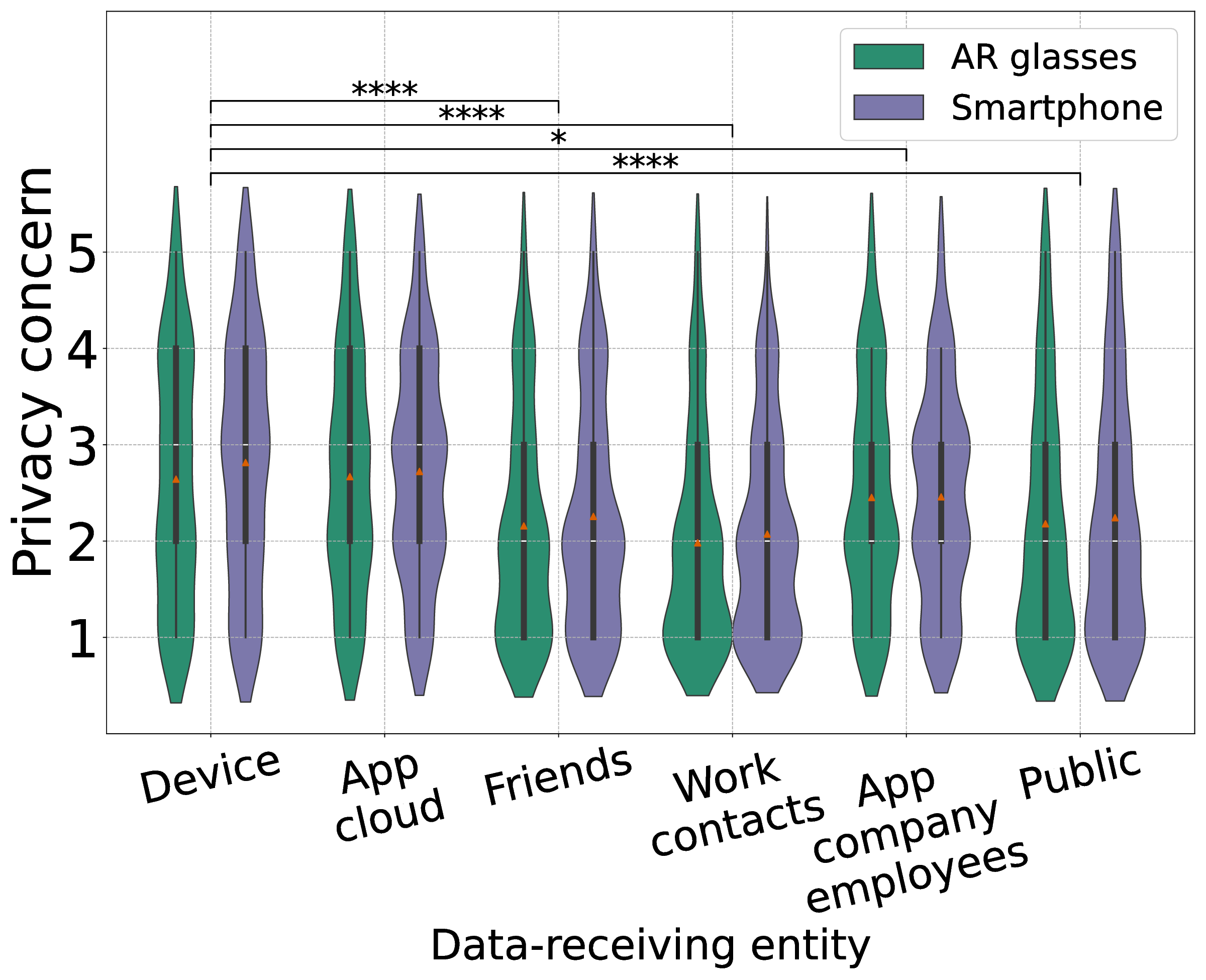}}
}
\subfigure[Personal identity.]{
    {\includegraphics[width=0.31\linewidth,keepaspectratio]{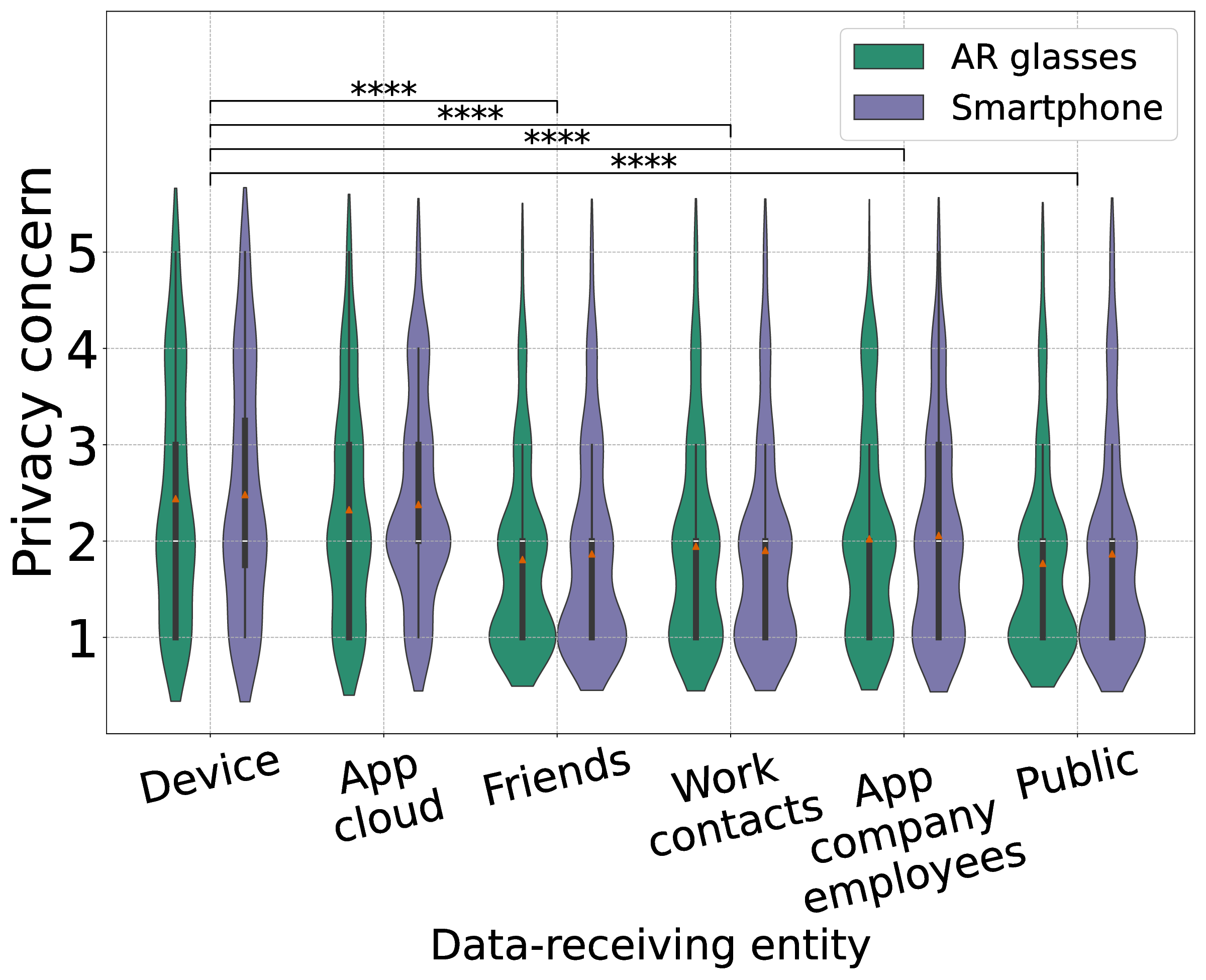}}
}
\subfigure[Sexual preference.]{
    {\includegraphics[width=0.31\linewidth,keepaspectratio]{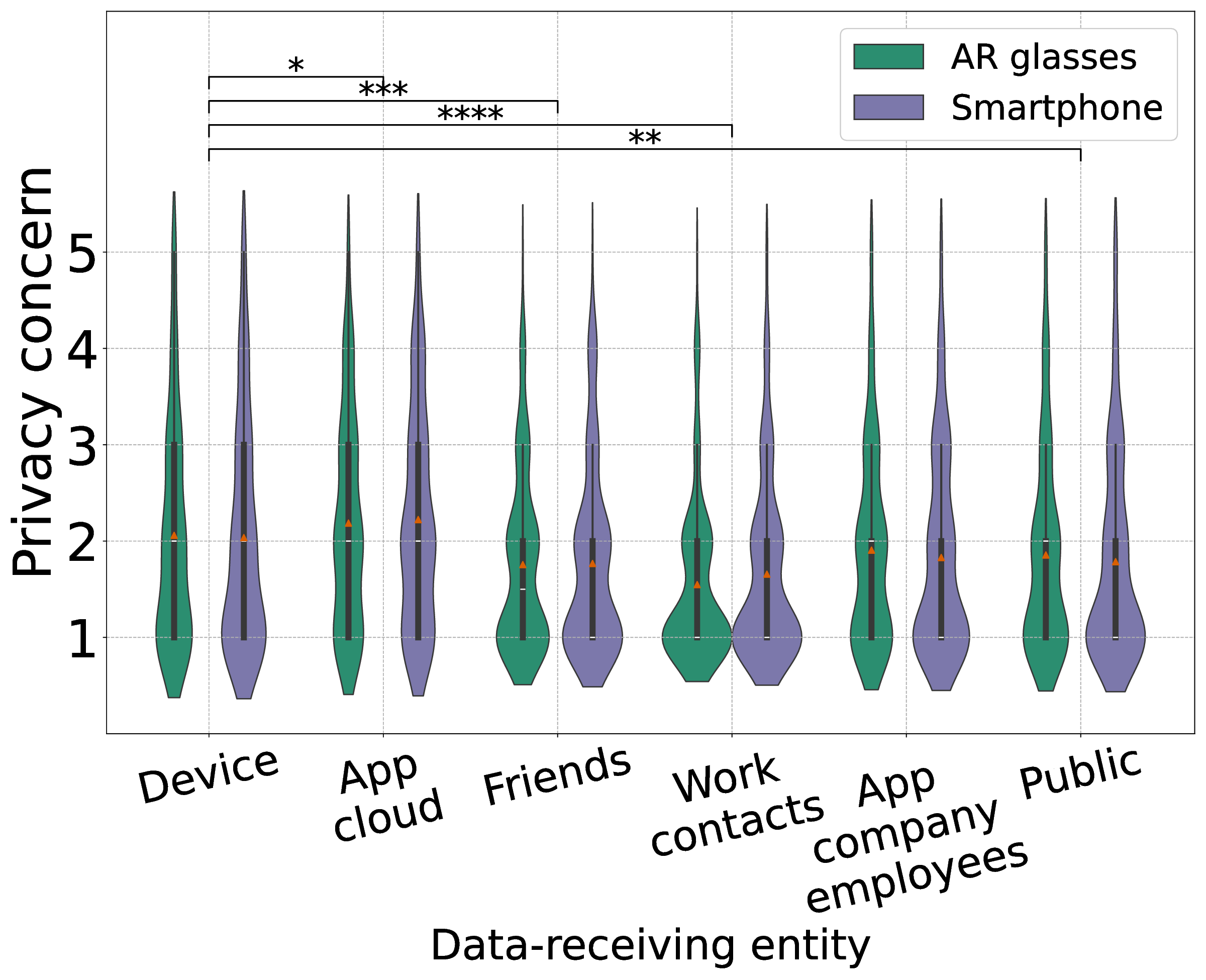}}
}
\subfigure[Stress.]{
    {\includegraphics[width=0.31\linewidth,keepaspectratio]{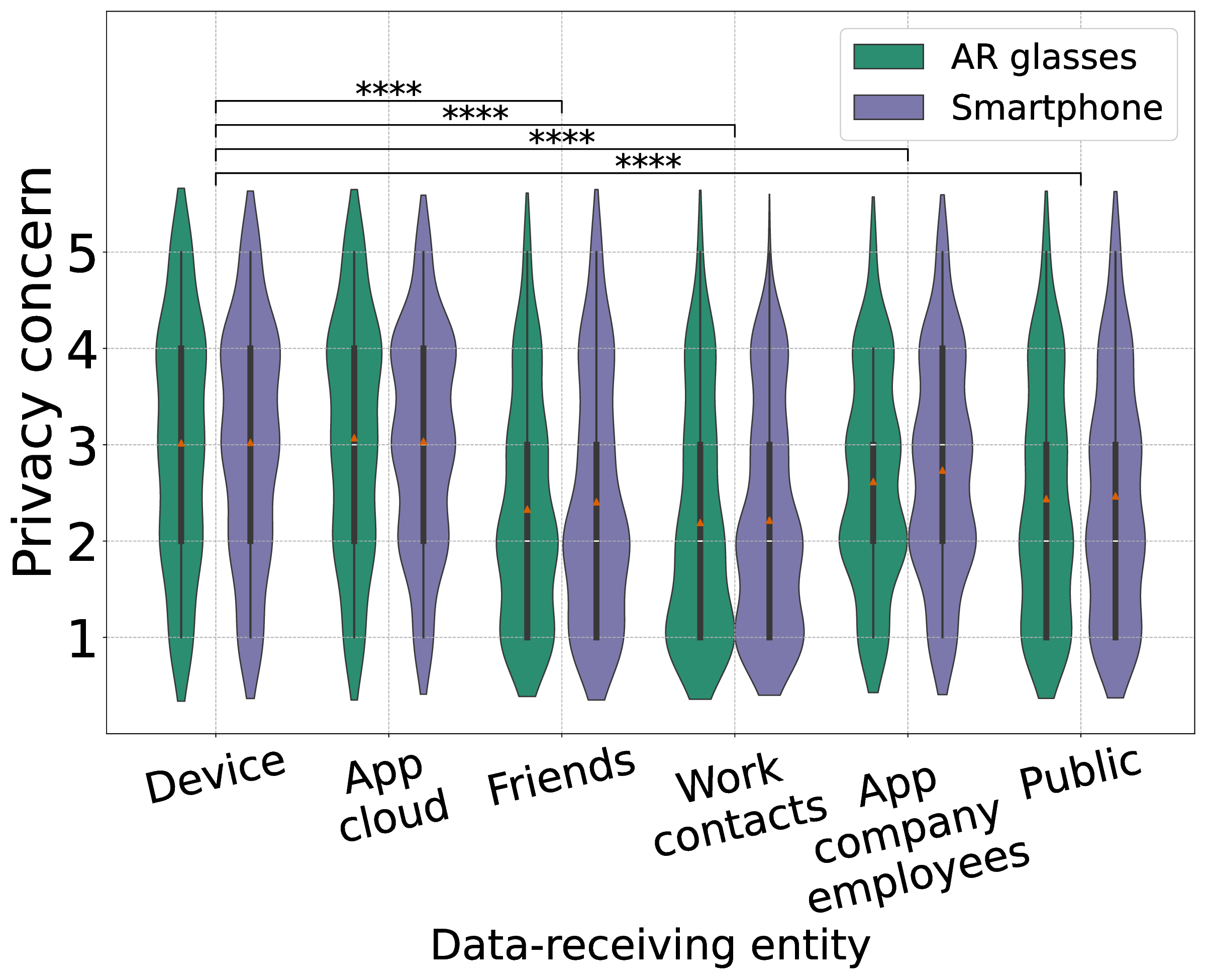}}
}
\caption{\label{fig_main_entity_results} Violin plots representing the relationship between privacy concerns and data-receiving entities for each user attribute.}
\alt{Violin plots representing the relationship between privacy concerns and data-receiving entities for our privacy concerns survey. Each subplot corresponds to an evaluated user attribute. The Y-axes in the plots represent privacy concerns ranging from 1 to 5, with 1 being very uncomfortable and 5 being very comfortable. The X-axes correspond to data for data-receiving entities, including the device itself, app cloud, friends, work contacts, app company employees, and the public, each with data for AR glasses and smartphones.}
\end{figure*}

We ran a regression analysis on each user attribute to explore privacy concerns. As participants had varying perceptions of beneficialness in the priming texts, we employed the mean values of beneficialness scores for each attribute according to country and device types separately as factors in the regressions. As there is no priming score for the no-priming condition since we did not provide those participants with priming texts, we used a value that is halfway between the beneficial and not-beneficial priming scores, as this value corresponds to ``neither beneficial nor harmful.'' We report our results in Figures~\ref{fig_main_priming_results},~\ref{fig_main_entity_results},~\ref{fig_main_retention_results}, and~\ref{fig_main_country_results} where privacy concerns range between 1 and 5, from ``very uncomfortable'' to ``very comfortable.'' 

\subsubsection{Priming type}
\label{subsubsec_priming}
We analyzed the relationship between priming and privacy concerns. Our regression analyses for all attributes on this factor constantly show that the more beneficially users are primed, the less concerned they are about their privacy ($p < .0001$). In addition, in the privacy concerns survey, the summary statistics of the beneficial, not beneficial, and no priming distributions are analogous to those in calibration surveys, especially regarding the order of these distributions. We report these relationships in Figure~\ref{fig_main_priming_results}. 

\subsubsection{Device type}
\label{subsubsec_devicetype}
\begin{figure*}[t!]
\centering
\subfigure[Activity.]{
    {\includegraphics[width=0.23\linewidth,keepaspectratio]{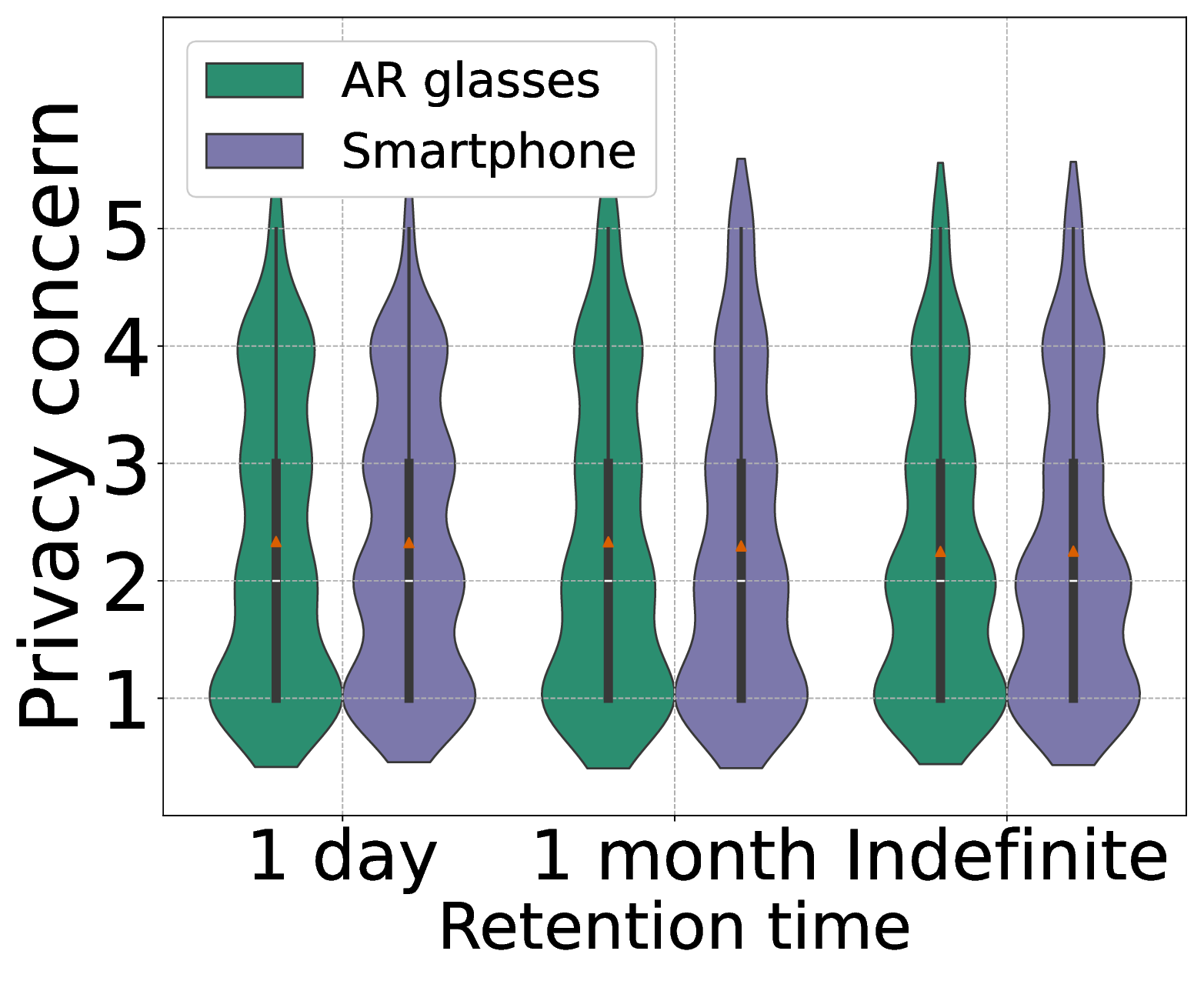}}
}
\subfigure[Alertness.]{
    {\includegraphics[width=0.23\linewidth,keepaspectratio]{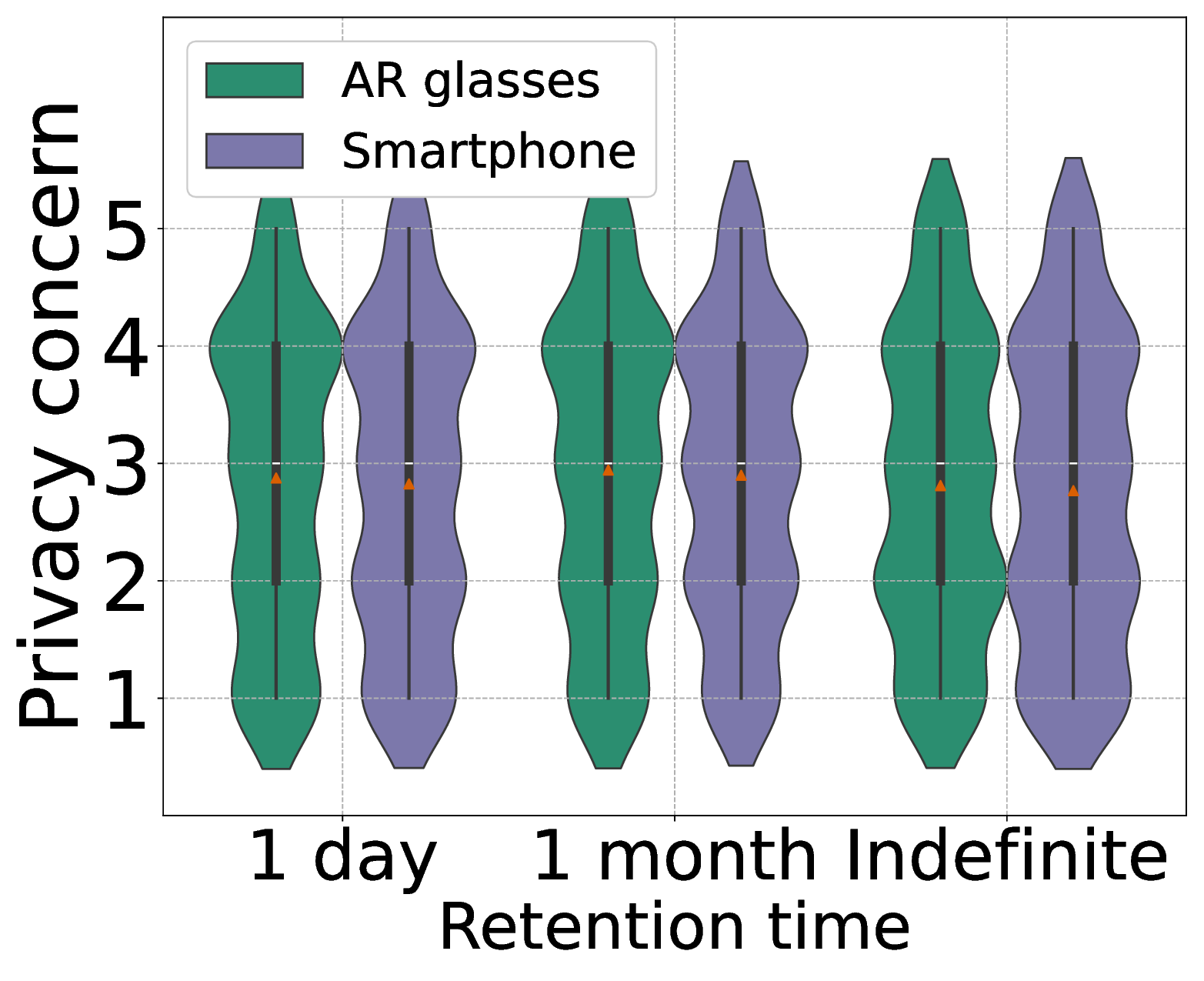}}
}
\subfigure[BMI.]{
    {\includegraphics[width=0.23\linewidth,keepaspectratio]{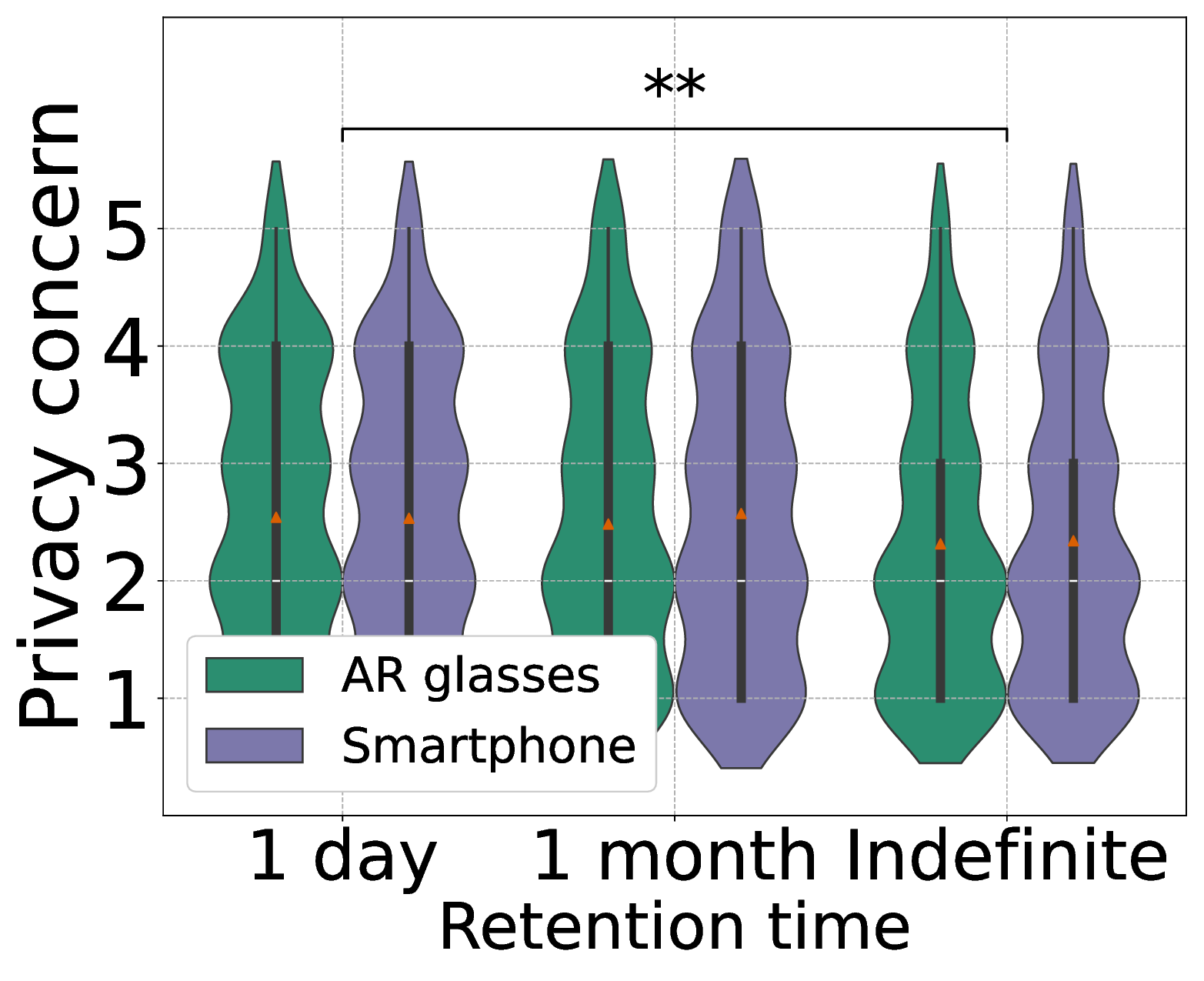}}
}
\subfigure[Cognitive load.]{
    {\includegraphics[width=0.23\linewidth,keepaspectratio]{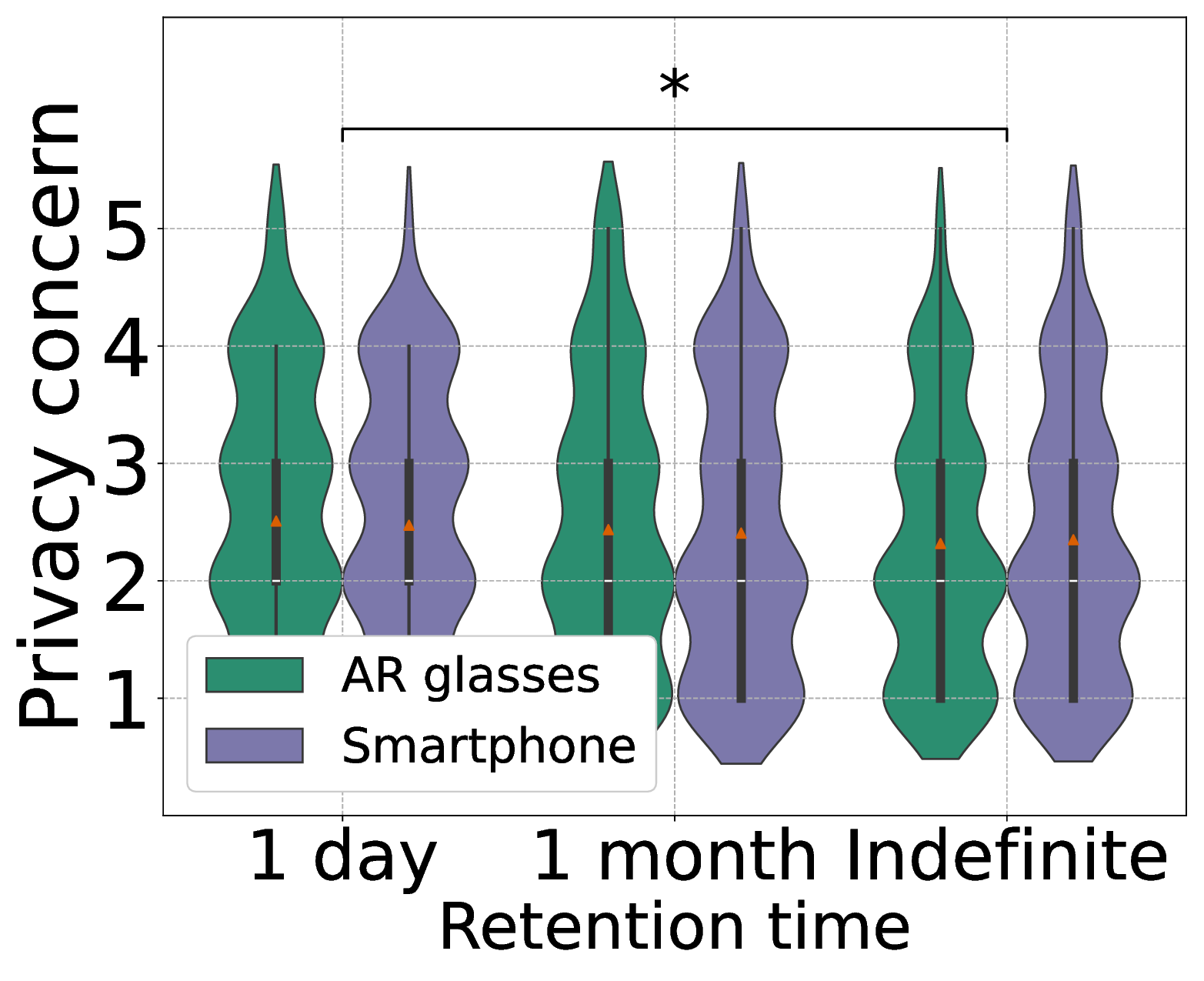}}
}
\subfigure[Depression.]{
    {\includegraphics[width=0.23\linewidth,keepaspectratio]{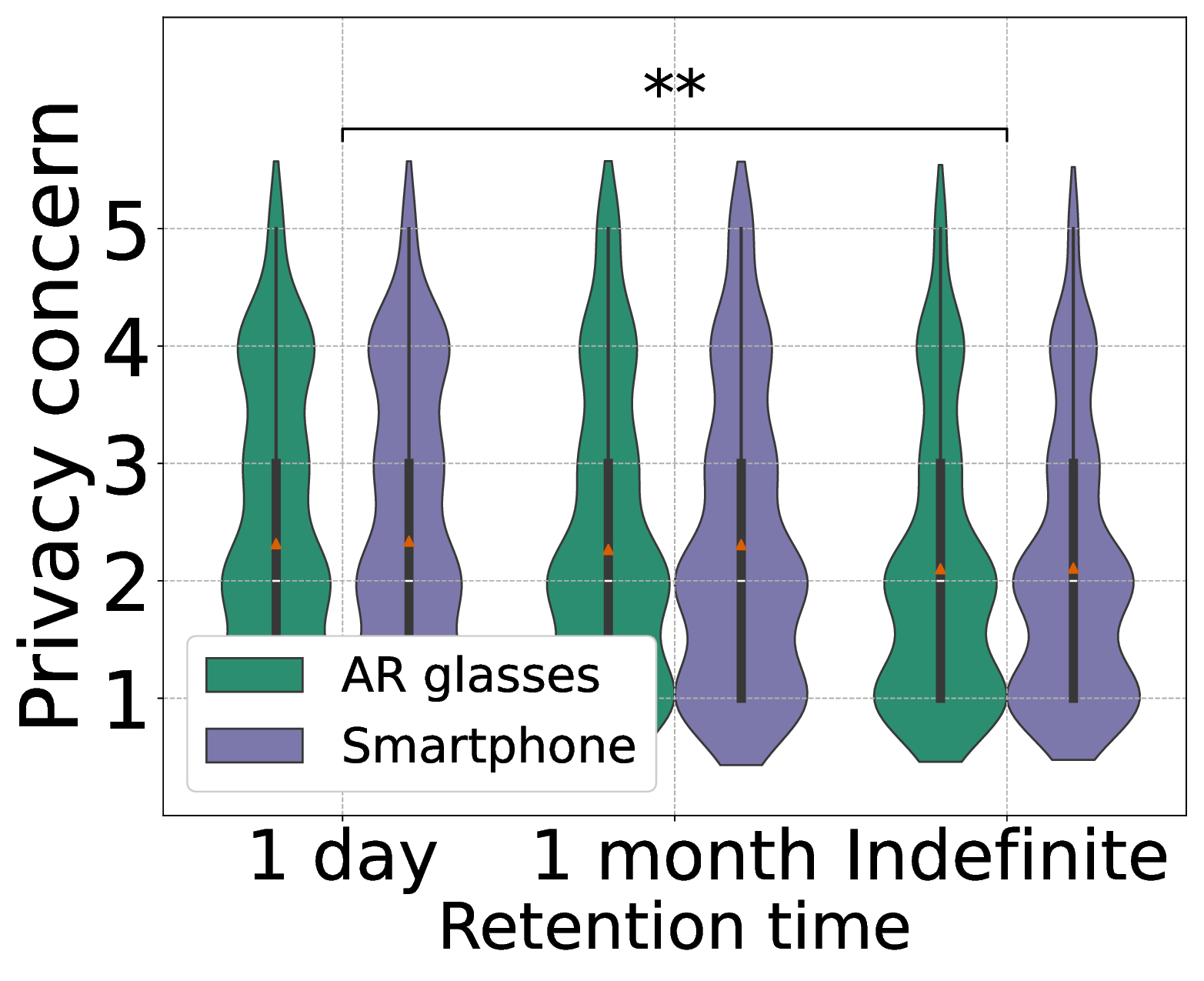}}
}
\subfigure[Gender.]{
    {\includegraphics[width=0.23\linewidth,keepaspectratio]{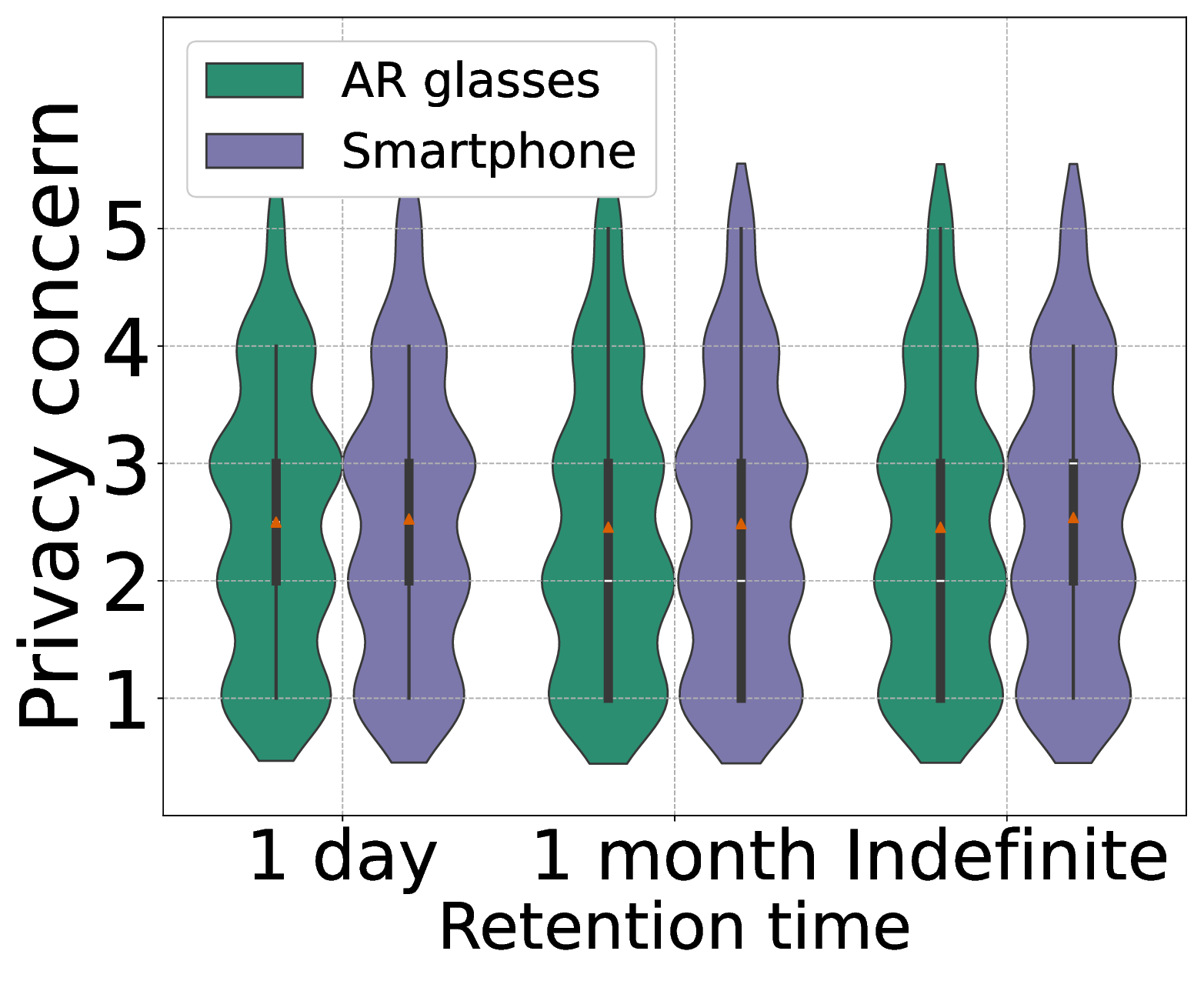}}
}
\subfigure[Heart condition.]{
    {\includegraphics[width=0.23\linewidth,keepaspectratio]{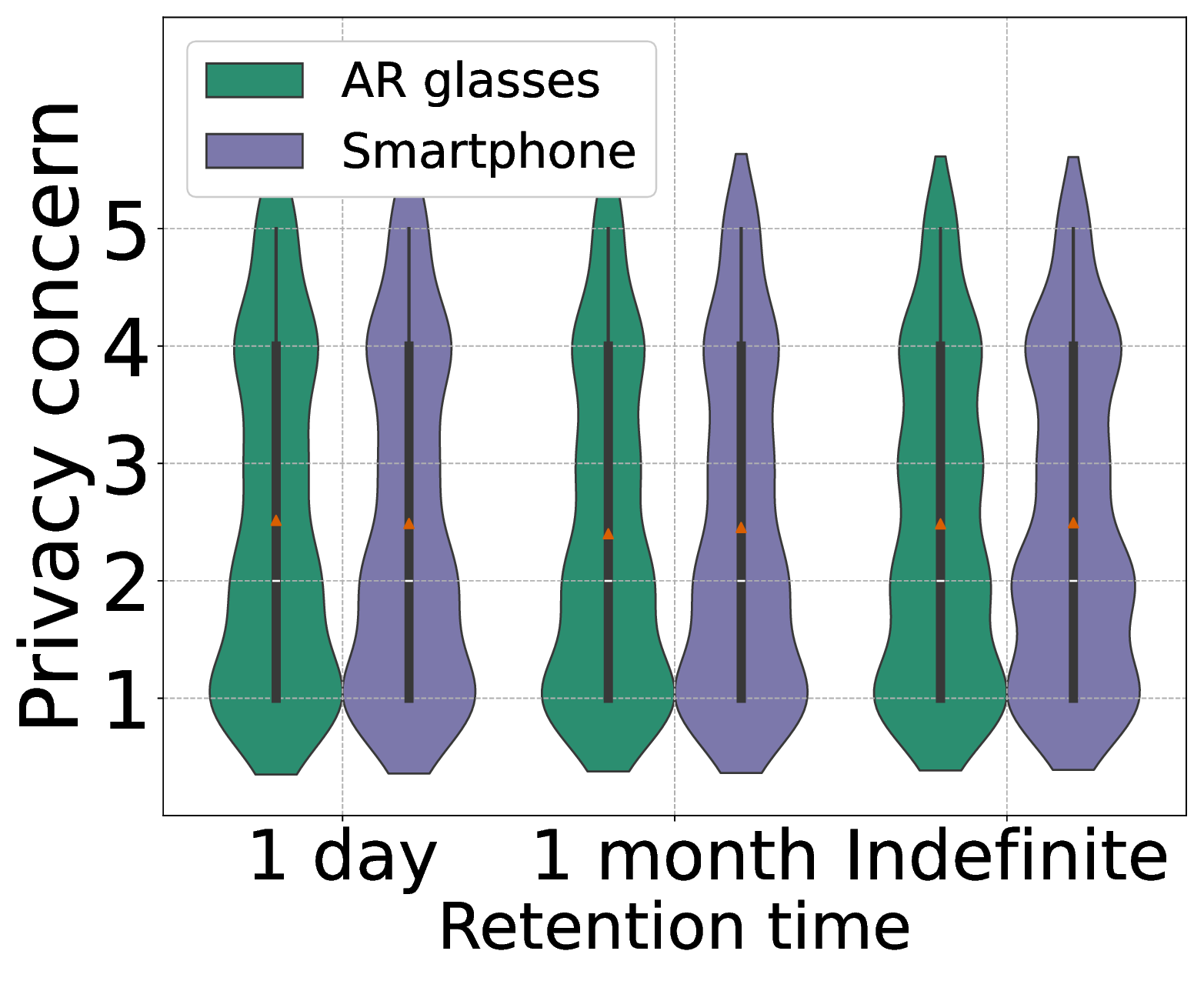}}
}
\subfigure[Location.]{
    {\includegraphics[width=0.23\linewidth,keepaspectratio]{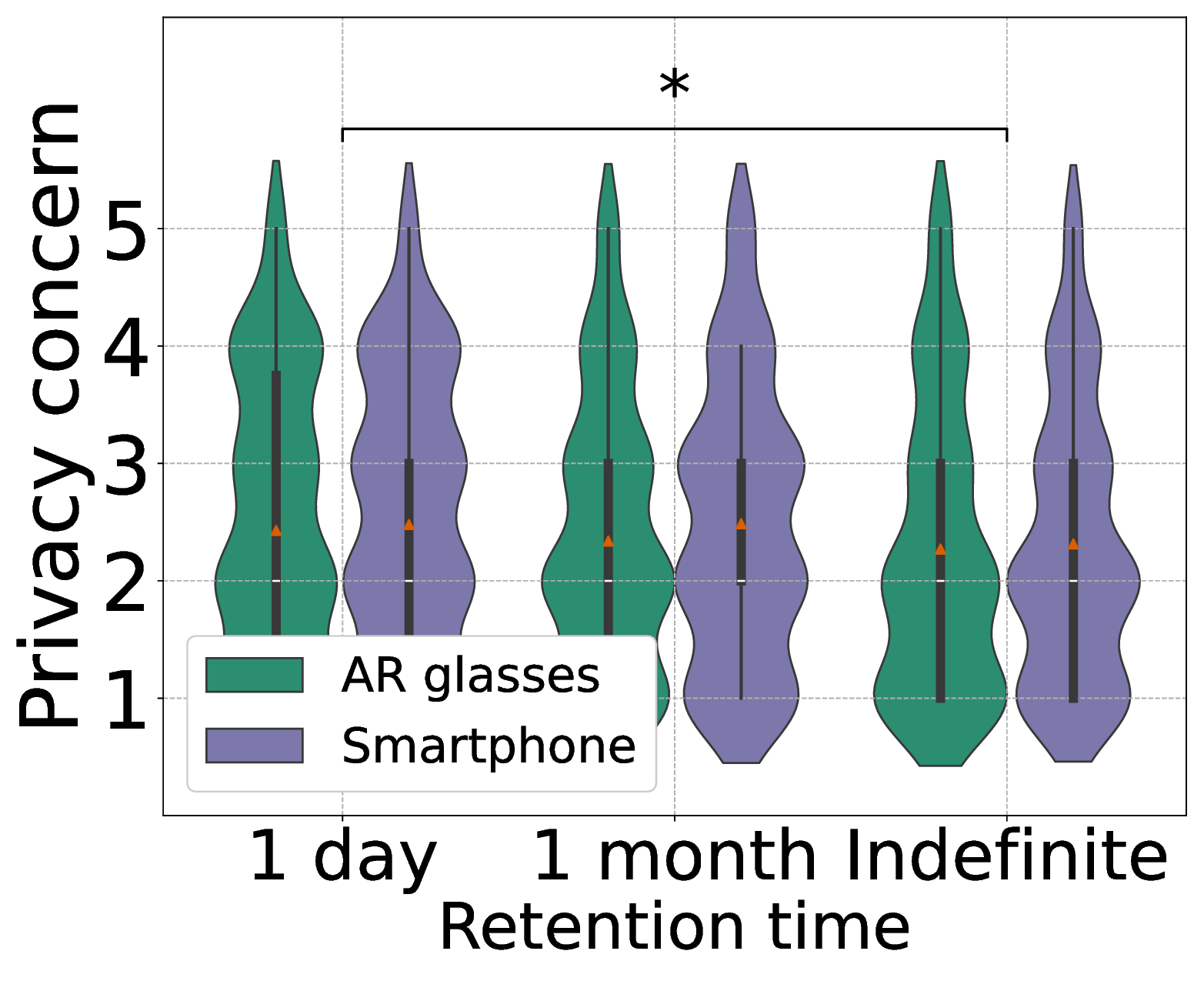}}
}
\subfigure[Personal identity.]{
    {\includegraphics[width=0.23\linewidth,keepaspectratio]{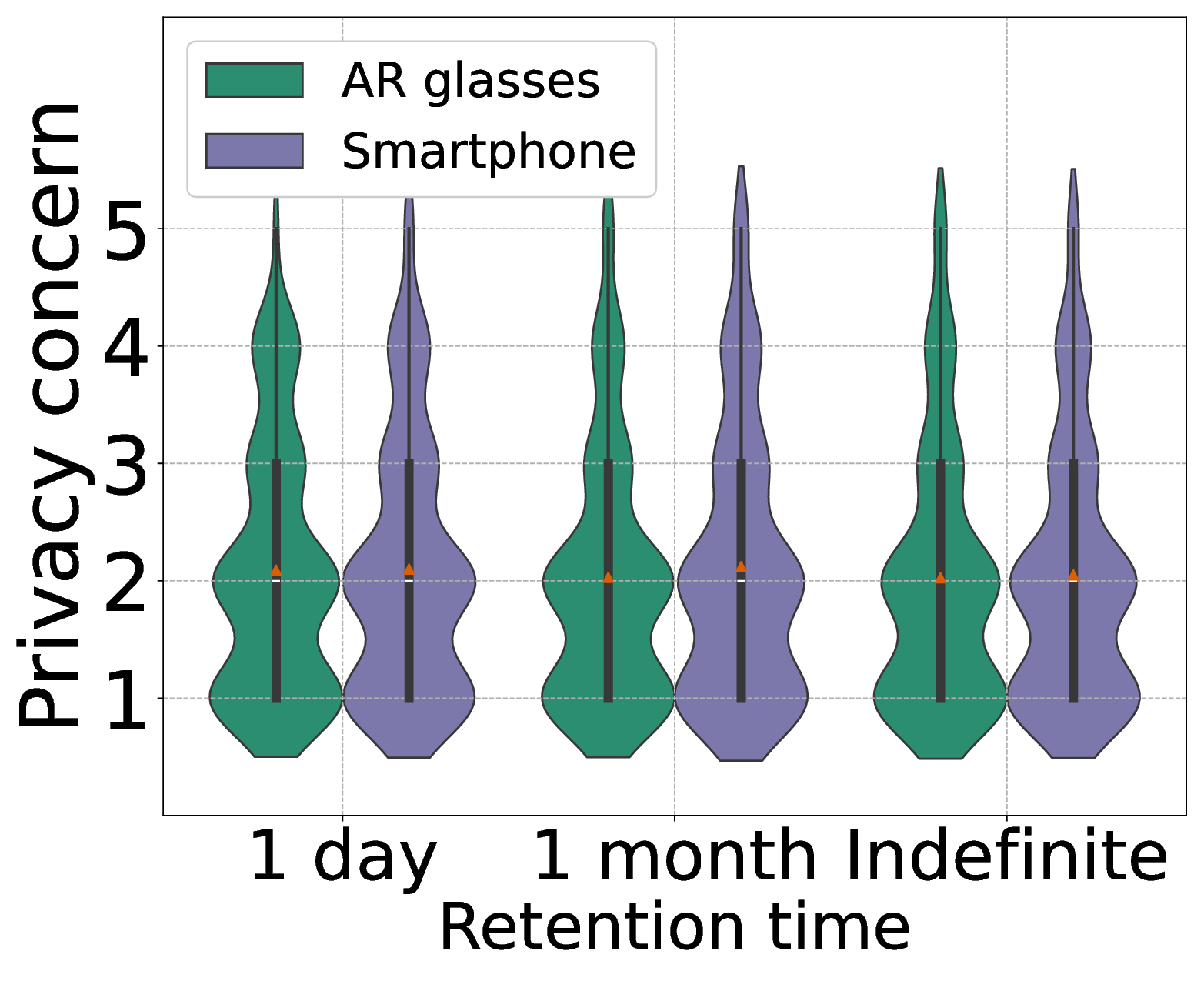}}
}
\subfigure[Sexual preference.]{
    {\includegraphics[width=0.23\linewidth,keepaspectratio]{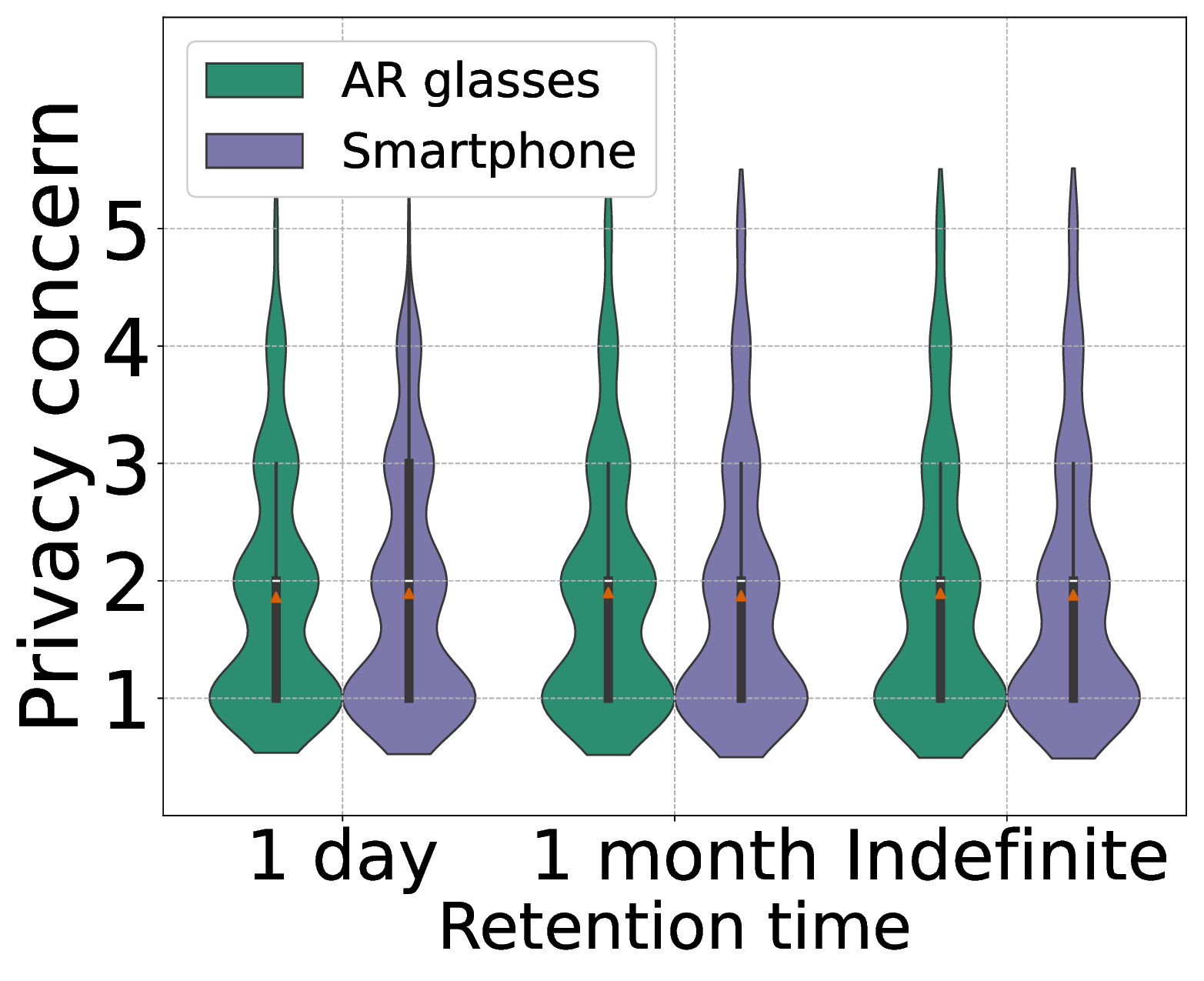}}
}
\subfigure[Stress.]{
    {\includegraphics[width=0.23\linewidth,keepaspectratio]{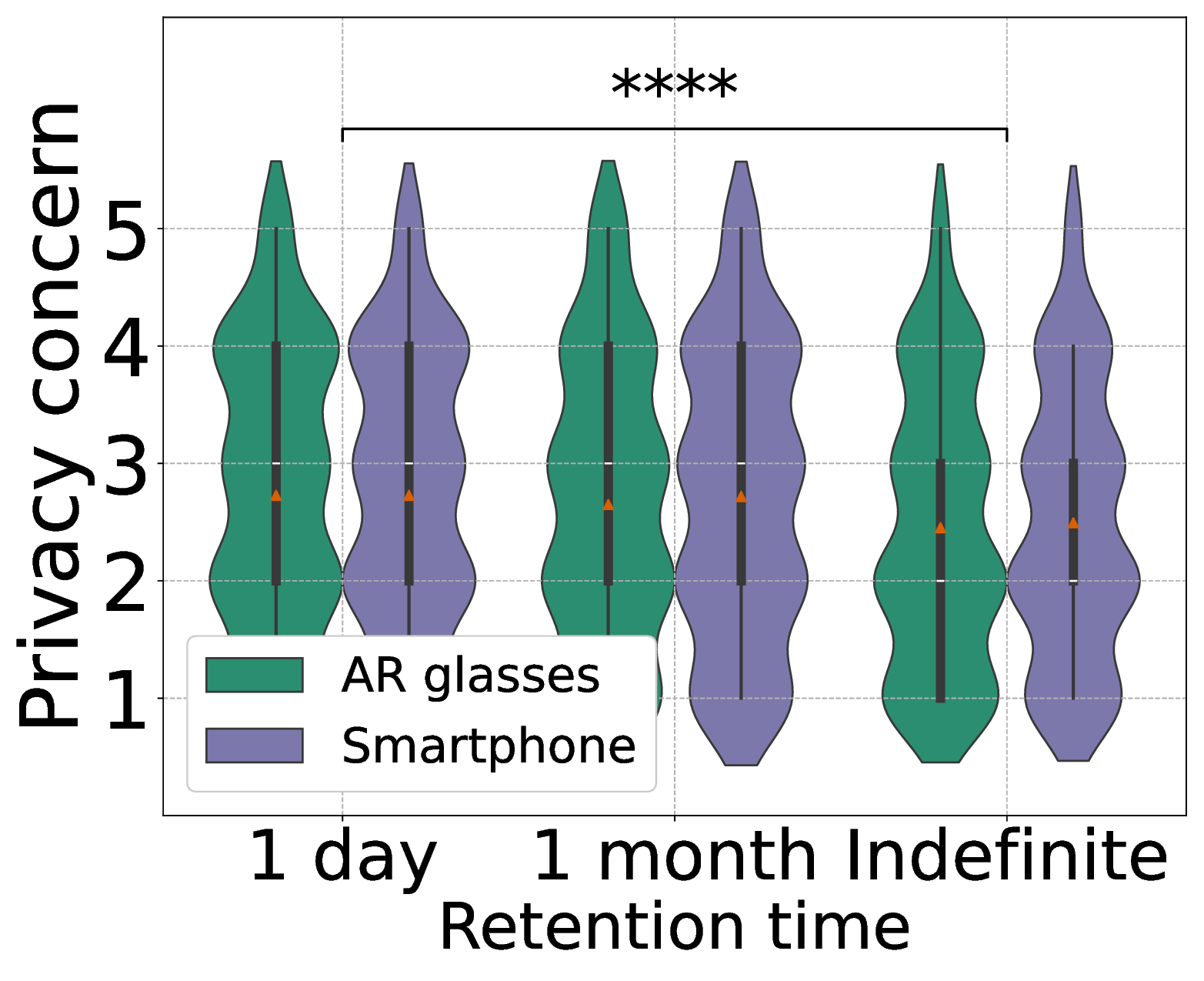}}
}
\caption{\label{fig_main_retention_results} Violin plots representing the relationship between privacy concerns and data retention times for each user attribute.}
\alt{Violin plots representing the relationship between privacy concerns and data retention times for our privacy concerns survey. Each subplot corresponds to an evaluated user attribute. The Y-axes in the plots represent privacy concerns ranging from 1 to 5, with 1 being very uncomfortable and 5 being very comfortable. The X-axes correspond to data for retention times of one day, one month, and indefinite, each with data for AR glasses and smartphones.}
\end{figure*}

To determine the relationship between AR device type and privacy concerns, we analyzed whether privacy concerns are influenced by whether the scenarios involve AR glasses or smartphones. We did not find evidence that privacy concerns are affected by the device type for any attribute ($p > .05$). We provide the relationship between privacy concerns and factors, including priming type, data-receiving entity, data retention times, and countries in AR glasses and smartphones, in Figures~\ref{fig_main_priming_results},~\ref{fig_main_entity_results},~\ref{fig_main_retention_results}, and~\ref{fig_main_country_results}, respectively.

\subsubsection{Data-receiving entity}
\label{subsubsec_datareceiving}
We analyzed the data-receiving entity factor to understand how data receivers affect user privacy concerns. This factor includes four conditions where humans can consume data (friends, work contacts, personal assistant app company employees, and the public) and two conditions where computers can consume data (the app cloud and the device). For the data remains in the device condition, as we did not explicitly tell participants about data sharing or that their device keeps the data, we assumed that the participants would view that as no data sharing happens. These conditions summarize the data-receiving entity condition in two broader aspects, namely, the storage (i.e., local and cloud) and human consumption. 

Our analyses show that participants have fewer privacy concerns when data remains in their devices than when humans can consume it. For instance, for the device and friends comparison, all the attributes yielded significant differences in the level of ($p < .0001$) apart from the sexual preference attribute ($p < .001$). For the device and work contacts comparison, we found that all the attributes yielded comparable significant differences ($p < .0001$). Furthermore, the results of the comparison between the device and the app company employees conditions showed a range of statistically meaningful insights for gender, heart condition, and location ($p < .05$), activity ($p < .01$), alertness and depression ($p < .001$), and BMI, cognitive load, personal identity, and stress ($p < .0001$). Similarly, our comparisons between the device and the public revealed significant differences mostly with ($p < .0001$), except for gender ($p < .001$) and sexual preference ($p < .01$). While we did not find statistically significant trends in every comparison for sexual preference between the local device and other humans, the trend is also similar, and participants are strongly concerned about privacy for this attribute, regardless of who accesses the data. Considering the lack of statistical significance (i.e., $p > .05$) between the local device and app cloud in privacy concerns, except for the sexual preference attribute ($p < .05$), it is reasonable to state that our participants generally preferred machine access to their data over the possibility of human access. We provide the relationship between privacy concerns and data-receiving entities in Figure~\ref{fig_main_entity_results}. 

\subsubsection{Data retention time}
\label{subsubsec_dataretention}
We also aimed to understand the relationship between data retention times and user privacy concerns, and these analyses yielded mixed results. When we compare one day of retention time (i.e., reference duration) with one month, we did not find any meaningful difference between privacy concerns. However, when the data retention period was indefinite, participants were more concerned, and analysis showed statistically meaningful differences at varying levels for several user attributes, including cognitive load and location ($p < .05$), BMI and depression ($p < .01$), and stress ($p < .0001$). Despite the general trend that participants are least concerned about their privacy for shorter retention periods, such a trend does not hold for all attributes. We provide these relationships in Figure~\ref{fig_main_retention_results}. 

\begin{figure*}[t!]
\centering
\subfigure[Activity.]{
    {\includegraphics[width=0.23\linewidth,keepaspectratio]{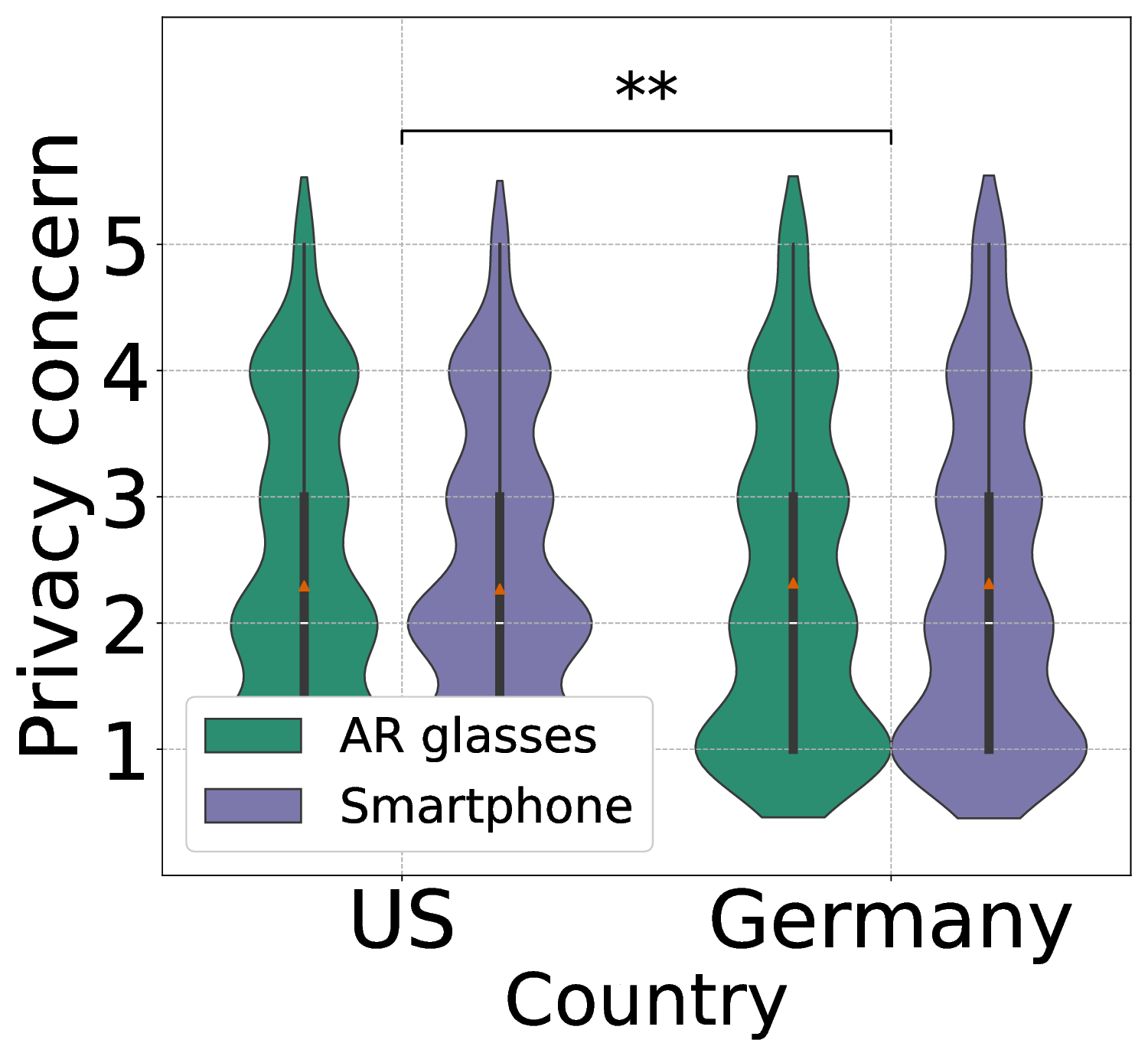}}
}
\subfigure[Alertness.]{
    {\includegraphics[width=0.23\linewidth,keepaspectratio]{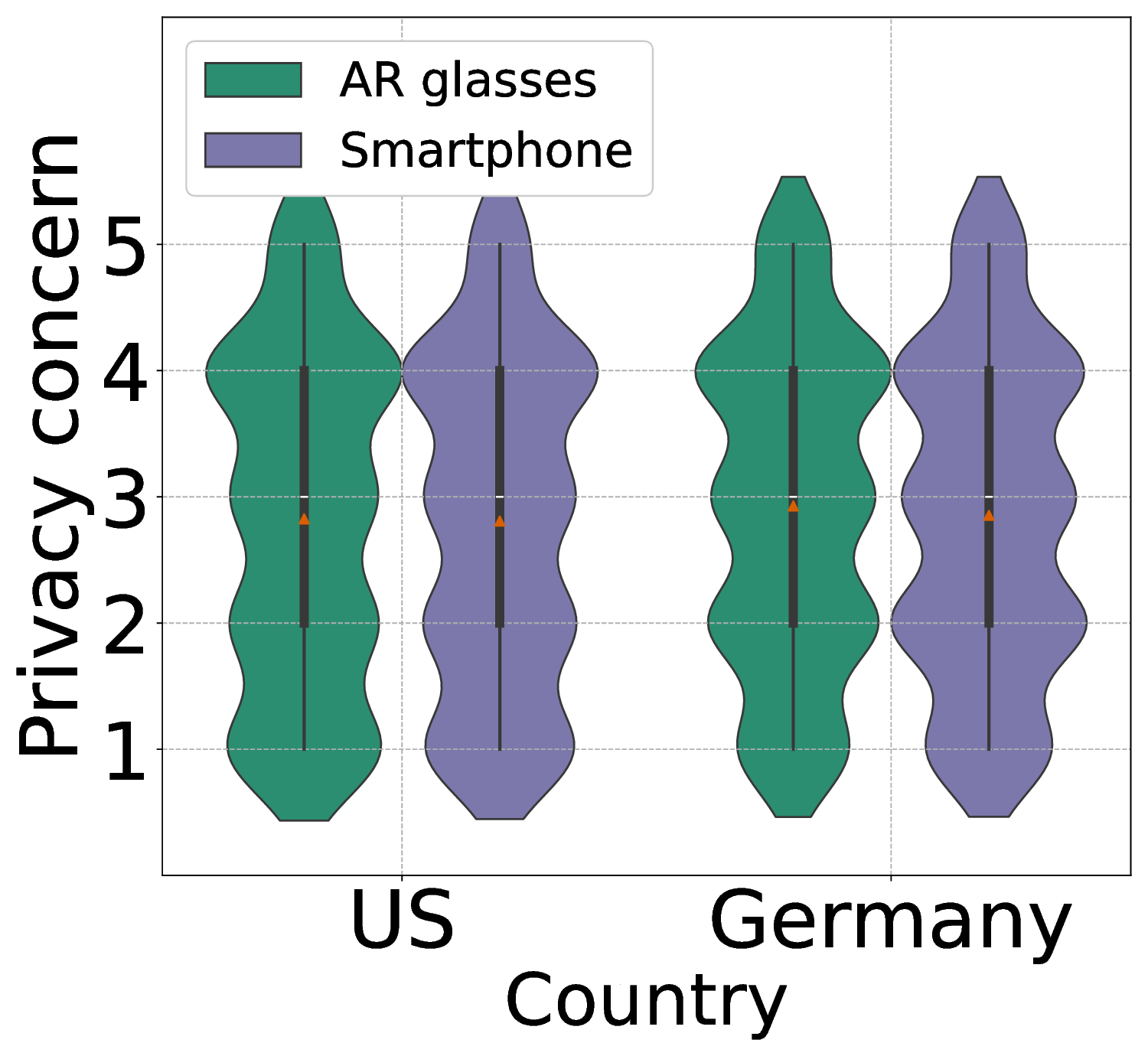}}
}
\subfigure[BMI.]{
    {\includegraphics[width=0.23\linewidth,keepaspectratio]{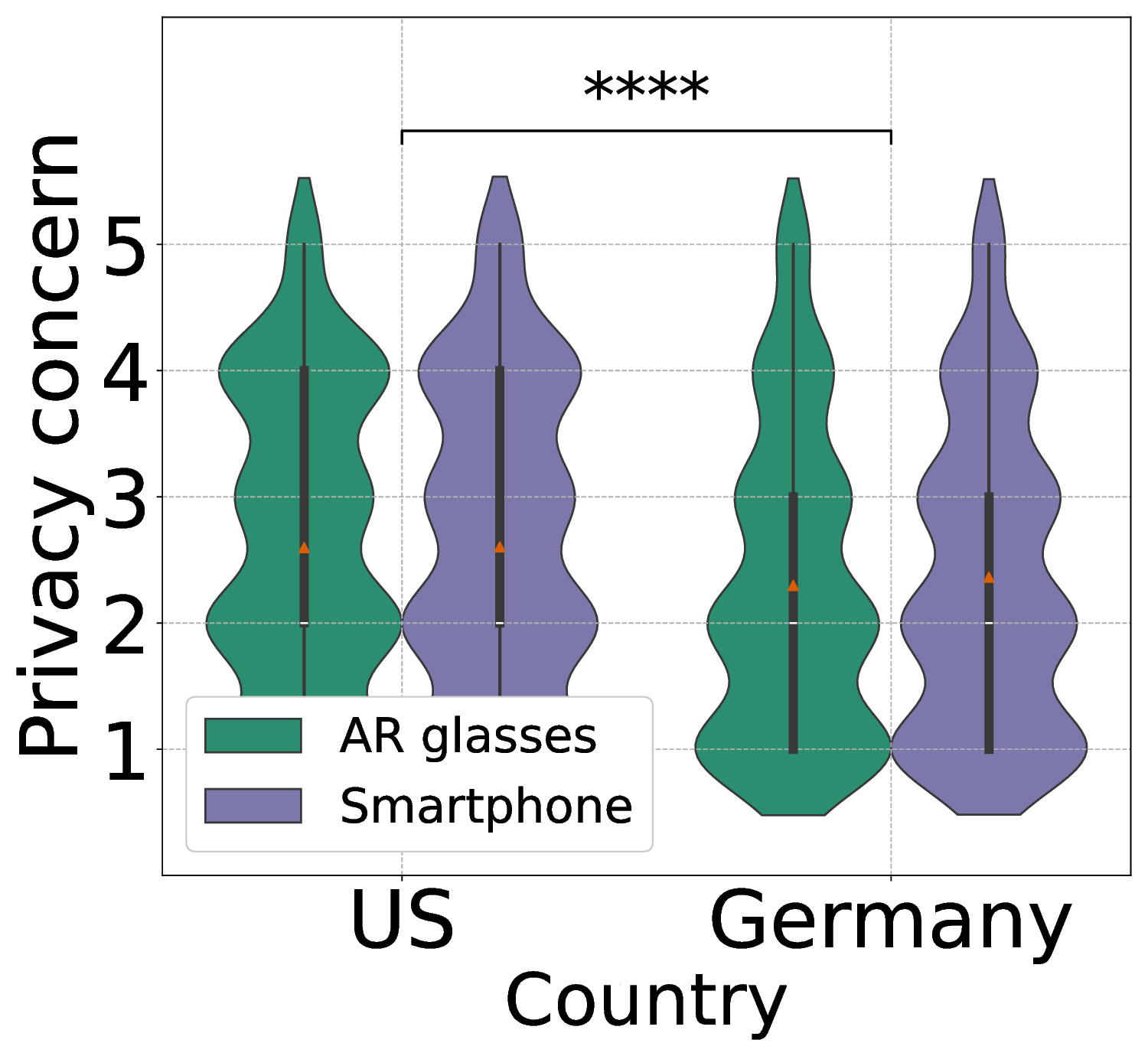}}
}
\subfigure[Cognitive load.]{
    {\includegraphics[width=0.23\linewidth,keepaspectratio]{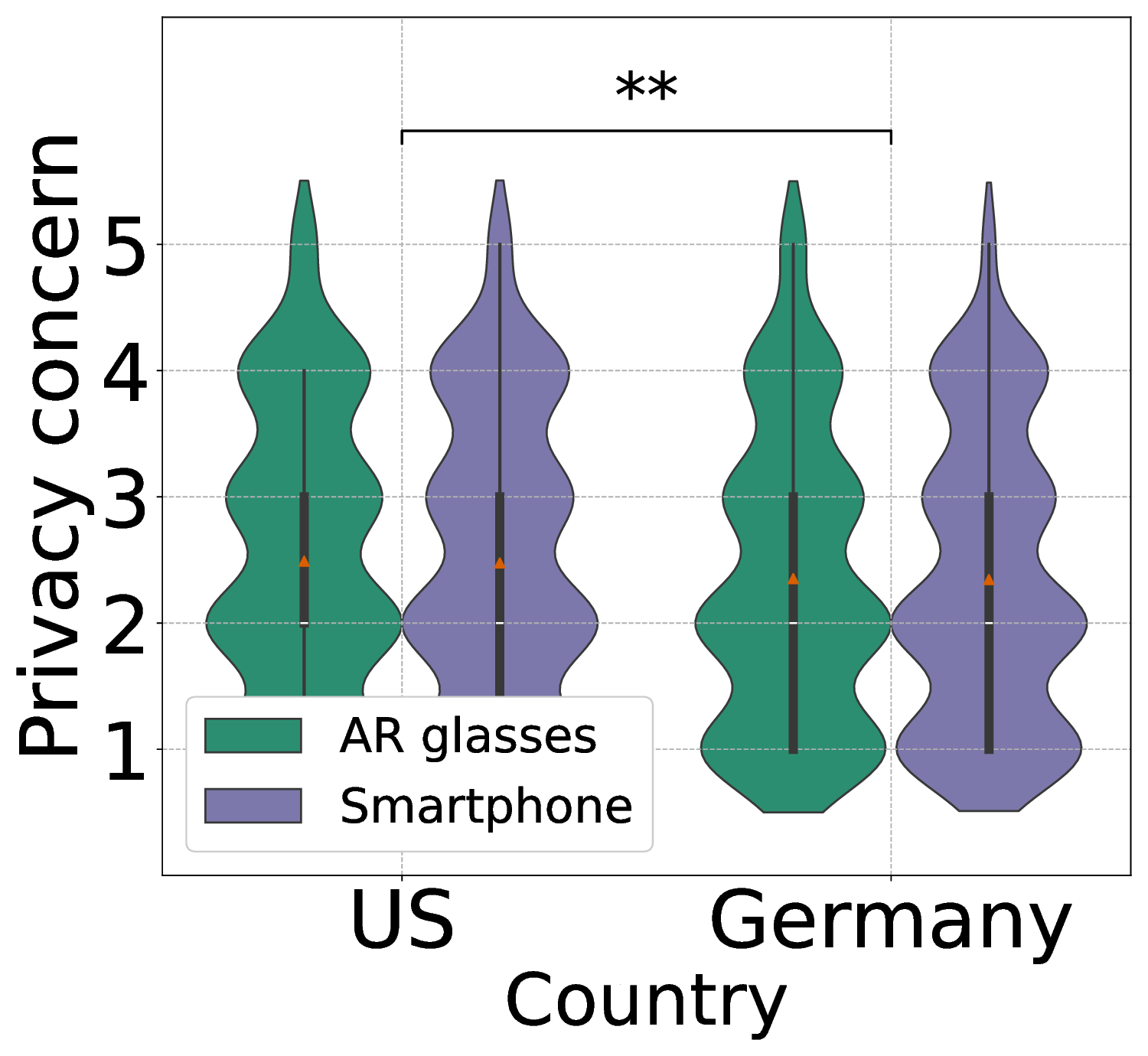}}
}
\subfigure[Depression.]{
    {\includegraphics[width=0.23\linewidth,keepaspectratio]{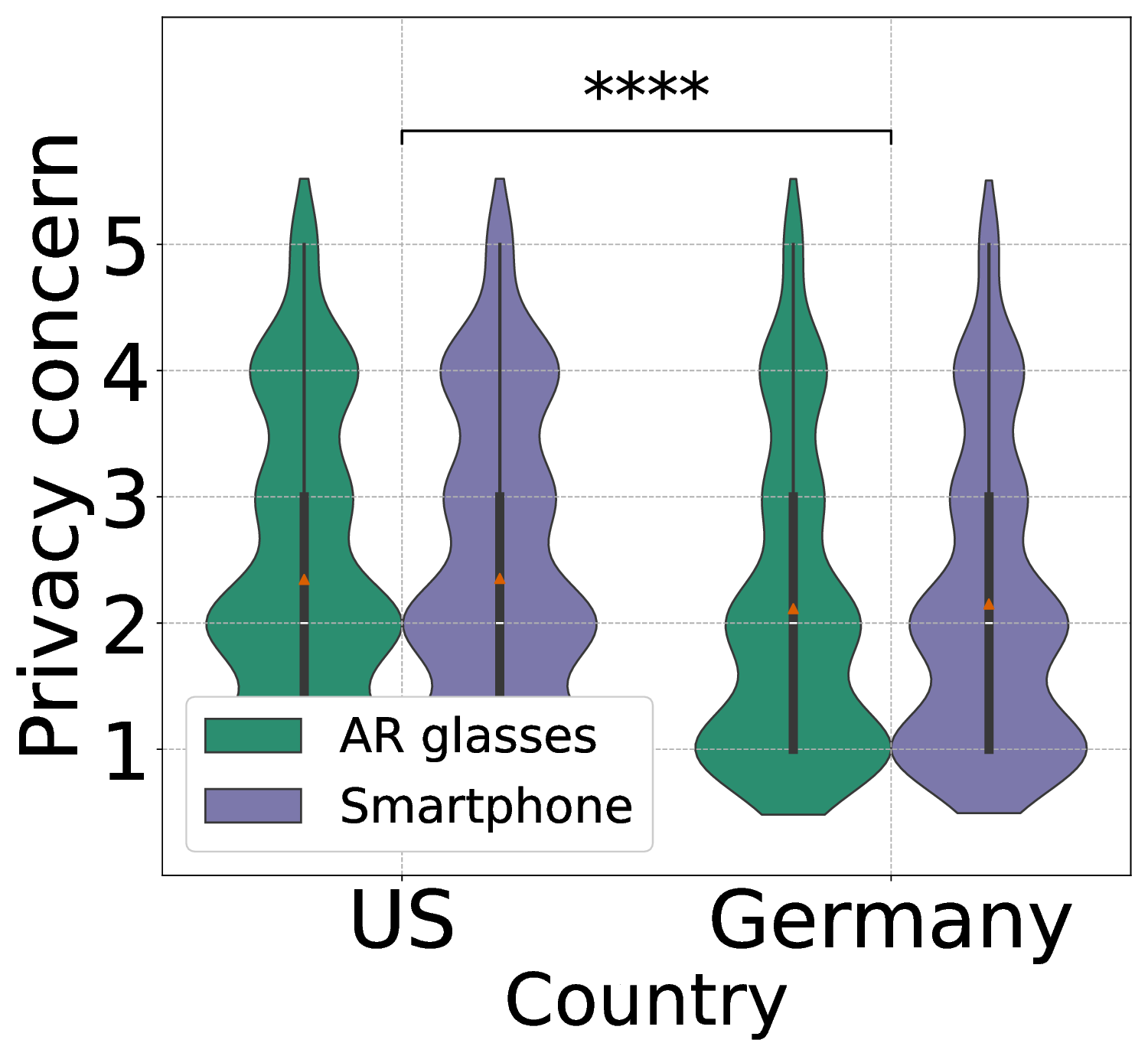}}
}
\subfigure[Gender.]{
    {\includegraphics[width=0.23\linewidth,keepaspectratio]{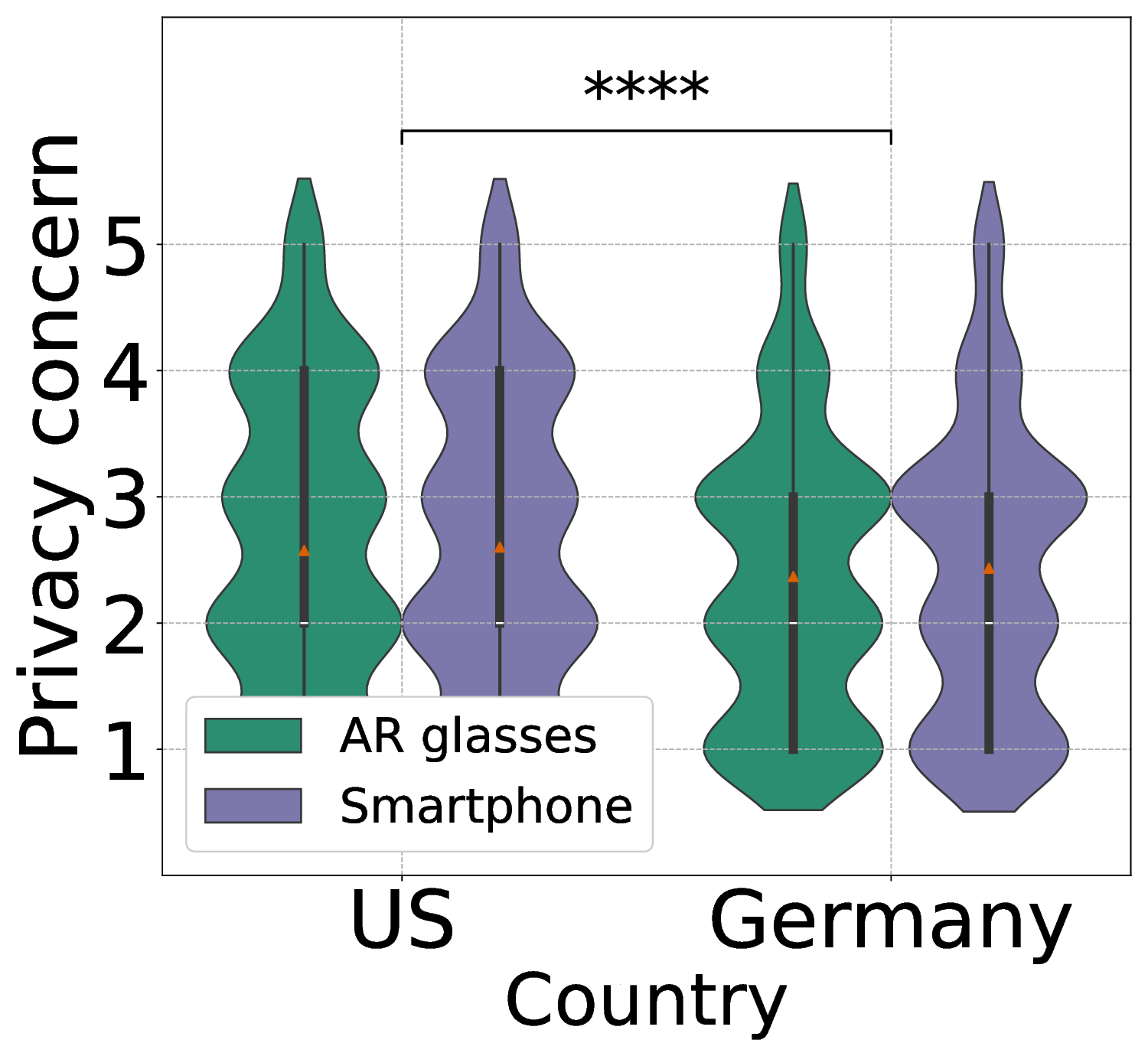}}
}
\subfigure[Heart condition.]{
    {\includegraphics[width=0.23\linewidth,keepaspectratio]{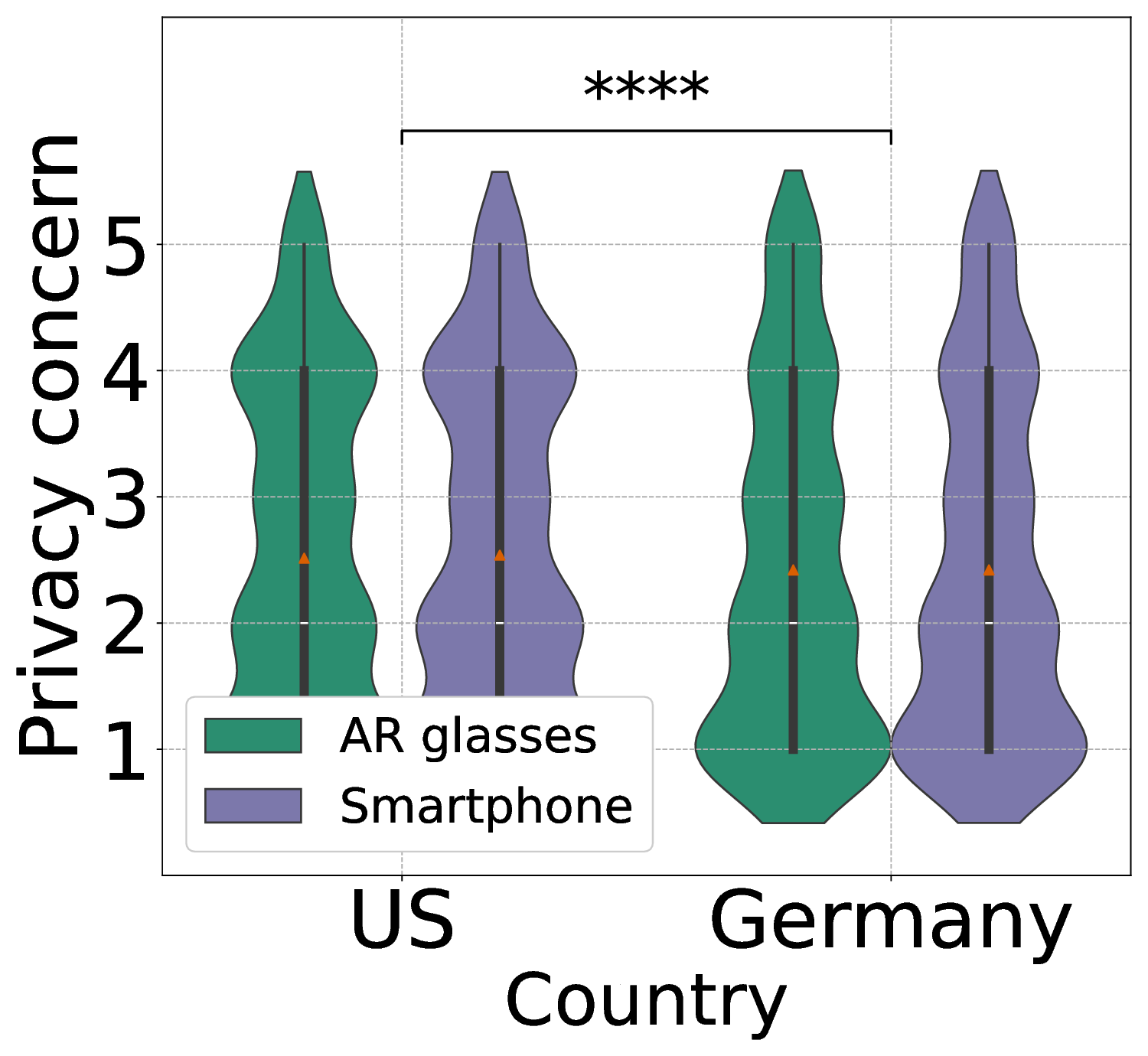}}
}
\subfigure[Location.]{
    {\includegraphics[width=0.23\linewidth,keepaspectratio]{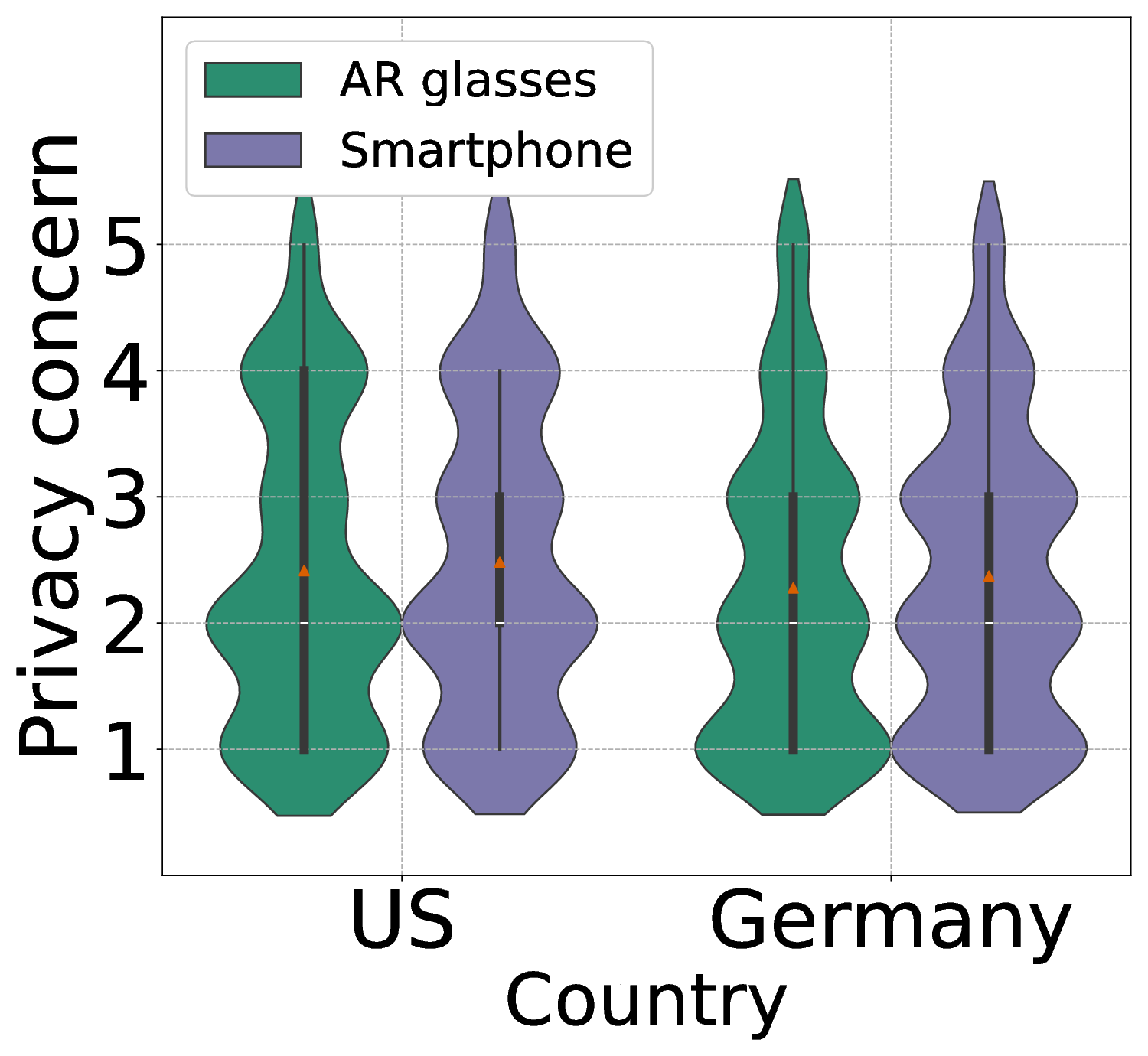}}
}
\subfigure[Personal identity.]{
    {\includegraphics[width=0.23\linewidth,keepaspectratio]{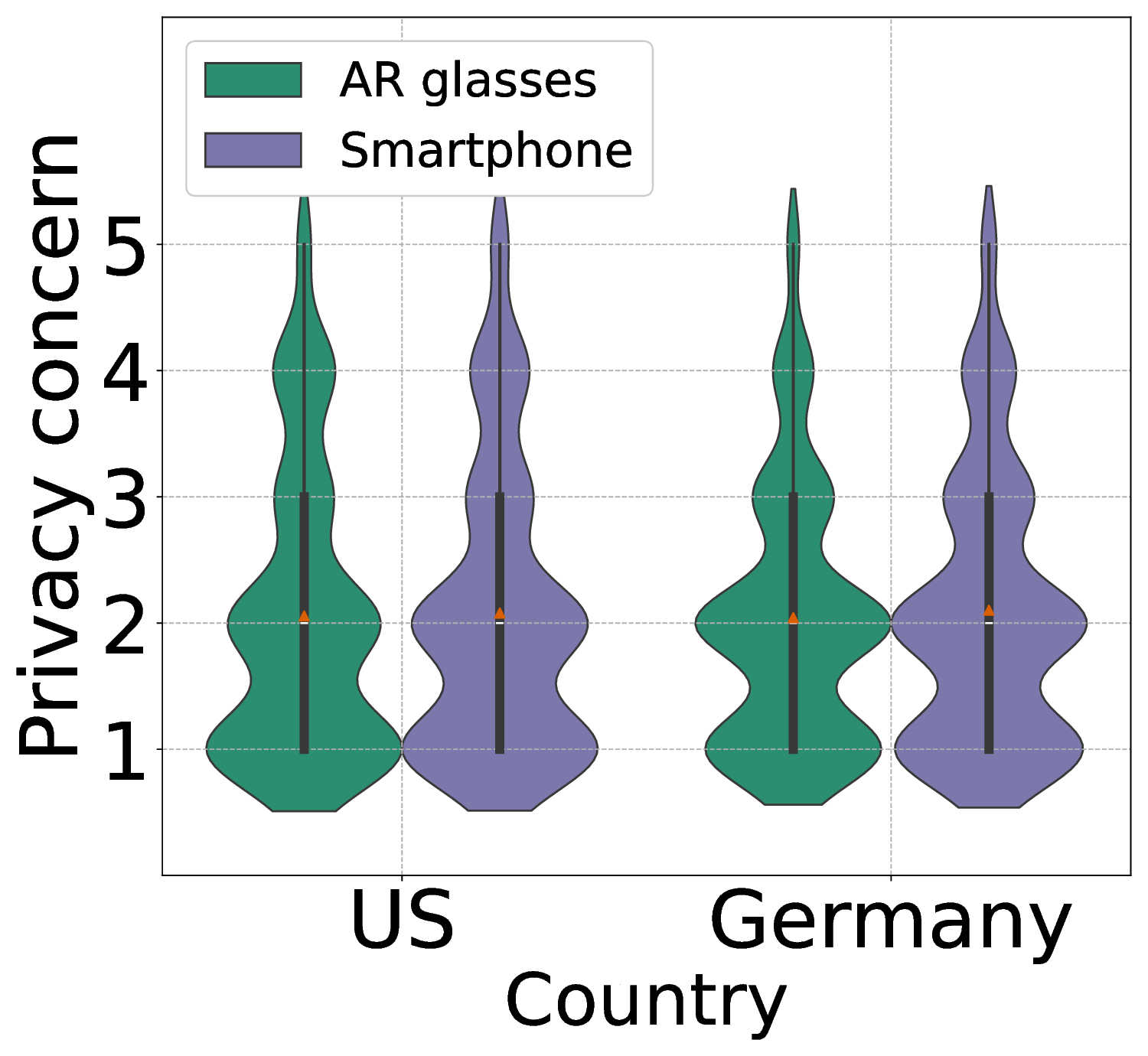}}
}
\subfigure[Sexual preference.]{
    {\includegraphics[width=0.23\linewidth,keepaspectratio]{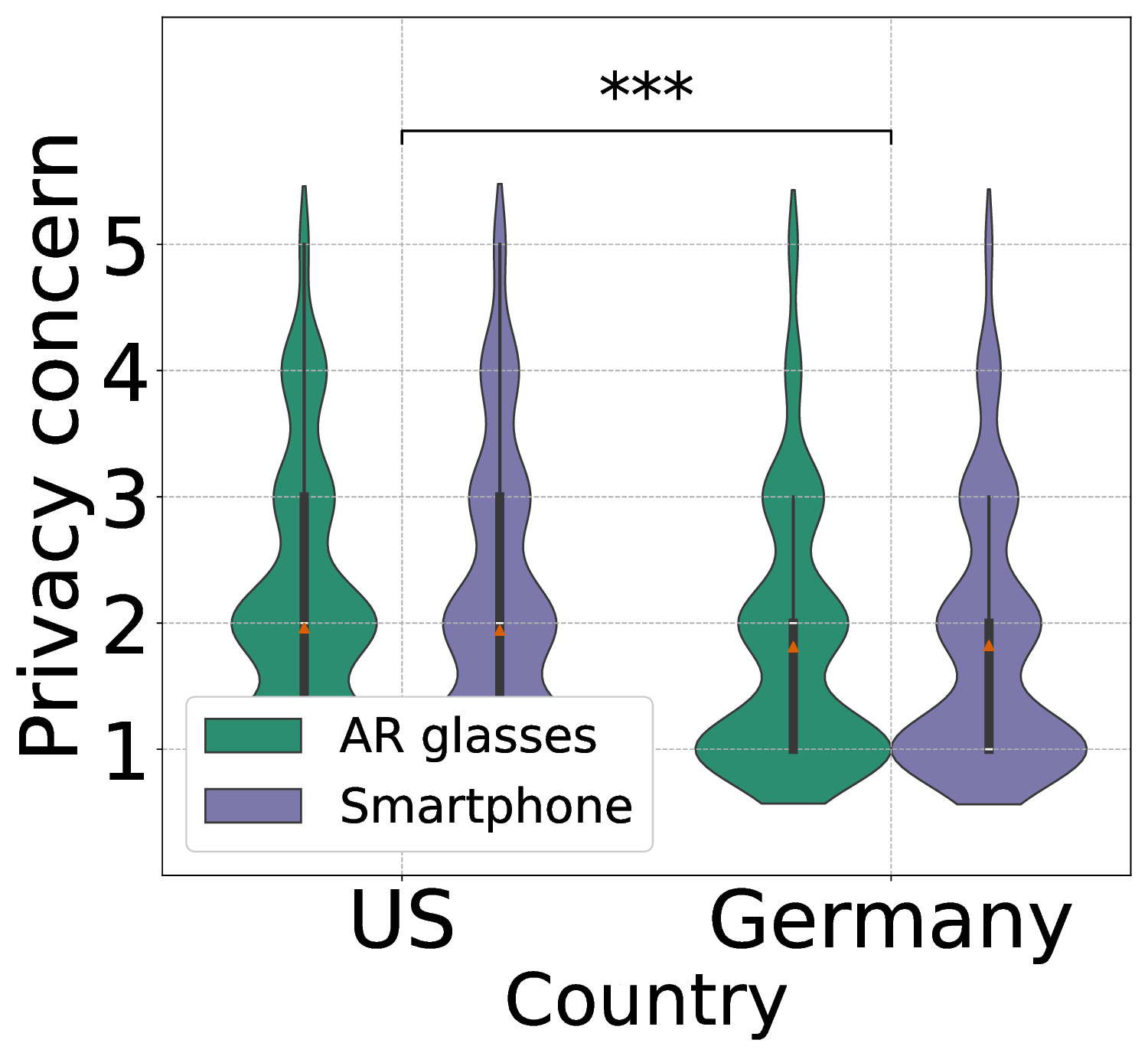}}
}
\subfigure[Stress.]{
    {\includegraphics[width=0.23\linewidth,keepaspectratio]{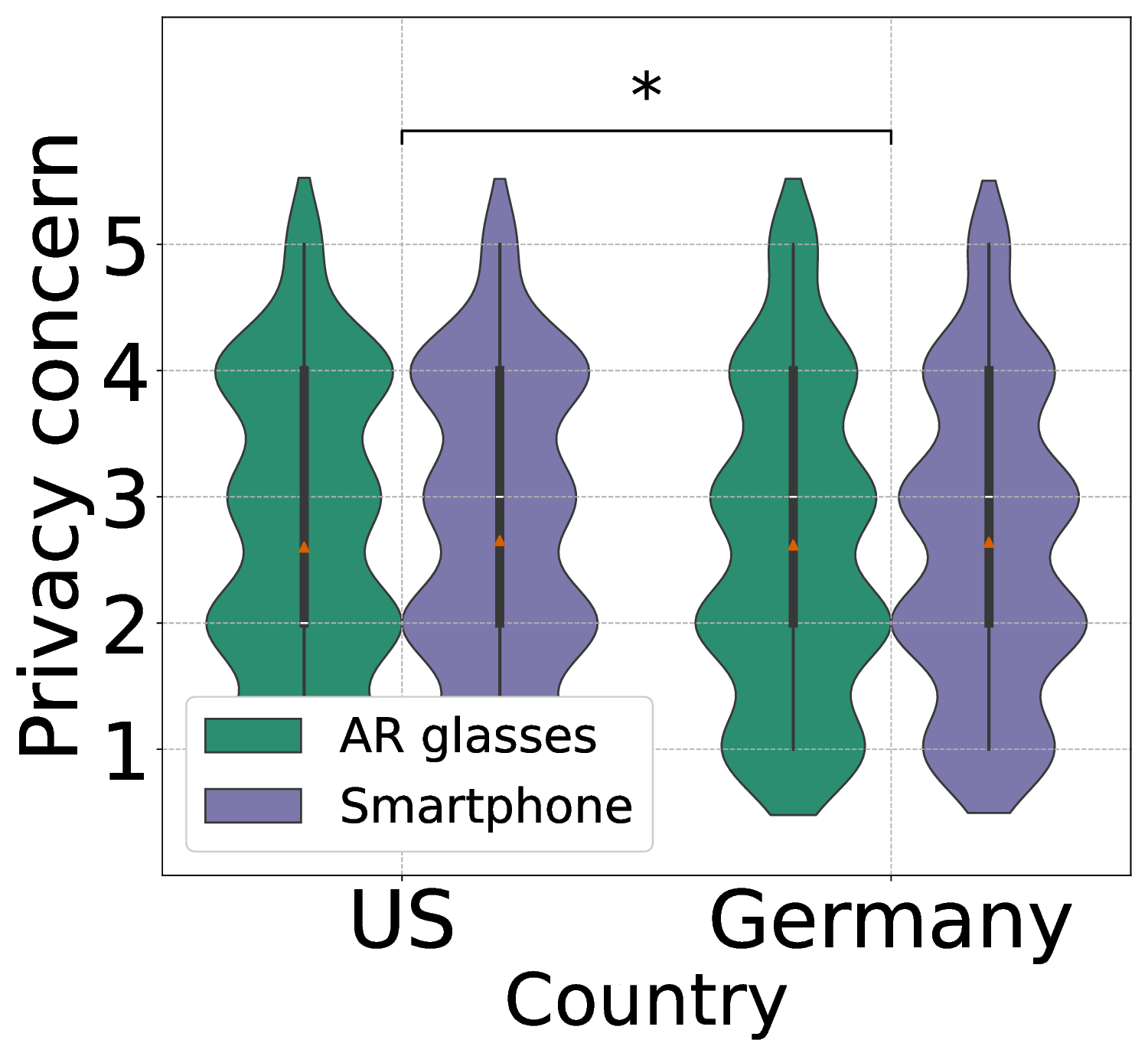}}
}
\caption{\label{fig_main_country_results} Violin plots representing the relationship between privacy concerns and countries for each user attribute.}
\alt{Violin plots representing the relationship between privacy concerns and countries for our privacy concerns survey. Each subplot corresponds to an evaluated user attribute. The Y-axes in the plots represent privacy concerns ranging from 1 to 5, with 1 being very uncomfortable and 5 being very comfortable. The X-axes correspond to data from the US and Germany, each with data for AR glasses and smartphones.}
\end{figure*}

\subsubsection{Country}
\label{subsubsec_country}
To understand the effect of cultural and sociological backgrounds on privacy concerns, we conducted our user studies in the US and Germany. We found that for most attributes, except for alertness, location, and personal identity, participants in Germany are significantly more concerned about privacy than those in the US. In particular, we found significant differences for heart condition, depression, gender, and BMI ($p < .0001$), sexual preference ($p < .001$), cognitive load and activity ($p < .01$), and stress ($p < .05$). However, the summary statistics of activity and stress are very similar, and when we ran the regression analyses that only included the country factor, we found that the country does not have a significant effect on privacy concerns for these two user attributes. 

Even though we found statistically significant results for privacy concerns between the US and Germany, when we approximate privacy concerns considering the country-wise distributions according to comfort levels, only the privacy concerns for BMI, gender, and heart condition diverge. For these three attributes, participants in the US lean towards being neutral, whereas participants in Germany lean towards being uncomfortable. While they fall slightly into different subcategories, these attributes are considered biological and medically relevant, and the German community is more sensitive to their privacy than the US community. We provide privacy concerns and country relationships in Figure~\ref{fig_main_country_results}. 

\subsubsection{Additional exploratory analyses}
To understand whether other factors influence privacy concerns, we analyzed how age and gender shape privacy concerns, which impacted privacy concerns in previous work~\cite{koelle_etal_2015}. We found that older participants are more concerned about their privacy compared to younger participants for activity, depression ($p < .01$), and heart condition and stress ($p < .0001$). Considering the greater likelihood that older participants may have health issues due to these attributes, our findings are interesting and worth investigating in more detail. We also found that men are less concerned about their privacy than women for all the attributes, the majority of them with ($p < .0001$). 

\subsubsection{Qualitative analyses}
We analyzed free-text responses to understand additional privacy concerns. The top three categories of additional concerns for the US for both AR glasses and smartphones are health-related information, social relationships/contacts, and video/audio recordings. 55 participants explicitly mentioned that their concerns regarding AR glasses are similar to smartphones or vice versa. The top categories for both AR devices in Germany are health-related information and video/audio recordings. Participants in Germany are also concerned about the environment/bystander information for AR glasses, and online tracking/messaging for smartphones. 73 participants in Germany explicitly mentioned similar concerns for both devices. It is also important to highlight that in the qualitative data for both AR devices, participants in both countries identified concern for health-related information as their top concern. Despite the small sample size in this analysis, participants in Germany also pointed out more concerns in this category than those in the US, which overlaps with the trends in our survey data. We provide the overview in Table~\ref{table_qualitative_coding}. 

\begin{table}[t!]
    \centering
    \caption{Additional concerns about different data types.}
    \alt{Table that provides results of the qualitative coding for additional concerns about different data types. The first column of the table corresponds to attributes. The second and third columns correspond to the number of occurrences for the corresponding attributes in AR glasses and smartphone concerns in the US. Similarly, the fourth and fifth columns correspond to the number of occurrences for the corresponding attributes in AR glasses and smartphone concerns in Germany.}
    \begin{tabular}{c|r|r|r|r}
        \toprule
        \multirow{2}{*}{\textbf{Attribute}} & \multicolumn{2}{c|}{$n^{US}_{p} = 504$} & \multicolumn{2}{c}{$n^{GER}_{p} = 511$} \\
        \cline{2-5}
         & AR glasses & Smartphone & AR glasses & Smartphone \\
        \hline
        Health & $54~(11\%)$ & $56~(11\%)$ & $65~(13\%)$ & $78~(15\%)$ \\
        \hline
        Social relationships/Contacts & $39~(8\%)$ & $40~(8\%)$ & $28~(5\%)$ & $23~(4\%)$ \\
        \hline
        Video/Audio & $35~(7\%)$ & $41~(8\%)$ & $49~(10\%)$ & $50~(10\%)$ \\
        \hline
        Financial & $24~(5\%)$ & $29~(6\%)$ & $15~(3\%)$ & $22~(4\%)$ \\
        \hline
        Online tracking/Messaging & $21~(4\%)$ & $39~(8\%)$ & $31~(6\%)$ & $44~(9\%)$ \\
        \hline
        Emotions & $21~(4\%)$ & $13~(3\%)$ & $12~(2\%)$ & $10~(2\%)$ \\
        \hline
        Routines & $19~(4\%)$ & $15~(3\%)$ & $23~(4\%)$ & $17~(3\%)$ \\
        \hline
        Environment/Bystander & $16~(3\%)$ & $7~(1\%)$ & $42~(8\%)$ & $15~(3\%)$ \\
        \hline
        Thoughts & $16~(3\%)$ & $14~(3\%)$ & $9~(2\%)$ & $5~(1\%)$ \\
        \hline
        Sexuality & $12~(2\%)$ & $12~(2\%)$ & $14~(3\%)$ & $16~(3\%)$ \\
        \hline
        Socio-economics & $12~(2\%)$ & $5~(1\%)$ & $10~(2\%)$ & $14~(3\%)$ \\
        \hline
        Physiological & $11~(2\%)$ & $3~(1\%)$ & $9~(2\%)$ & $5~(1\%)$ \\
        \hline
        Gaze & $9~(2\%)$ & $2~(<1\%)$ & $13~(2\%)$ & $5~(1\%)$ \\
        \hline
        Political views & $8~(2\%)$ & $5~(1\%)$ & $10~(2\%)$ & $9~(2\%)$ \\
        \hline
        Psychological & $3~(1\%)$ & $1~(<1\%)$ & $6~(1\%)$ & $3~(1\%)$ \\
        \hline \hline
        Overlap & $23~(5\%)$ & $32~(6\%)$ & $36~(7\%)$ & $37~(7\%)$ \\
        \bottomrule
    \end{tabular}
    \label{table_qualitative_coding}
\end{table}

\section{Discussion}
This section summarizes and interprets our findings, provides recommendations, and discusses our limitations with suggestions for future research.

\subsection{Summary and interpretation of the findings}
Addressing our \textbf{RQ1}, we found that user privacy concerns differ based on the various factors we had: data-receiving entity, priming type, and data retention times. Considering the findings of previous works that users were not highly concerned about eye tracking, especially compared to other types of data such as passwords, banking information~\cite{lee2016information}, face images, and facial expressions~\cite{gallardo2023speculative}, we assumed that most participants in previous studies might have expressed their comfort about eye-tracking data without fully understanding possible inferences. Suspecting this, we asked our participants about their concerns regarding the user attributes that eye movements can reveal in AR. Firstly, we found a general tendency toward being uncomfortable with the data collection that enables inferences of different user attributes, as for most attributes, comfort rarely reaches neutral levels. Furthermore, participants indicated their discomfort when the use case does not benefit them, other humans can consume the data, and the retention period is long. Our participants do not have reservations based on AR device type, which is unexpected considering that smartphones are more common in everyday life than AR glasses. Answering the \textbf{RQ2}, we found that participants in Germany are more concerned about their privacy than those in the US. 

\textbf{Participants are less concerned about their privacy if the use case scenario is beneficial for them.} Our findings concerning the priming are consistent with what previous research found~\cite{harborth_and_frik_2021}, even though we conducted more detailed analyses focusing on particular attributes. This finding means that if a use case scenario somehow benefits users in AR, it is likely that the acceptance of AR technology will be higher with increased user comfort, which also overlaps with the psychological phenomenon of privacy calculus theory~\cite{culnan_and_armstrong_1999} and confirms that users also weigh the perceived risks and technological benefits they receive in emerging technologies like AR. However, in practice, users may also be confronted with applications and services that appear to be beneficial, while primarily aiming to collect personal data for monetization. In such situations, assessing genuine utility and privacy trade-offs from the users' perspective may become challenging, especially when these services lack transparency on their multi-purpose. As a result, relying solely on users' agency is unrealistic to protect privacy. Therefore, supporting users with privacy-preserving functionality and potential regulatory oversight is essential in addition to precise transparency requirements. 

\textbf{We did not find meaningful evidence that user privacy concerns differ based on AR device type.} While Koelle et al.\ previously found that users are more sensitive to privacy violations enabled by AR glasses than those enabled by established devices, such as smartphones~\cite{koelle_etal_2015}, we did not find that to be the case for our user attributes and use case scenarios. One likely explanation might be our participants' limited real-world AR glasses experience. For instance, most participants of our calibration and privacy concerns surveys in both countries have never owned AR glasses. Without owning such devices, the participants' privacy concerns may be based on abstract assumptions rather than real experiences. 

It is also possible that the use cases we asked about were already more concerning than any device-related effect, leading to no differences. On the other hand, participants might be familiar with the attributes presented by more established devices than AR glasses, like wristbands and fitness trackers, which result in similar concerns between AR glasses and smartphones. In this regard, participants might have perceived that the functionalities of AR glasses and smartphones have been converging in terms of sensor-based capabilities, such as gaze estimation through front-facing cameras of the smartphones~\cite{apple_ios18_eyetracking_may2024}, and they did not perceive AR glasses data collection as a new privacy threat, even though they may introduce novel privacy issues in practice. As a result, their privacy concerns did not change based on the device type. Both interpretations require further research for validation, especially with qualitative methods to understand the potential reasons more in depth. 

\textbf{Participants are more comfortable when computers consume their data.} Our analyses yield that when humans can consume the collected data, participants are more concerned about privacy than when their data remains in their devices. The findings of previous work on smartphones~\cite{felt_etal_2012} and AR glasses~\cite{lee2016information} are similar, as they found that users are less concerned when their data is sent to an app server than consumed by humans, such as the public or friends. This finding also partly overlaps with the findings of previous research on human data consumption that elderly users have privacy concerns when other people (e.g., their children) can watch the moves that they make through smart glasses~\cite{McNaney_etal_2014}. With different computers and human access to the data, while we intended to understand privacy concerns towards differing data exposure and control levels, we did not ask our participants to explain their reasoning about their privacy concerns directly. However, we also acknowledge that participants might have found computers less judgmental than humans, as human data consumption may introduce a subjective layer that can lead to unintended social consequences, such as misunderstandings and privacy exposure. We did not find enough evidence that privacy concerns differ based on whether data remains on local devices or is stored in the app cloud. While both of these conditions are characterized by computer access to the data, some users may trust the system more when it is entirely local than cloud storage, even if the data is encrypted and said to be only accessible with credentials. Considering the potential ease of third-party access to data through an app cloud rather than local devices, one can investigate user privacy concerns and trust relationships to this end further. 

\textbf{Analyses on the data retention times yield mixed results.} We initially assumed that there would be a constant cut-off in privacy concerns between the shortest and longest retention times, similar to previous research~\cite{leon_etal_2013}, across all the attributes. While such a cut-off partly exists for some of our attributes, we did not find the same trend for all attributes. We explain this finding as the relationship between privacy concerns and retention time being highly coupled with the user attribute type, which should be assessed individually and contextually. In addition, our study considers that there is always data retention; however, inferring user attributes on the fly without storing the raw data is also possible as long as one does not use models that require temporal data. Users might perceive such a setup as less concerning, leading to the aforementioned cut-off, which is worth further investigating. Still, due to computational requirements, an efficient real-time inference mechanism may not always be possible in AR, and it might lead to an unpleasant user experience due to the cost of data privacy. Furthermore, prior work has shown that modeling user attributes, such as cognitive load~\cite{Hou_etal_2025}, activity~\cite{ozdel_etal_2024}, or personal identity~\cite{Liu_etal_2020}, can benefit from temporal data sequences and machine learning models that utilize such data, which often necessitates data retention for a certain time. Therefore, considering all, it is also essential to investigate whether there is an optimal trade-off for user experience and privacy in AR, similar to prior work in VR~\cite{li_etal_usenix_2021, 10464194} and its relationship with data retention. 

\textbf{Privacy concerns differ between countries.} We found that participants in Germany are more concerned about the majority of the attributes than those in the US, which overlaps with the findings of the prior research on cross-cultural privacy attitudes~\cite{Wilkowska_etal_2021, Harbach_etal_2016, Hossain_etal_2014}. The reason might be historical, cultural, and socio-economic differences. As no other AR privacy study has analyzed the differences between the US and Germany, our findings present a baseline for that device class. 

A recent survey in Germany implies that around 70\% of young individuals do not feel fit for digital technologies at work~\cite{vod_stiftung_2023}. Considering the American society's greater focus on entrepreneurship and technological innovations than Germany~\cite{hommes2011research}, the US participants might have shaped their privacy norms according to these factors and might be more open to the utility of the new technologies, such as AR, than those in Germany. Another reason for differing privacy concerns might be that these countries have different privacy regulations. For instance, the General Data Protection Regulation (GDPR)~\cite{EuropeanParliament2016a} is prominent in Germany. In contrast, despite the similarities with the GDPR, sector- and region-specific regulations exist in the US, such as the Health Insurance Portability and Accountability Act~\cite{hipaa_bib} and California Consumer Privacy Act~\cite{ccpa_bib}, respectively. As participants might already be familiar with different regulations and norms, they might have perceived our use cases about AR data practices differently. In addition to the cultural and regulatory differences, another possible explanation for the fewer privacy concerns of the US participants may be due to the digital resignation. Prior research has indicated that individuals in the US report an inevitability regarding digital privacy protection~\cite{draper_etal_2024}, not because they are unaware of the risks, but because they have little control over what information digital actors know about them~\cite{Draper_and_Turow_2019}. Such a resignation might have led to fewer concerns involving sensitive inferences from eye-tracking data for our participants in the US than those in Germany. 

\subsection{Recommendations}
Understanding users' privacy concerns is a challenging process, considering numerous factors that may affect them. By considering different factors, such as data-receiving entities, beneficialness of the use case scenarios, AR device types, and characteristics of the different communities, we provide recommendations to AR practitioners, policymakers, and educators, which may help design viable human-centered privacy solutions for AR that are informed by our data and analyses. 

\subsubsection{Enhancing transparency and granularity of privacy, consent, and permission mechanisms}
Participants' privacy concerns did not follow the same trend for all the evaluated user attributes, implying that device manufacturers and applications should explicitly inform users about what attributes collected eye movements can reveal. Based on this, users should be able to limit certain inferences. For instance, a user should be able to consent to gaze-based authentication without consenting to use their eye movements to infer attributes like sexual preference or to provide targeted advertising. Menéndez González and Bozkir\ recently stated that many AR devices do not have particular privacy policies but instead follow the general privacy policies of the manufacturer (e.g., Microsoft HoloLens 2, Varjo XR-3)~\cite{MenendezGonzalez2024}. In these policies and further technical descriptions from the manufacturers, key terms, such as the types of collected data, details of processing, and data sharing practices, are often described in ambiguous ways, which hints at the current shortcomings in GDPR's implementation in Europe in the context of AR and eye-tracking technologies. According to the authors, even the explicit policies for AR HMDs (e.g., Magic Leap 2, Meta Quest Pro) only discuss implications revolving around raw eye-tracking data, such as eye images or gaze, and fail to touch on possible inferences. Coupling our findings with the current state of the AR device landscape, we suggest manufacturers enhance transparency by providing fine-granular yet conceivable explanations and policies on eye movements in AR and facilitating these when obtaining users' consent. While designing such fine-granular and conceivable explanations that users can understand the implications of eye-tracking data in AR requires further research, it is necessary so that users can have informed decisions on their actions in AR. 

Furthermore, we recommend that researchers and practitioners explain not only the possibility but also the likelihood of inferences being made, as AR glasses may increase the risk of inference of sensitive attributes (e.g., sexual preference), even though this may be possible with smartphone-based AR. Doing so may cause people to assess risk differently than if they are informed only of the (sometimes practically hypothetical) possibility of an inference. 

As the beneficialness perceptions of use case scenarios differ, AR applications should also be transparent with fine-grained permission mechanisms in AR when accessing eye-tracking data. For instance, users might be comfortable with their AR device inferring their alertness levels to help them drive safely. However, they might be uncomfortable, or even irritated, with the inference of the same attribute if they receive constant traffic warnings. Hence, we argue for more granular permission mechanisms for AR applications, especially regarding the type of use and inference, as we have constantly found that beneficialness levels affect privacy concerns. How to design such granular permission mechanisms is an open question and complex, especially in commercial contexts, such as advertising, where inferring user attributes can provide personalization and potentially revenue. Commercial entities might not prefer to adopt such fine-granular permission mechanisms without legal and economic incentives, as this may reduce their monetization. 

Even with transparent and granular consent and permission mechanisms, users may remain skeptical about whether the organizations respect these mechanisms, and they might form their trust accordingly. In terms of long-term credibility and obtaining users' trust for AR systems in handling behavioral data, such as eye tracking, manufacturers, organizations, and practitioners should also invest in visible and verifiable practices that build users' confidence over time. This way, transparent and granular privacy, consent, and permission mechanisms can go hand-in-hand with such verifications to facilitate responsible data processing practices for AR. 

\subsubsection{Privacy-by-design settings for AR}
Considering the different privacy concern implications we found and some of the previous works stating eye-tracking data use is not the most concerning use~\cite{lee2016information, gallardo2023speculative}, especially for the users who are not aware of what inferences are possible with eye movements, regulations might be needed to protect them from possible harm since people's own assessments of potential harm might not sufficiently consider the possible inferences. To this end, we suggest utilizing privacy-by-design settings, as also mentioned in previous research~\cite{Langheinrich_2001_privacybydesign, Gressel_etal_2023, abraham_etal_chi24}. Therefore, if users do not particularly opt in for inferences that could support them in many ways, their data is not used for inferences. While this can negatively affect the user experience to a certain extent in AR, as users may not be able to take advantage of the personalized AR functionality, it will facilitate better privacy protection. To this end, the GDPR in Europe includes principles such as privacy-by-design. In practice, implementing such principles for AR systems with behavioral data processing, like eye tracking, remains complex. While we do not necessarily suggest a brand-new regulation for privacy-by-design, adhering to existing principles by facilitating transparency in AR design choices is essential. 

When users opt in for particular functionality for personalization in AR, as we found our participants' relative comfort if their data remains in their AR devices, we recommend exploring whether inference tasks for which data is now shared can instead be accomplished via federated learning. Federated learning enables ML model processing on local devices (e.g., AR glasses or smartphones) without sharing the data, but rather by sharing the model~\cite{mcmahan2017communication}. While further research is needed to validate this, particularly from a user perception and inference performance perspective, federated learning-enabled AR has great potential to protect data privacy while enabling preferred inferences, especially considering that federated learning gives the minimal possibility for sensitive user data to be spread around different entities. This way may maximize the utility of personalized AR interfaces while minimizing data privacy concerns. However, as we found out, users might have more reservations about particular attributes than others based on data practices such as use case beneficialness, and practitioners should consider such contextual privacy concerns also in the federated learning setting. 

\subsubsection{Improving literacy on AR technology}
As AR glasses are not as prevalent as smartphones in everyday life, many users may not be aware of the possibility of collecting fine-grained information about them with AR glasses. In addition, in our experiments, most of our participants do not own AR glasses. Considering these, AR literacy and privacy awareness are essential for the vast use of AR glasses, which can happen by educating users on the AR devices' capabilities, collected data types, and potential data use. While such a literacy may lead to different privacy concerns than we found, in this way, users can make more informed decisions balancing their privacy and technological utility. 

To this end, we suggest several concrete directions. Firstly, manufacturers and practitioners can develop targeted interactive and educational resources to facilitate the understanding of the device capabilities and possibilities with data collected from them. For instance, especially with the Apple Vision Pro, Apple has been offering potential users of their spatial computer to experience and interact with their device for a significant amount of time with a specialist introducing the device~\cite{apple_vision_pro_demo}. While not limited to this specific example, such and similar strategies can help increase public exposure to different AR devices, and compared to traditional advertisements and short videos, they may foster greater understanding of and familiarity with these technologies. In addition, public awareness campaigns that regulators and manufacturers might organize can play another important role in raising literacy on AR technology and the implications of behavioral data collection, such as eye-tracking data. Lastly, to facilitate transparency in terms of privacy, approaches similar to privacy nutrition labels~\cite{EmamiNaeini_etal_2022_ieeesp, Li_etal_2022_chi} can be utilized for AR technology, which can inform users about what inferences are made, how they are used, and how long corresponding datasets are retained, and these practices can potentially contribute to improving users' AR literacy. 

\subsection{Limitations and future works}
Our study led to novel insights about eye-tracked AR, with some limitations. Firstly, as we crowdsourced the survey data, the technology literacy of our participants may be higher than that of the general population, which likely includes less frequent users of the Internet and smart devices. Secondly, prior work on XR recently stated that researchers and user study participants have been from White, Educated, Industrialized, Rich, and Democratic (WEIRD) populations~\cite{9646525}. Even though we collected data from two countries, which is advantageous compared to most prior works with data from a single country, our user studies also fall into the WEIRD category, as we collected data from the US and Germany. Furthermore, the demographics of our dataset in Germany are less representative than the US dataset, which could limit the generalizability of our findings. Regarding our participants, despite their technology literacy, they might not represent typical AR users and, therefore, might not be knowledgeable enough about the AR devices and their capabilities. In addition, while we incorporated beneficialness scores into our privacy concerns analyses, participants might perceive different parts of the same use case and priming text as both beneficial and harmful. Lastly, while we collected privacy concerns data for eye-tracked AR, we lack actual behavioral data, which might be misaligned with the reported privacy concerns, often identified as the privacy paradox~\cite{colnago_etal_2023_popets, solove2021myth}. 

Considering our limitations and results, future research should focus on extending our work with more communities beyond WEIRD populations and recruiting participants who use AR technology more regularly. The latter may also lead to understanding whether there is any misalignment (e.g., privacy paradox) between privacy concerns and behaviors if researchers can collect behavioral data while using AR technologies, especially if different contexts, such as beneficialness, are considered. Another vital future avenue is analyzing privacy concerns with more user attributes (e.g., personal preferences) and sensor data types beyond eye tracking, potentially also with qualitative studies to understand users' concerns in a more detailed and nuanced way. Furthermore, as we found that privacy concerns are contextual and depend on many factors, such as data-receiving entities or the beneficialness of the use cases, it is also essential to study which of the specific actors or organizations can facilitate such factors more comfortably for users. For instance, users' privacy concerns may differ if a large commercial company or a non-profit academic institution offers a particular AR service and product. Lastly, while we did not present our use case scenarios to our participants with specific privacy policies or detailed explanations to deliberately understand their baseline privacy concerns, participants may report different concerns or comfort in the presence of a privacy policy. Understanding these distinctions can provide more nuanced insights into users' privacy concerns and trust, which requires future research. Addressing these future directions should help generalize findings on privacy concerns and behaviors in a broader context. 

\section{Conclusion}
In this work, we studied user privacy concerns about eye-tracked AR with four crowdsourcing studies in the US and Germany, including various use cases and data practices. We found a general tendency of participants to be uncomfortable with the user attributes we asked about, which eye movements may reveal. We found no indication that participants exhibited any excitement or fear regarding AR glasses. However, participants are concerned about their privacy if other humans can access their data or if a use case does not benefit them. Our country-specific analysis also concludes that participants in Germany are more concerned about their privacy than those in the US. Considering all, we recommend privacy-by-design settings with more granular permission mechanisms and educating users about novel AR settings when eye-tracking data is collected and inferences about user attributes are made. 

\section{Appendix A: Additional details on experimental design}
In this section, we provide exhaustive details on the experimental design, such as recruitment materials, introductory texts, different types of priming texts we used, and other questionnaire details, which are as follows.

\subsection{US study}
\subsubsection{Recruitment texts}
\textit{\\Recruitment text (Calibration survey)}
\begin{quote}
``Study participants wanted!

Researchers from Carnegie Mellon University are looking for participants for a 10-minute survey study on personal assistant apps! You will be answering several questions regarding your potential use of personal assistant apps. You will receive 2.50 USD for your participation.

Individuals aged 18 or older, who are located in the US can participate in this study.''
\end{quote}

\textit{\\Recruitment text (Privacy concerns survey)}
\begin{quote}
``Study participants wanted!

Researchers from Carnegie Mellon University are looking for participants for an 11-minute survey study on augmented reality glasses! You will be answering several questions regarding your potential use of augmented reality glasses and smartphones. You will receive 2.75 USD for your participation.

Individuals aged 18 or older, who are located in the US can participate in this study.''
\end{quote}

\subsubsection{Introductory scripts and briefing}
\begin{quote}
In this survey, you are expected to answer some questions about possible uses of augmented reality (AR) glasses and smartphones. You do not necessarily need to have any experience with AR glasses.

\textit{AR Glasses briefing:} Imagine that you are an owner of augmented reality (AR) smart glasses which are worn on your head and can perceive you and your surroundings and understand where you look with its integrated sensors. In addition to these sensing possibilities, your AR glasses can display holograms in your field of view. Imagine that you install a personal assistant app in your AR glasses to manage your daily tasks.

\textit{Smartphone briefing:} Imagine that you own a smartphone that can perceive you and your surroundings with its integrated sensors, and you install a personal assistant app in your smartphone to manage your daily tasks.
\end{quote}

\subsubsection{Priming texts \& Question templates}
In Table~\ref{lbl_attribute_tbl}, priming texts for beneficial and not-beneficial priming conditions are provided. After priming texts for each attribute, depending on the condition, the following question is asked in the calibration survey. 

How beneficial or harmful would you find a personal assistant app on your [AR glasses/smartphone] that does this?

$\square$ Very harmful $\square$ Somewhat harmful $\square$ Neither beneficial nor harmful $\square$ Somewhat beneficial $\square$ Very beneficial

The privacy concerns survey question is as follows. 

How would you feel if your personal assistant app on your [AR glasses/smartphone] can determine your [attribute] using data collected by your [AR glasses/smartphone], share the collected data with [data-receiving entity], and store the collected data for [data retention time]?

$\square$ Very uncomfortable $\square$ Uncomfortable $\square$ Neutral $\square$ Comfortable $\square$ Very comfortable

\subsubsection{Demographic questionnaire}
We provide demographic questions that we asked to characterize the sample pools in the following, adapted from previous work~\cite{leon_etal_2013}. \\
Briefing: In this part of our study, we will ask you about your demographic information.

\begin{itemize}
  \item How old are you? (In years) \underline{\hspace{0.5cm}}
  \item What is your gender? $\square$ Woman $\square$ Man $\square$ Non-binary $\square$ Prefer not to disclose $\square$ Prefer to self-describe as \underline{\hspace{0.5cm}}
  \item Which of the following describes best your primary occupation? $\square$ Administrative support (e.g., secretary, assistant) $\square$ Art, writing, or journalism (e.g., author, reporter, sculptor) $\square$ Business, management, or financial (e.g., manager, accountant, banker) $\square$ Computer engineer or IT professional (e.g., systems administrator, programmer, IT consultant) $\square$ Education (e.g., teacher) $\square$ Engineer in other fields (e.g., civil engineer, bio-engineer) $\square$ Homemaker $\square$ Legal (e.g., lawyer, law clerk) $\square$ Medical (e.g., doctor, nurse, dentist) $\square$ Retired $\square$ Scientist (e.g., researcher, professor) $\square$ Service (e.g., retail clerks, server) $\square$ Skilled labor (e.g., electrician, plumber, carpenter) $\square$ Student $\square$ Unemployed $\square$ Decline to answer $\square$ Other (Please specify): \underline{\hspace{0.5cm}}
  \item Which of the following best describes your highest achieved education level? $\square$ No high school $\square$ Some high school – no degree $\square$ High school graduate $\square$ Some college - no degree $\square$ Associates/2-year degree $\square$ Bachelors/4-year degree $\square$ Graduate degree - Masters, PhD, professional, medicine, etc.
  \item Have you ever owned augmented reality glasses? $\square$ Yes $\square$ No
  \item If you currently use augmented reality glasses, approximately how many hours do you spend wearing augmented reality glasses each day? $\square$ \underline{\hspace{0.5cm}} $\square$ Not applicable
  \item Do you own a smartphone? $\square$ Yes $\square$ No
  \item If yes, approximately how many hours do you spend using your smartphone each day? $\square$ \underline{\hspace{0.5cm}} $\square$ Not applicable
  \item Approximately how many hours do you spend on the Internet each day? $\square$ \underline{\hspace{0.5cm}} $\square$ Not applicable
\end{itemize}

\subsection{German study}
\subsubsection{Recruitment texts}
\textbf{\\Recruitment text (Calibration survey)}
\begin{quote}
``Studienteilnehmer gesucht!

Forscher der Technischen Universit\"at M\"unchen suchen Teilnehmer*innen für eine 10-minütige Umfragestudie zu persönlichen Assistenten-Apps! Sie werden mehrere Fragen zu Ihrer möglichen Nutzung von persönlichen Assistenten-Apps beantworten. Für Ihre Teilnahme erhalten Sie ~2,50 EUR.

An dieser Studie können Personen ab 18 Jahren, mit Wohnsitz in Deutschland teilnehmen.''
\end{quote}

\textbf{\\Recruitment text (Privacy concerns survey)}
\begin{quote}
``Studienteilnehmer gesucht!

Forscher der Technischen Universit\"at M\"unchen suchen Teilnehmer*innen für eine 11-minütige Umfragestudie zu Augmented-Reality-Brillen! Sie werden mehrere Fragen zu Ihrer möglichen Nutzung von Augmented-Reality-Brillen und Smartphones beatworten. Für Ihre Teilnahme erhalten Sie ~2,75 EUR.

An der Studie können Personen ab 18 Jahren, mit Wohnsitz in Deutschland teilnehmen.''
\end{quote}

\subsubsection{Introductory scripts and briefing}
\begin{quote}
Vielen Dank für Ihr Interesse an unserer Forschung und an dieser Studie!
In dieser Umfrage werden Sie einige Fragen zu den Einsatzmöglichkeiten von Augmented-Reality (AR)-Brillen und Smartphones beantworten. Sie müssen nicht unbedingt Erfahrung mit AR-Brillen haben.

\textit{AR Glasses briefing:} Stellen Sie sich vor, Sie sind Besitzer einer Augmented-Reality (AR)-Brille, die Sie auf dem Kopf tragen. Diese AR-Brille kann mit ihren integrierten Sensoren Sie und Ihre Umgebung wahrnehmen und verstehen, wohin Sie schauen. Zusätzlich zu diesen\\ Wahrnehmungsmöglichkeiten kann Ihre AR-Brille auch Hologramme in Ihrem Blickfeld anzeigen. Stellen Sie sich vor, Sie installieren eine persönliche Assistenten-App in Ihrer AR-Brille, um Ihre täglichen Aufgaben zu erledigen.

\textit{Smartphone briefing:} Stellen Sie sich vor, Sie besitzen ein Smartphone, das Sie und Ihre Umgebung mit seinen integrierten Sensoren wahrnehmen kann, und Sie installieren eine persönliche Assistenten-App auf Ihrem Smartphone, um Ihre täglichen Aufgaben zu erledigen.
\end{quote}

\subsubsection{Priming texts \& Question template}
In Table~\ref{tab_priming_texts_Germany}, priming texts for beneficial and not-beneficial priming conditions are provided. After the following priming texts for each attribute, depending on the condition, the following question is asked in the calibration survey. 

Wie nützlich oder schädlich würden Sie eine persönliche Assistenten-App auf [Ihrer AR Brille/Ihrem Smartphone] finden, die dies tut?

$\square$ Sehr schädlich $\square$ Mäßig schädlich $\square$ Weder nützlich noch schädlich $\square$ Mäßig nützlich $\square$ Sehr nützlich

\begin{table*}[h!]
    \caption{User attributes and German versions of the priming texts for beneficial and not-beneficial priming.}
    \alt{Table that includes evaluated user attributes, beneficial and not-beneficial priming texts for the German study. The first column includes eleven user attributes, whereas the second and third columns include beneficial and not-beneficial priming texts for corresponding attributes in German.}
    \centering
    \footnotesize
    \begin{tabular}{|p{1.5cm}|p{7.5cm}|p{7.5cm}|}
        \hline
        \textbf{Attribute} & \textbf{Beneficial priming text} & \textbf{Not-beneficial priming text} \\
        \hline
        activity & Angenommen, Ihre persönliche Assistenten-App auf [Ihrer AR-Brille/Ihrem Smartphone] gibt Ihnen während der Arbeit oder des Studiums Hinweise, die Ihnen helfen, Ihre Arbeit zu optimieren. & Angenommen, Ihre persönliche Assistenten-App auf [Ihrer AR-Brille/Ihrem Smartphone] beobachtet Ihre Aktivitäten während der Arbeit oder des Studiums und meldet Ihrem Arbeitgeber oder Ihrer Universität Ihr Produktivitätsniveau. \\
        \hline
        alertness & Angenommen, Ihre persönliche Assistenten-App auf [Ihrer AR-Brille/Ihrem Smartphone] warnt Sie vor potenziellen Gefahren, die Sie sonst beim Autofahren oder Radfahren nicht bemerken würden. & Angenommen, Ihre persönliche Assistenten-App auf [Ihrer AR-Brille/Ihrem Smartphone] warnt Sie ständig, dass Sie beim Autofahren müde sind. \\
        \hline
        body mass index & Angenommen, Ihre persönliche Assistenten-App auf [Ihrer AR-Brille/Ihrem Smartphone] schlägt Ihnen auf der Grundlage Ihres Body-Mass-Index gesunde Ernährung und Sportübungen vor. & Angenommen, Ihre persönliche Assistenten-App auf [Ihrer AR-Brille/Ihrem Smartphone] schlägt Ihnen Snacks vor, die Sie aufgrund Ihrer ungesunden Ernährungsgewohnheiten mögen könnten. \\
        \hline
        cognitive load & Angenommen, Ihre persönliche Assistenten-App auf [Ihrer AR-Brille/Ihrem Smartphone] erkennt, dass Sie während einer Aufgabe kognitiv überlastet sind, und schlägt Ihnen vor, eine Weile abzuschalten. & Angenommen, Ihre persönliche Assistenten-App auf [Ihrer AR-Brille/Ihrem Smartphone] versucht, Ihre kognitive Belastung während Ihrer täglichen Routine ständig zu maximieren. \\
        \hline
        depression & Angenommen, Ihre persönliche Assistenten-App auf [Ihrer AR-Brille/Ihrem Smartphone] stellt fest, dass Sie depressiv sind, und bietet Ihnen Lösungen zur Verbesserung Ihres Wohlbefindens an. & Angenommen, Ihre persönliche Assistenten-App auf [Ihrer AR-Brille/Ihrem Smartphone] schlägt Ihnen auf der Grundlage von Inhalten aus dem Internet immer wieder vor, wie Sie Ihre psychische Gesundheit verbessern können. \\
        \hline
        gender & Angenommen, Ihre persönliche Assistenten-App auf [Ihrer AR-Brille/Ihrem Smartphone] schlägt Ihnen Produkte vor, die Sie aufgrund Ihres Geschlechts interessieren könnten. & Angenommen, Ihre persönliche Assistenten-App auf [Ihrer AR-Brille/Ihrem Smartphone] liefert ihnen während der Arbeit Produktvorschläge auf der Grundlage von Geschlechterstereotypen. \\
        \hline
        heart \mbox{condition} & Angenommen, Ihre persönliche Assistenten-App auf [Ihrer AR-Brille/Ihrem Smartphone] informiert Sie darüber, dass Sie möglicherweise ein Herzleiden haben, und schlägt vor, dass Sie einen Arzt aufsuchen sollten. & Angenommen, Ihre persönliche Assistenten-App auf [Ihrer AR-Brille/Ihrem Smartphone] informiert Ihre Versicherung und Ihren Arbeitgeber über Ihr bereits bestehendes Herzleiden. \\
        \hline
        location & Angenommen, Ihre persönliche Assistenten-App auf [Ihrer AR-Brille/Ihrem Smartphone] bietet Ihnen eine freihändige Navigationshilfe. & Angenommen, Ihre persönliche Assistenten-App auf [Ihrer AR-Brille/Ihrem Smartphone] verfolgt Sie anhand Ihres Standorts in Innenräumen, z. B. in einem Café, im Büro oder zu Hause, um Ihnen auf der Grundlage Ihres Standorts gezielt Werbung anzuzeigen. \\
        \hline
        personal identity & Angenommen, Ihre persönliche Assistenten-App auf [Ihrer AR-Brille/Ihrem Smartphone] ermöglicht Ihnen die Authentifizierung ohne Passwort und personalisiert das Gerät für Sie. & Angenommen, Ihre persönliche Assistenten-App auf [Ihrer AR-Brille/Ihrem Smartphone] ermöglicht Ihnen die Authentifizierung ohne Passwort und personalisiert das Gerät für Sie, was jedoch das Risiko eines Identitätsdiebstahls erhöht.\\
        \hline
        sexual \mbox{preference} & Angenommen, Ihre persönliche Assistenten-App auf [Ihrer AR-Brille/Ihrem Smartphone] bietet Ihnen eine Einschätzung darüber an, ob die Person, an der Sie interessiert sind, Sie attraktiv findet. & Angenommen, Ihre persönliche Assistenten-App auf [Ihrer AR-Brille/Ihrem Smartphone] schlägt Ihnen während der Arbeit immer wieder Veranstaltungen vor, die auf Ihrer sexuellen Orientierung basieren. \\
        \hline
        stress & Angenommen, Ihre persönliche Assistenten-App auf [Ihrer AR-Brille/Ihrem Smartphone] schlägt Ihnen vor, eine Pause von einer stressigen Tätigkeit einzulegen und Aktivitäten zu unternehmen, die Ihnen Spaß machen. & Angenommen, Ihre persönliche Assistenten-App auf [Ihrer AR-Brille/Ihrem Smartphone] erinnert Sie ständig an Punkte auf Ihrer Aufgabenliste. \\
        \hline
    \end{tabular}
    \label{tab_priming_texts_Germany}
\end{table*}

The privacy concerns survey question is as follows. 

Wie würden Sie sich fühlen, wenn Ihre persönliche Assistenten-App auf [Ihrer AR Brille/Ihrem Smartphone] anhand der von [Ihrer AR Brille/Ihrem Smartphone] gesammelten Daten [attribute] feststellen, die gesammelten Daten mit [data-receiving entity] teilen und die gesammelten Daten [data retention time] speichern könnte?

$\square$ Sehr unwohl $\square$ Unwohl $\square$ Neutral $\square$ Wohl $\square$ Sehr wohl

\subsubsection{Demographic questionnaire}
In the following, we provide demographic questions that we asked to characterize our study's sample pools, which are adapted from previous work~\cite{leon_etal_2013}. \\
Briefing: In diesem Teil unserer Studie werden wir Sie nach Ihren demografischen Daten fragen.

\begin{itemize}
  \item Wie alt sind Sie? (In Jahren) \underline{\hspace{0.5cm}}
  \item Was ist Ihr Geschlecht? $\square$ Frau $\square$ Mann $\square$ Nicht-binär $\square$ Möchte keine Angaben machen $\square$ Bevorzuge die Selbstbeschreibung als \underline{\hspace{0.5cm}}
  \item Welche der folgenden Angaben beschreibt am besten Ihre Hauptbeschäftigung? $\square$ Administrative Unterstützung (z. B. Sekretär-in, Assistent-in) $\square$ Kunst, Schreiben oder Journalismus (z. B. Autor, Reporter, Bildhauer) $\square$ Wirtschaft, Management oder Finanzen (z. B. Manager, Buchhalter, Banker) $\square$ Computeringenieur oder IT-Fachmann (z. B. Systemadministrator, Programmierer, IT-Berater) $\square$ Bildung (z. B. Lehrer-in) $\square$ Ingenieur in anderen Bereichen (z. B. Bauingenieur, Bio-Ingenieur) $\square$ Hausfrau oder Hausmann $\square$ Jurist (z. B. Rechtsanwalt, Rechtsreferendar) $\square$ Medizinisch (z. B. Arzt, Krankenschwester, Zahnarzt) $\square$ Im Ruhestand $\square$ Wissenschaftler (z. B. Forscher, Professor) $\square$ Dienstleistung (z. B. Einzelhandelsangestellte, Server) $\square$ Fachkräfte (z. B. Elektriker, Klempner, Zimmerleute) $\square$ Studierende $\square$ Arbeitslos $\square$ Antwort ablehnen $\square$ Sonstiges (bitte angeben): \underline{\hspace{0.5cm}}
  \item Welche der folgenden Angaben beschreibt Ihren höchsten Bildungsabschluss am besten? $\square$ Ohne allgemeinbildenden Schulabschluss $\square$ Haupt-(Volks-)schulabschluss $\square$ Mittlerer Abschluss $\square$ Fachhochschul- oder Hochschulreife $\square$ Abgeschlossene Berufsausbildung $\square$ Fachhochschulabschluss $\square$ Hochschulabschluss - Bachelor/4-Jahres-Abschluss $\square$ Hochschulabschluss – Master $\square$ Promotion
  \item Haben Sie schon einmal eine Augmented-Reality-Brille besessen? $\square$ Ja $\square$ Nein
  \item Wenn Sie derzeit eine Augmented-Reality-Brille verwenden, wie viele Stunden verbringen Sie dann täglich mit einer Augmented-Reality-Brille? $\square$ \underline{\hspace{0.5cm}} $\square$ Nicht zutreffend
  \item Besitzen Sie ein Smartphone? $\square$ Ja $\square$ Nein
  \item Wenn ja, wie viele Stunden verbringen Sie täglich mit Ihrem Smartphone? $\square$ \underline{\hspace{0.5cm}} $\square$ Nicht zutreffend
  \item Wie viele Stunden verbringen Sie ungefähr täglich im Internet? $\square$ \underline{\hspace{0.5cm}} $\square$ Nicht zutreffend
\end{itemize}

\section{Appendix B: Detailed statistics}
\begin{figure}[ht]
  \centering
   \includegraphics[width=0.45\linewidth, keepaspectratio]{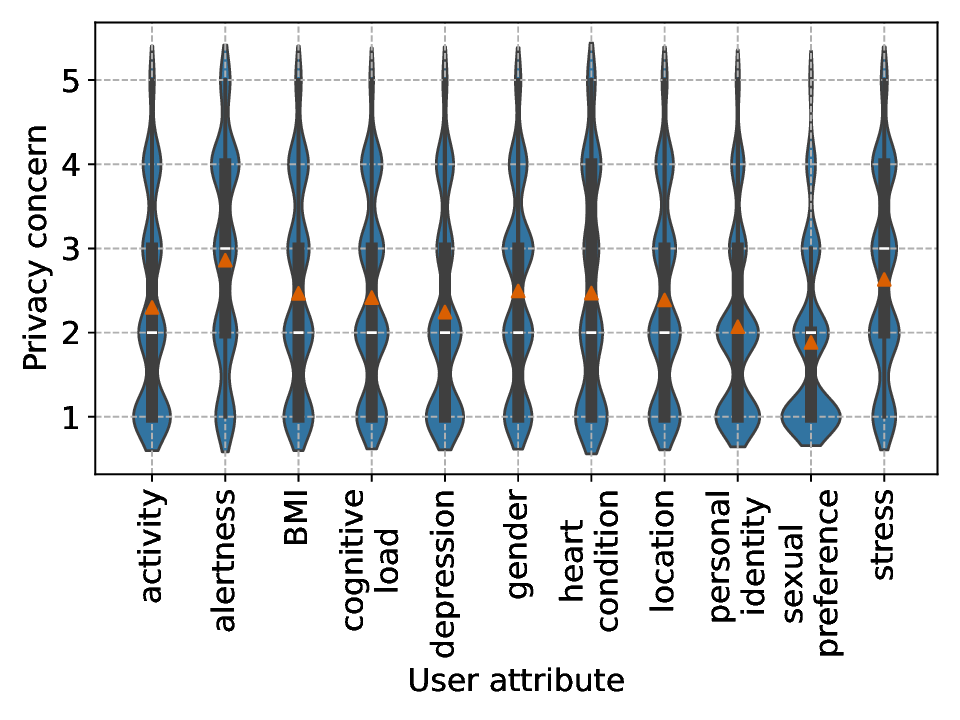}
  \caption{Violin plot representing the relationship between privacy concerns and user attribute, all the factors and conditions combined.}
  \alt{Violin plots representing the relationship between privacy concerns and user attributes for our main survey. The Y-axis in the plot represents privacy concerns ranging from 1 to 5, with 1 being very uncomfortable and 5 being very comfortable. The X-axis corresponds to eleven user attributes.}
  \label{fig_results_combined_per_attribute}%
\end{figure}

This section provides detailed statistics regarding regression analyses and mean and standard deviations from the analyses. In Figure~\ref{fig_results_combined_per_attribute}, we provide privacy concerns results of all conditions per attribute. We indicate mean values with orange triangles. 

While we did not conduct statistical tests between the privacy concerns scores of user attributes as we do not explore this in our RQs, we explored summary statistics for each attribute. When we combine all conditions per user attribute and sort them, we found the following order of attributes from the most concerning to the least concerning in terms of privacy: sexual preference ($M = 1.88 \pm 1.05$), personal identity ($M = 2.07 \pm 1.09$), depression ($M = 2.24 \pm 1.20$), activity ($M = 2.30 \pm 1.23$), location ($M = 2.39 \pm 1.20$), cognitive load ($M = 2.42 \pm 1.16$), BMI ($M = 2.47 \pm 1.22$), heart condition ($M = 2.47 \pm 1.34$), gender ($M = 2.49 \pm 1.17$), stress ($M = 2.63 \pm 1.20$), and alertness ($M = 2.86 \pm 1.27$). As observed in the main regression analysis in our paper, participants' overall comfort regarding their privacy rarely reaches the neutral level. Tables between~\ref{tab_ordered_regression_result_activity}-\ref{tab_ordered_regression_result_stress} report detailed regression statistics that we conducted for each attribute. In the regression tables, we indicate retention time and data-receiving entity as RT and DR, respectively. Later, in Tables between~\ref{tab_main_survey_country_mean_std}-\ref{tab_main_surveys_priming_mean_std}, we report detailed mean and standard deviation values from our distributions in privacy concerns survey. Lastly, Table~\ref{tab_priming_survey_mean_std} provides mean and standard deviation values from our calibration surveys. 

\begin{table}[ht]
\centering
\caption{Ordinal regression results for activity.}
\alt{Ordinal regression table for activity attribute. The first column represents variables in our regression analyses. The columns from second to seventh represent various regression analysis statistics.}
\label{tab_ordered_regression_result_activity}
\small
\begin{tabular}{@{}lcccccc@{}}
\toprule
& \multicolumn{5}{c}{Coefficients} & \\
\cmidrule(r){2-7}
Variable & Coeff & SE & z & $p$-value & [0.025 & 0.975] \\
\midrule
Age & -0.0091 & 0.003 & -2.691 & 0.007 & -0.016 & -0.002 \\
Device & -0.0817 & 0.084 & -0.978 & 0.328 & -0.245 & 0.082 \\
Priming & 1.1128 & 0.048 & 22.956 & 0.000 & 1.018 & 1.208 \\
Internet use & -0.0083 & 0.013 & -0.615 & 0.539 & -0.035 & 0.018 \\
AR glasses ownership & -0.5942 & 0.146 & -4.073 & 0.000 & -0.880 & -0.308 \\
RT (1 month) & -0.0004 & 0.102 & -0.004 & 0.997 & -0.201 & 0.200 \\
RT (Indefinite) & -0.1767 & 0.103 & -1.718 & 0.086 & -0.378 & 0.025 \\
DR (friends) & -1.1449 & 0.147 & -7.779 & 0.000 & -1.433 & -0.856 \\
DR (work contacts) & -1.0086 & 0.149 & -6.777 & 0.000 & -1.300 & -0.717 \\
DR (app company employees) & -0.4323 & 0.142 & -3.039 & 0.002 & -0.711 & -0.153 \\
DR (app cloud) & -0.1398 & 0.142 & -0.984 & 0.325 & -0.418 & 0.139 \\
DR (public) & -0.9137 & 0.145 & -6.292 & 0.000 & -1.198 & -0.629 \\
Gender (man) & 0.2870 & 0.089 & 3.228 & 0.001 & 0.113 & 0.461 \\
Gender (non-binary) & -0.0428 & 0.309 & -0.139 & 0.890 & -0.649 & 0.563 \\
Gender (self-described) & -1.5260 & 0.649 & -2.353 & 0.019 & -2.797 & -0.255 \\
Gender (do not say) & -0.0768 & 0.765 & -0.100 & 0.920 & -1.577 & 1.423 \\
Country (Germany) & -0.3224 & 0.098 & -3.273 & 0.001 & -0.515 & -0.129 \\
Smartphone use & 0.0931 & 0.016 & 5.835 & 0.000 & 0.062 & 0.124 \\
Comfort (1/2) & 1.1628 & 0.302 & 3.844 & 0.000 & 0.570 & 1.756 \\
Comfort (2/3) & 0.3288 & 0.040 & 8.150 & 0.000 & 0.250 & 0.408 \\
Comfort (3/4) & 0.1130 & 0.048 & 2.337 & 0.019 & 0.018 & 0.208 \\
Comfort (4/5) & 0.7189 & 0.053 & 13.583 & 0.000 & 0.615 & 0.823 \\
\bottomrule
\end{tabular}
\end{table}

\begin{table}[ht]
\centering
\caption{Ordinal regression results for alertness.}
\alt{Ordinal regression table for activity alertness. The first column represents variables in our regression analyses. The columns from second to seventh represent various regression analysis statistics.}
\label{tab_ordered_regression_result_alertness}
\small
\begin{tabular}{@{}lcccccc@{}}
\toprule
& \multicolumn{5}{c}{Coefficients} & \\
\cmidrule(r){2-7}
Variable & Coeff & SE & z & $p$-value & [0.025 & 0.975] \\
\midrule
Age & -0.0020 & 0.003 & -0.613 & 0.540 & -0.008 & 0.004 \\
Device & -0.0625 & 0.080 & -0.781 & 0.435 & -0.219 & 0.094 \\
Priming & 0.8722 & 0.067 & 12.954 & 0.000 & 0.740 & 1.004 \\
Internet use & -0.0011 & 0.013 & -0.089 & 0.929 & -0.026 & 0.024 \\
AR glasses ownership & -0.3034 & 0.139 & -2.186 & 0.029 & -0.575 & -0.031 \\
RT (1 month) & 0.1108 & 0.099 & 1.123 & 0.261 & -0.083 & 0.304 \\
RT (Indefinite) & -0.0858 & 0.099 & -0.866 & 0.386 & -0.280 & 0.108 \\
DR (friends) & -0.8238 & 0.140 & -5.904 & 0.000 & -1.097 & -0.550 \\
DR (work contacts) & -1.2958 & 0.143 & -9.088 & 0.000 & -1.575 & -1.016 \\
DR (app company employees) & -0.5104 & 0.138 & -3.692 & 0.000 & -0.781 & -0.239 \\
DR (app cloud) & 0.0749 & 0.139 & 0.540 & 0.589 & -0.197 & 0.347 \\
DR (public) & -0.6329 & 0.140 & -4.526 & 0.000 & -0.907 & -0.359 \\
Gender (man) & 0.3649 & 0.086 & 4.257 & 0.000 & 0.197 & 0.533 \\
Gender (non-binary) & -0.0811 & 0.288 & -0.282 & 0.778 & -0.645 & 0.483 \\
Gender (self-described) & -1.7617 & 0.613 & -2.872 & 0.004 & -2.964 & -0.559 \\
Gender (do not say) & 0.0013 & 0.710 & 0.002 & 0.999 & -1.390 & 1.393 \\
Country (Germany) & -0.0380 & 0.093 & -0.409 & 0.683 & -0.220 & 0.144 \\
Smartphone use & 0.0670 & 0.015 & 4.327 & 0.000 & 0.037 & 0.097 \\
Comfort (1/2) & 1.1655 & 0.360 & 3.235 & 0.001 & 0.459 & 1.872 \\
Comfort (2/3) & 0.2292 & 0.043 & 5.363 & 0.000 & 0.145 & 0.313 \\
Comfort (3/4) & -0.0031 & 0.044 & -0.071 & 0.944 & -0.089 & 0.083 \\
Comfort (4/5) & 0.6273 & 0.040 & 15.586 & 0.000 & 0.548 & 0.706 \\
\bottomrule
\end{tabular}
\end{table}

\begin{table}[ht]
\centering
\caption{Ordinal regression results for BMI.}
\alt{Ordinal regression table for BMI attribute. The first column represents variables in our regression analyses. The columns from second to seventh represent various regression analysis statistics.}
\label{tab_ordered_regression_result_bmi}
\small
\begin{tabular}{@{}lcccccc@{}}
\toprule
& \multicolumn{5}{c}{Coefficients} & \\
\cmidrule(r){2-7}
Variable & Coeff & SE & z & $p$-value & [0.025 & 0.975] \\
\midrule
Age & -0.0037 & 0.003 & -1.152 & 0.249 & -0.010 & 0.003 \\
Device & 0.0022 & 0.081 & 0.027 & 0.978 & -0.156 & 0.160 \\
Priming & 0.5880 & 0.085 & 6.930 & 0.000 & 0.422 & 0.754 \\
Internet use & -0.0138 & 0.013 & -1.056 & 0.291 & -0.039 & 0.012 \\
AR glasses ownership & -0.4784 & 0.140 & -3.427 & 0.001 & -0.752 & -0.205 \\
RT (1 month) & 0.0277 & 0.099 & 0.279 & 0.780 & -0.167 & 0.222 \\
RT (Indefinite) & -0.2866 & 0.099 & -2.908 & 0.004 & -0.480 & -0.093 \\
DR (friends) & -1.3382 & 0.143 & -9.358 & 0.000 & -1.619 & -1.058 \\
DR (work contacts) & -1.3178 & 0.144 & -9.132 & 0.000 & -1.601 & -1.035 \\
DR (app company employees) & -0.7059 & 0.139 & -5.063 & 0.000 & -0.979 & -0.433 \\
DR (app cloud) & -0.2670 & 0.139 & -1.922 & 0.055 & -0.539 & 0.005 \\
DR (public) & -1.0891 & 0.142 & -7.689 & 0.000 & -1.367 & -0.811 \\
Gender (man) & 0.6485 & 0.087 & 7.492 & 0.000 & 0.479 & 0.818 \\
Gender (non-binary) & -0.9259 & 0.330 & -2.810 & 0.005 & -1.572 & -0.280 \\
Gender (self-described) & -0.1512 & 0.558 & -0.271 & 0.787 & -1.246 & 0.943 \\
Gender (do not say) & 0.3346 & 0.664 & 0.504 & 0.614 & -0.966 & 1.635 \\
Country (Germany) & -0.4448 & 0.095 & -4.680 & 0.000 & -0.631 & -0.259 \\
Smartphone use & 0.0640 & 0.015 & 4.162 & 0.000 & 0.034 & 0.094 \\
Comfort (1/2) & -0.4046 & 0.378 & -1.070 & 0.284 & -1.146 & 0.336 \\
Comfort (2/3) & 0.2794 & 0.038 & 7.309 & 0.000 & 0.204 & 0.354 \\
Comfort (3/4) & 0.0318 & 0.046 & 0.699 & 0.484 & -0.057 & 0.121 \\
Comfort (4/5) & 0.6171 & 0.051 & 12.148 & 0.000 & 0.518 & 0.717 \\
\bottomrule
\end{tabular}
\end{table}

\begin{table}[ht]
\centering
\caption{Ordinal regression results for cognitive load.}
\alt{Ordinal regression table for cognitive load attribute. The first column represents variables in our regression analyses. The columns from second to seventh represent various regression analysis statistics.}
\label{tab_ordered_regression_result_cognload}
\small
\begin{tabular}{@{}lcccccc@{}}
\toprule
& \multicolumn{5}{c}{Coefficients} & \\
\cmidrule(r){2-7}
Variable & Coeff & SE & z & $p$-value & [0.025 & 0.975] \\
\midrule
Age & -0.0060 & 0.003 & -1.844 & 0.065 & -0.012 & 0.000 \\
Device & -0.0231 & 0.081 & -0.287 & 0.774 & -0.181 & 0.135 \\
Priming & 0.5933 & 0.064 & 9.315 & 0.000 & 0.468 & 0.718 \\
Internet use & -0.0068 & 0.013 & -0.518 & 0.605 & -0.033 & 0.019 \\
AR glasses ownership & -0.5800 & 0.143 & -4.054 & 0.000 & -0.860 & -0.300 \\
RT (1 month) & -0.0683 & 0.099 & -0.687 & 0.492 & -0.263 & 0.126 \\
RT (Indefinite) & -0.2384 & 0.099 & -2.408 & 0.016 & -0.432 & -0.044 \\
DR (friends) & -1.0702 & 0.142 & -7.555 & 0.000 & -1.348 & -0.793 \\
DR (work contacts) & -1.4316 & 0.146 & -9.795 & 0.000 & -1.718 & -1.145 \\
DR (app company employees) & -0.6258 & 0.138 & -4.525 & 0.000 & -0.897 & -0.355 \\
DR (app cloud) & -0.1583 & 0.138 & -1.144 & 0.253 & -0.430 & 0.113 \\
DR (public) & -0.9884 & 0.142 & -6.979 & 0.000 & -1.266 & -0.711 \\
Gender (man) & 0.2784 & 0.086 & 3.237 & 0.001 & 0.110 & 0.447 \\
Gender (non-binary) & -0.7739 & 0.312 & -2.477 & 0.013 & -1.386 & -0.162 \\
Gender (self-described) & -2.1674 & 0.685 & -3.165 & 0.002 & -3.509 & -0.825 \\
Gender (do not say) & -1.3594 & 0.867 & -1.567 & 0.117 & -3.060 & 0.341 \\
Country (Germany) & -0.2788 & 0.095 & -2.923 & 0.003 & -0.466 & -0.092 \\
Smartphone use & 0.0929 & 0.016 & 5.922 & 0.000 & 0.062 & 0.124 \\
Comfort (1/2) & -0.4560 & 0.339 & -1.345 & 0.179 & -1.121 & 0.209 \\
Comfort (2/3) & 0.3529 & 0.037 & 9.590 & 0.000 & 0.281 & 0.425 \\
Comfort (3/4) & 0.1297 & 0.044 & 2.952 & 0.003 & 0.044 & 0.216 \\
Comfort (4/5) & 0.8004 & 0.054 & 14.728 & 0.000 & 0.694 & 0.907 \\
\bottomrule
\end{tabular}
\end{table}

\begin{table}[ht]
\centering
\caption{Ordinal regression results for depression.}
\alt{Ordinal regression table for depression attribute. The first column represents variables in our regression analyses. The columns from second to seventh represent various regression analysis statistics.}
\label{tab_ordered_regression_result_depression}
\small
\begin{tabular}{@{}lcccccc@{}}
\toprule
& \multicolumn{5}{c}{Coefficients} & \\
\cmidrule(r){2-7}
Variable & Coeff & SE & z & $p$-value & [0.025 & 0.975] \\
\midrule
Age & -0.0107 & 0.003 & -3.287 & 0.001 & -0.017 & -0.004 \\
Device & -0.1194 & 0.085 & -1.397 & 0.162 & -0.287 & 0.048 \\
Priming & 0.6583 & 0.095 & 6.909 & 0.000 & 0.472 & 0.845 \\
Internet use & -0.0156 & 0.013 & -1.165 & 0.244 & -0.042 & 0.011 \\
AR glasses ownership & -0.3924 & 0.143 & -2.740 & 0.006 & -0.673 & -0.112 \\
RT (1 month) & 0.0052 & 0.100 & 0.051 & 0.959 & -0.192 & 0.202 \\
RT (Indefinite) & -0.3102 & 0.101 & -3.081 & 0.002 & -0.508 & -0.113 \\
DR (friends) & -1.1141 & 0.143 & -7.805 & 0.000 & -1.394 & -0.834 \\
DR (work contacts) & -1.3636 & 0.149 & -9.143 & 0.000 & -1.656 & -1.071 \\
DR (app company employees) & -0.4598 & 0.139 & -3.307 & 0.001 & -0.732 & -0.187 \\
DR (app cloud) & 0.0969 & 0.139 & 0.699 & 0.485 & -0.175 & 0.369 \\
DR (public) & -0.8751 & 0.141 & -6.209 & 0.000 & -1.151 & -0.599 \\
Gender (man) & 0.2489 & 0.087 & 2.852 & 0.004 & 0.078 & 0.420 \\
Gender (non-binary) & -0.7279 & 0.319 & -2.284 & 0.022 & -1.352 & -0.103 \\
Gender (self-described) & -1.1674 & 0.631 & -1.849 & 0.065 & -2.405 & 0.070 \\
Gender (do not say) & 0.5700 & 0.737 & 0.773 & 0.439 & -0.875 & 2.015 \\
Country (Germany) & -0.5951 & 0.097 & -6.148 & 0.000 & -0.785 & -0.405 \\
Smartphone use & 0.1088 & 0.016 & 7.008 & 0.000 & 0.078 & 0.139 \\
comfort (1/2) & 0.1745 & 0.408 & 0.428 & 0.669 & -0.625 & 0.974 \\
comfort (2/3) & 0.3460 & 0.036 & 9.487 & 0.000 & 0.274 & 0.417 \\
comfort (3/4) & -0.1612 & 0.054 & -2.993 & 0.003 & -0.267 & -0.056 \\
comfort (4/5) & 0.6294 & 0.057 & 11.073 & 0.000 & 0.518 & 0.741 \\
\bottomrule
\end{tabular}
\end{table}

\begin{table}[ht]
\centering
\caption{Ordinal regression results for gender.}
\alt{Ordinal regression table for gender attribute. The first column represents variables in our regression analyses. The columns from second to seventh represent various regression analysis statistics.}
\label{tab_ordered_regression_result_gender}
\small
\begin{tabular}{@{}lcccccc@{}}
\toprule
& \multicolumn{5}{c}{Coefficients} & \\
\cmidrule(r){2-7}
Variable & Coeff & SE & z & $p$-value & [0.025 & 0.975] \\
\midrule
Age & -0.0014 & 0.003 & -0.443 & 0.658 & -0.008 & 0.005 \\
Device & -0.0500 & 0.081 & -0.618 & 0.537 & -0.209 & 0.109 \\
Priming & 1.1547 & 0.133 & 8.675 & 0.000 & 0.894 & 1.416 \\
Internet use & -0.0039 & 0.013 & -0.298 & 0.766 & -0.029 & 0.022 \\
AR glasses ownership & -0.3362 & 0.140 & -2.405 & 0.016 & -0.610 & -0.062 \\
RT (1 month) & 0.0085 & 0.098 & 0.087 & 0.931 & -0.184 & 0.201 \\
RT (Indefinite) & -0.0142 & 0.098 & -0.144 & 0.885 & -0.207 & 0.178 \\
DR (friends) & -0.6640 & 0.140 & -4.740 & 0.000 & -0.939 & -0.389 \\
DR (work contacts) & -0.7351 & 0.142 & -5.172 & 0.000 & -1.014 & -0.457 \\
DR (app company employees) & -0.3297 & 0.139 & -2.372 & 0.018 & -0.602 & -0.057 \\
DR (app cloud) & -0.1060 & 0.140 & -0.759 & 0.448 & -0.379 & 0.168 \\
DR (public) & -0.5391 & 0.140 & -3.842 & 0.000 & -0.814 & -0.264 \\
Gender (man) & 0.3955 & 0.085 & 4.630 & 0.000 & 0.228 & 0.563 \\
Gender (non-binary) & -1.2200 & 0.324 & -3.763 & 0.000 & -1.855 & -0.585 \\
Gender (self-described) & -0.9507 & 0.553 & -1.718 & 0.086 & -2.035 & 0.134 \\
Gender (do not say) & -1.1286 & 0.760 & -1.485 & 0.138 & -2.618 & 0.361 \\
Country (Germany) & -0.4780 & 0.095 & -5.022 & 0.000 & -0.664 & -0.291 \\
Smartphone use & 0.0303 & 0.015 & 1.980 & 0.048 & 0.000 & 0.060 \\
Comfort (1/2) & 1.2135 & 0.434 & 2.795 & 0.005 & 0.362 & 2.064 \\
Comfort (2/3) & 0.2142 & 0.039 & 5.468 & 0.000 & 0.137 & 0.291 \\
Comfort (3/4) & 0.2879 & 0.039 & 7.434 & 0.000 & 0.212 & 0.364 \\
Comfort (4/5) & 0.5240 & 0.056 & 9.423 & 0.000 & 0.415 & 0.633 \\
\bottomrule
\end{tabular}
\end{table}

\begin{table}[ht]
\centering
\caption{Ordinal regression results for heart condition.}
\alt{Ordinal regression table for heart condition attribute. The first column represents variables in our regression analyses. The columns from second to seventh represent various regression analysis statistics.}
\label{tab_ordered_regression_result_heart}
\small
\begin{tabular}{@{}lcccccc@{}}
\toprule
& \multicolumn{5}{c}{Coefficients} & \\
\cmidrule(r){2-7}
Variable & Coeff & SE & z & $p$-value & [0.025 & 0.975] \\
\midrule
Age & -0.0145 & 0.003 & -4.366 & 0.000 & -0.021 & -0.008 \\
Device & -0.0067 & 0.082 & -0.082 & 0.934 & -0.167 & 0.153 \\
Priming & 0.7862 & 0.042 & 18.519 & 0.000 & 0.703 & 0.869 \\
Internet use & -0.0004 & 0.013 & -0.027 & 0.979 & -0.026 & 0.026 \\
AR glasses ownership & -0.5313 & 0.142 & -3.746 & 0.000 & -0.809 & -0.253 \\
RT (1 month) & -0.0521 & 0.101 & -0.515 & 0.606 & -0.250 & 0.146 \\
RT (Indefinite) & -0.0185 & 0.100 & -0.185 & 0.853 & -0.215 & 0.178 \\
DR (friends) & -0.8984 & 0.143 & -6.274 & 0.000 & -1.179 & -0.618 \\
DR (work contacts) & -0.8922 & 0.144 & -6.209 & 0.000 & -1.174 & -0.611 \\
DR (app company employees) & -0.3602 & 0.141 & -2.557 & 0.011 & -0.636 & -0.084 \\
DR (app cloud) & -0.0465 & 0.142 & -0.327 & 0.744 & -0.325 & 0.232 \\
DR (public) & -0.8292 & 0.142 & -5.829 & 0.000 & -1.108 & -0.550 \\
Gender (man) & 0.2683 & 0.087 & 3.076 & 0.002 & 0.097 & 0.439 \\
Gender (non-binary) & -0.8199 & 0.327 & -2.510 & 0.012 & -1.460 & -0.180 \\
Gender (self-described) & -1.7315 & 0.727 & -2.382 & 0.017 & -3.156 & -0.307 \\
Gender (do not say) & -0.9931 & 0.871 & -1.140 & 0.254 & -2.701 & 0.715 \\
Country (Germany) & -0.3939 & 0.096 & -4.113 & 0.000 & -0.582 & -0.206 \\
Smartphone use & 0.0689 & 0.016 & 4.322 & 0.000 & 0.038 & 0.100 \\
comfort (1/2) & 0.2396 & 0.298 & 0.803 & 0.422 & -0.345 & 0.824 \\
comfort (2/3) & 0.1461 & 0.042 & 3.458 & 0.001 & 0.063 & 0.229 \\
comfort (3/4) & -0.1915 & 0.053 & -3.637 & 0.000 & -0.295 & -0.088 \\
comfort (4/5) & 0.5131 & 0.047 & 10.889 & 0.000 & 0.421 & 0.605 \\
\bottomrule
\end{tabular}
\end{table}

\begin{table}[ht]
\centering
\caption{Ordinal regression results for location.}
\alt{Ordinal regression table for location attribute. The first column represents variables in our regression analyses. The columns from second to seventh represent various regression analysis statistics.}
\label{tab_ordered_regression_result_location}
\small
\begin{tabular}{@{}lcccccc@{}}
\toprule
& \multicolumn{5}{c}{Coefficients} & \\
\cmidrule(r){2-7}
Variable & Coeff & SE & z & $p$-value & [0.025 & 0.975] \\
\midrule
Age & -0.0022 & 0.003 & -0.667 & 0.505 & -0.009 & 0.004 \\
Device & 0.0643 & 0.081 & 0.792 & 0.428 & -0.095 & 0.223 \\
Priming & 0.7107 & 0.048 & 14.690 & 0.000 & 0.616 & 0.805 \\
Internet use & -0.0043 & 0.013 & -0.338 & 0.735 & -0.030 & 0.021 \\
AR glasses ownership & -0.5394 & 0.139 & -3.872 & 0.000 & -0.813 & -0.266 \\
RT (1 month) & 0.0170 & 0.099 & 0.171 & 0.864 & -0.178 & 0.212 \\
RT (Indefinite) & -0.2394 & 0.100 & -2.387 & 0.017 & -0.436 & -0.043 \\
DR (friends) & -0.9716 & 0.143 & -6.787 & 0.000 & -1.252 & -0.691 \\
DR (work contacts) & -1.1752 & 0.145 & -8.132 & 0.000 & -1.458 & -0.892 \\
DR (app company employees) & -0.3538 & 0.138 & -2.567 & 0.010 & -0.624 & -0.084 \\
DR (app cloud) & -0.0856 & 0.139 & -0.615 & 0.539 & -0.358 & 0.187 \\
DR (public) & -0.8967 & 0.143 & -6.276 & 0.000 & -1.177 & -0.617 \\
Gender (man) & 0.3520 & 0.087 & 4.054 & 0.000 & 0.182 & 0.522 \\
Gender (non-binary) & -0.1375 & 0.297 & -0.463 & 0.644 & -0.720 & 0.445 \\
Gender (self-described) & -1.8038 & 0.618 & -2.918 & 0.004 & -3.015 & -0.592 \\
Gender (do not say) & -0.1484 & 0.608 & -0.244 & 0.807 & -1.341 & 1.044 \\
Country (Germany) & -0.0693 & 0.096 & -0.725 & 0.469 & -0.257 & 0.118 \\
Smartphone use & 0.0812 & 0.016 & 5.222 & 0.000 & 0.051 & 0.112 \\
Comfort (1/2) & 0.4492 & 0.308 & 1.457 & 0.145 & -0.155 & 1.053 \\
Comfort (2/3) & 0.3437 & 0.037 & 9.268 & 0.000 & 0.271 & 0.416 \\
Comfort (3/4) & 0.1402 & 0.045 & 3.109 & 0.002 & 0.052 & 0.229 \\
Comfort (4/5) & 0.5999 & 0.055 & 10.997 & 0.000 & 0.493 & 0.707 \\
\bottomrule
\end{tabular}
\end{table}

\begin{table}[ht]
\centering
\caption{Ordinal regression results for personal identity.}
\alt{Ordinal regression table for personal identity attribute. The first column represents variables in our regression analyses. The columns from second to seventh represent various regression analysis statistics.}
\label{tab_ordered_regression_result_identity}
\small
\begin{tabular}{@{}lcccccc@{}}
\toprule
& \multicolumn{5}{c}{Coefficients} & \\
\cmidrule(r){2-7}
Variable & Coeff & SE & z & $p$-value & [0.025 & 0.975] \\
\midrule
Age & -0.0015 & 0.003 & -0.450 & 0.653 & -0.008 & 0.005 \\
Device & 0.1340 & 0.083 & 1.610 & 0.107 & -0.029 & 0.297 \\
Priming & 0.4842 & 0.083 & 5.846 & 0.000 & 0.322 & 0.647 \\
Internet use & 0.0007 & 0.013 & 0.054 & 0.957 & -0.025 & 0.027 \\
AR glasses ownership & -0.5547 & 0.142 & -3.911 & 0.000 & -0.833 & -0.277 \\
RT (1 month) & 0.0243 & 0.101 & 0.240 & 0.810 & -0.174 & 0.222 \\
RT (Indefinite) & -0.0959 & 0.102 & -0.944 & 0.345 & -0.295 & 0.103 \\
DR (friends) & -1.1339 & 0.146 & -7.779 & 0.000 & -1.420 & -0.848 \\
DR (work contacts) & -0.8666 & 0.146 & -5.956 & 0.000 & -1.152 & -0.581 \\
DR (app company employees) & -0.6381 & 0.143 & -4.454 & 0.000 & -0.919 & -0.357 \\
DR (app cloud) & -0.1388 & 0.141 & -0.984 & 0.325 & -0.415 & 0.138 \\
DR (public) & -1.1285 & 0.146 & -7.745 & 0.000 & -1.414 & -0.843 \\
Gender (man) & 0.4833 & 0.088 & 5.490 & 0.000 & 0.311 & 0.656 \\
Gender (non-binary) & -0.0966 & 0.300 & -0.322 & 0.747 & -0.684 & 0.491 \\
Gender (self-described) & -2.2187 & 0.784 & -2.829 & 0.005 & -3.756 & -0.682 \\
Gender (do not say) & -0.5422 & 0.672 & -0.807 & 0.420 & -1.858 & 0.774 \\
Country (Germany) & 0.0653 & 0.096 & 0.681 & 0.496 & -0.123 & 0.253 \\
Smartphone use & 0.0635 & 0.016 & 3.990 & 0.000 & 0.032 & 0.095 \\
comfort (1/2) & 0.0690 & 0.357 & 0.193 & 0.847 & -0.630 & 0.768 \\
comfort (2/3) & 0.4789 & 0.033 & 14.334 & 0.000 & 0.413 & 0.544 \\
comfort (3/4) & -0.0279 & 0.055 & -0.502 & 0.615 & -0.137 & 0.081 \\
comfort (4/5) & 0.5469 & 0.071 & 7.675 & 0.000 & 0.407 & 0.687 \\
\bottomrule
\end{tabular}
\end{table}

\begin{table}[ht]
\centering
\caption{Ordinal regression results for sexual preference.}
\alt{Ordinal regression table for sexual preference attribute. The first column represents variables in our regression analyses. The columns from second to seventh represent various regression analysis statistics.}
\label{tab_ordered_regression_result_sexpref}
\small
\begin{tabular}{@{}lcccccc@{}}
\toprule
& \multicolumn{5}{c}{Coefficients} & \\
\cmidrule(r){2-7}
Variable & Coeff & SE & z & $p$-value & [0.025 & 0.975] \\
\midrule
Age & -8.678e-05 & 0.003 & -0.026 & 0.979 & -0.007 & 0.007 \\
Device & -0.0572 & 0.084 & -0.679 & 0.497 & -0.222 & 0.108 \\
Priming & 0.6110 & 0.102 & 5.963 & 0.000 & 0.410 & 0.812 \\
Internet use & -0.0031 & 0.014 & -0.226 & 0.821 & -0.030 & 0.024 \\
AR glasses ownership & -0.2564 & 0.142 & -1.802 & 0.072 & -0.535 & 0.023 \\
RT (1 month) & 0.0787 & 0.103 & 0.765 & 0.444 & -0.123 & 0.280 \\
RT (Indefinite) & -0.0183 & 0.104 & -0.176 & 0.860 & -0.222 & 0.185 \\
DR (friends) & -0.4972 & 0.146 & -3.408 & 0.001 & -0.783 & -0.211 \\
DR (work contacts) & -0.7963 & 0.151 & -5.259 & 0.000 & -1.093 & -0.500 \\
DR (app company employees) & -0.2296 & 0.146 & -1.577 & 0.115 & -0.515 & 0.056 \\
DR (app cloud) & 0.3443 & 0.144 & 2.396 & 0.017 & 0.063 & 0.626 \\
DR (public) & -0.3970 & 0.147 & -2.709 & 0.007 & -0.684 & -0.110 \\
Gender (man) & 0.4431 & 0.090 & 4.943 & 0.000 & 0.267 & 0.619 \\
Gender (non-binary) & -0.6794 & 0.347 & -1.956 & 0.050 & -1.360 & 0.001 \\
Gender (self-described) & -1.7190 & 0.781 & -2.201 & 0.028 & -3.250 & -0.188 \\
Gender (do not say) & -0.6176 & 0.853 & -0.724 & 0.469 & -2.290 & 1.055 \\
Country (Germany) & -0.3752 & 0.101 & -3.703 & 0.000 & -0.574 & -0.177 \\
Smartphone use & 0.0671 & 0.016 & 4.213 & 0.000 & 0.036 & 0.098 \\
comfort (1/2) & 1.2708 & 0.375 & 3.390 & 0.001 & 0.536 & 2.006 \\
comfort (2/3) & 0.2820 & 0.038 & 7.469 & 0.000 & 0.208 & 0.356 \\
comfort (3/4) & 0.1469 & 0.056 & 2.630 & 0.009 & 0.037 & 0.256 \\
comfort (4/5) & 0.4469 & 0.086 & 5.179 & 0.000 & 0.278 & 0.616 \\
\bottomrule
\end{tabular}
\end{table}

\begin{table}[ht]
\centering
\caption{Ordinal regression results for stress.}
\alt{Ordinal regression table for stress attribute. The first column represents variables in our regression analyses. The columns from second to seventh represent various regression analysis statistics.}
\label{tab_ordered_regression_result_stress}
\small
\begin{tabular}{@{}lcccccc@{}}
\toprule
& \multicolumn{5}{c}{Coefficients} & \\
\cmidrule(r){2-7}
Variable & Coeff & SE & z & $p$-value & [0.025 & 0.975] \\
\midrule
Age & -0.0142 & 0.003 & -4.374 & 0.000 & -0.021 & -0.008 \\
Device & 0.0230 & 0.080 & 0.287 & 0.774 & -0.134 & 0.180 \\
Priming & 0.8710 & 0.096 & 9.069 & 0.000 & 0.683 & 1.059 \\
Internet use & -0.0111 & 0.013 & -0.850 & 0.396 & -0.037 & 0.015 \\
AR glasses ownership & -0.3680 & 0.143 & -2.580 & 0.010 & -0.648 & -0.088 \\
RT (1 month) & -0.0506 & 0.099 & -0.510 & 0.610 & -0.245 & 0.144 \\
RT (Indefinite) & -0.4059 & 0.099 & -4.109 & 0.000 & -0.600 & -0.212 \\
DR (friends) & -1.1112 & 0.142 & -7.817 & 0.000 & -1.390 & -0.833 \\
DR (work contacts) & -1.3378 & 0.144 & -9.259 & 0.000 & -1.621 & -1.055 \\
DR (app company employees) & -0.5686 & 0.138 & -4.111 & 0.000 & -0.840 & -0.298 \\
DR (app cloud) & 0.0288 & 0.140 & 0.206 & 0.837 & -0.245 & 0.303 \\
DR (public) & -0.8332 & 0.140 & -5.971 & 0.000 & -1.107 & -0.560 \\
Gender (man) & 0.3230 & 0.086 & 3.770 & 0.000 & 0.155 & 0.491 \\
Gender (non-binary) & -0.5709 & 0.298 & -1.919 & 0.055 & -1.154 & 0.012 \\
Gender (self-described) & -1.9402 & 0.641 & -3.029 & 0.002 & -3.196 & -0.685 \\
Gender (do not say) & 0.3445 & 0.770 & 0.447 & 0.655 & -1.165 & 1.854 \\
Country (Germany) & -0.2297 & 0.095 & -2.426 & 0.015 & -0.415 & -0.044 \\
Smartphone use & 0.0961 & 0.016 & 6.183 & 0.000 & 0.066 & 0.127 \\
Comfort (1/2) & 0.2976 & 0.425 & 0.701 & 0.483 & -0.535 & 1.130 \\
Comfort (2/3) & 0.3613 & 0.038 & 9.456 & 0.000 & 0.286 & 0.436 \\
Comfort (3/4) & 0.1122 & 0.042 & 2.653 & 0.008 & 0.029 & 0.195 \\
Comfort (4/5) & 0.6794 & 0.047 & 14.483 & 0.000 & 0.587 & 0.771 \\
\bottomrule
\end{tabular}
\end{table}

\begin{table*}[htbp]
    \caption{Mean and standard deviation values of privacy concerns from our privacy concerns survey when concerns from different countries are considered (Figure~\ref{fig_main_country_results}). AR and SP indicate AR glasses and smartphone, respectively.}
    \alt{Mean and standard deviation values of privacy concerns for different devices, namely AR glasses and smartphones, organized according to countries. The first column represents the user attributes we have. The rest of the columns include the summary statistics.}
    \centering
    \begin{tabular}{|c|c|*{5}{c|}}
        Attribute & AR (US) & SP (US) & AR (Germany) & SP (Germany) \\
        \toprule
        activity & $ 2.29\pm 1.23 $  & $2.27 \pm 1.16$ & $2.32 \pm 1.25$ & $2.31 \pm 1.27$ \\
        alertness & $2.82 \pm 1.31 $  & $2.81 \pm 1.28$ & $ 2.93\pm 1.24$ & $ 2.85\pm 1.24$ \\
        BMI & $2.60 \pm 1.22 $  & $2.60 \pm 1.24$ & $2.30 \pm 1.21$ & $2.36 \pm 1.20$ \\
        cognitive load & $ 2.49\pm 1.17 $  & $2.48 \pm 1.17$ & $2.35 \pm 1.16$ & $2.34 \pm 1.13$ \\
        depression & $ 2.34\pm 1.20 $  & $2.35 \pm 1.20$ & $2.11 \pm 1.20$ & $2.15 \pm 1.17$ \\
        gender & $2.57 \pm 1.21 $  & $2.60 \pm 1.20$ & $2.37 \pm 1.12$ & $2.43 \pm 1.15$ \\
        heart condition & $ 2.51\pm 1.33 $  & $2.54 \pm 1.33$ & $2.42 \pm 1.36$ & $2.42 \pm 1.35$ \\
        location & $ 2.41\pm 1.22 $  & $2.48 \pm 1.19$ & $2.28 \pm 1.20$ & $2.37 \pm 1.16$ \\
        personal identity & $2.05 \pm 1.14 $  & $2.08 \pm 1.13 $ & $2.04 \pm 1.02$ & $2.10 \pm 1.07$ \\
        sexual preference & $1.96 \pm 1.07 $  & $ 1.94\pm 1.11 $ & $1.81 \pm 1.00 $ & $1.82 \pm 1.01$ \\
        stress & $ 2.60\pm 1.22 $  & $ 2.65\pm 1.20$ & $ 2.62\pm 1.21$ & $2.64 \pm 1.17$ \\
        \bottomrule
    \end{tabular}
    \label{tab_main_survey_country_mean_std}
\end{table*}

\begin{table*}[htbp]
    \caption{Mean and standard deviation values of privacy concerns from our privacy concerns survey when different retention times are considered (Figure~\ref{fig_main_retention_results}). AR and SP correspond to AR glasses and smartphone, respectively.}
    \alt{Mean and standard deviation values of privacy concerns for different data retention times, namely one day, one month, and indefinitely, organized according to different devices. The first column represents the user attributes we have. The rest of the columns include the summary statistics.}
    \centering
    \small
    \begin{tabular}{|c|c|*{7}{c|}}
        Attribute & 1 day (AR) & 1 day (SP) & 1 month (AR) & 1 month (SP) & Indefinite (AR) & Indefinite (SP) \\
        \toprule
        activity & $ 2.33\pm 1.25$  & $ 2.32\pm 1.16$ & $ 2.33\pm 1.28$ & $ 2.30\pm 1.27$ & $ 2.25\pm 1.20$ & $ 2.25\pm 1.21$ \\
        alertness & $ 2.88\pm 1.28$  & $2.82 \pm 1.27$ & $2.94 \pm 1.27$ & $2.90 \pm 1.22$ & $2.81 \pm 1.27$ & $2.77 \pm 1.29$ \\
        BMI & $ 2.54\pm 1.22$  & $2.53 \pm 1.22$ & $2.48 \pm 1.26$ & $2.57 \pm 1.27$ & $2.31 \pm 1.18 $ & $2.34 \pm 1.18$ \\
        cognitive load & $2.51 \pm 1.17$  & $2.47 \pm 1.12$ & $2.44 \pm 1.22$ & $2.41 \pm 1.19$ & $2.32 \pm 1.10$ & $2.35 \pm 1.15$ \\
        depression & $ 2.32\pm 1.23$  & $2.34 \pm 1.23$ & $2.27 \pm 1.23$ & $2.31 \pm 1.22$ & $2.10 \pm 1.16$ & $2.11 \pm 1.12$ \\
        gender & $ 2.50\pm 1.14$  & $ 2.53\pm 1.17$ & $2.46 \pm 1.19$ & $2.49 \pm 1.18$ & $2.46 \pm 1.17$ & $2.54 \pm 1.18$ \\
        heart condition & $2.51 \pm 1.39$  & $2.49 \pm 1.37$ & $2.40 \pm 1.33$ & $2.45 \pm 1.36$ & $2.48 \pm 1.32$ & $2.49 \pm 1.30$ \\
        location & $ 2.43\pm 1.23$  & $ 2.48\pm 1.19$ & $ 2.34\pm 1.17$ & $ 2.49\pm 1.18$ & $ 2.27\pm 1.23$ & $ 2.31\pm 1.16$ \\
        personal identity & $2.09 \pm 1.07$  & $2.10 \pm 1.08$ & $2.03 \pm 1.07$ & $2.12 \pm 1.13 $ & $2.03 \pm 1.10$ & $2.05 \pm 1.09$ \\
        sexual preference & $1.86 \pm 0.99$  & $1.89 \pm 1.02$ & $1.90 \pm 1.03$ & $1.87 \pm 1.07$ & $1.89 \pm 1.08$ & $1.88 \pm 1.10$ \\
        stress & $ 2.73\pm 1.22$  & $2.73 \pm 1.19$ & $2.65 \pm 1.23$ & $2.72 \pm 1.22$ & $2.45 \pm 1.17$ & $2.49 \pm 1.14$ \\
        \bottomrule
    \end{tabular}
    \label{tab_main_surveys_retentiontimes_mean_std}
\end{table*}

\begin{sidewaystable}
    \caption{Mean and standard deviation values of privacy concerns from our privacy concerns survey when different data-receiving entities are considered (Figure~\ref{fig_main_entity_results}). AR, SP, WC, and ACE indicate AR glasses, smartphone, work contacts, and app company employees, respectively.}
    \alt{Mean and standard deviation values of privacy concerns for different data receiving entities, namely device itself, app cloud, friends, work contacts, app company employees, and the public, organized according to different devices. The first column represents the user attributes we have. The rest of the columns include the summary statistics.}
    \centering
    \scriptsize
    \begin{tabular}{|c|c|*{13}{c|}}
        Attribute & Device (AR) & Device (SP) & Cloud (AR) & Cloud (SP) & Friends (AR) & Friends (SP) & WC (AR) & WC (SP) & ACE (AR) & ACE (SP) & Public (AR) & Public (SP) \\
        \toprule
        activity & $ 2.66\pm 1.32 $  & $2.62 \pm 1.32$ & $ 2.66\pm 1.24$ & $ 2.55\pm 1.24$ & $2.06 \pm 1.13$ & $ 2.07\pm 1.08$ & $ 2.02\pm 1.16$ & $ 2.05\pm 1.15$ & $ 2.32\pm 1.20$ & $ 2.28\pm 1.19$ & $ 2.13\pm 1.23$ & $ 2.18\pm 1.20$ \\
        alertness & $3.23 \pm 1.24 $  & $ 3.21\pm 1.19$ & $ 3.31\pm 1.16$ & $ 3.19\pm 1.14$ & $ 2.67\pm 1.21$ & $ 2.67\pm 1.28$ & $2.40 \pm 1.24$ & $2.31 \pm 1.25$ & $2.92 \pm 1.24$ & $ 2.83\pm 1.17$ & $ 2.73\pm 1.34$ & $ 2.77\pm 1.32 $\\
        BMI & $2.93 \pm 1.31 $  & $ 2.99\pm 1.24$ & $2.82 \pm 1.16$ & $ 2.81\pm 1.19$ & $ 2.12\pm 1.11$ & $2.17 \pm 1.11$ & $2.12 \pm 1.18$ & $2.17 \pm 1.16$ & $2.46 \pm 1.17$ & $2.51 \pm 1.19$ & $2.24 \pm 1.16$ & $2.27 \pm 1.22$ \\
        cognitive load & $ 2.86\pm 1.19 $  & $2.81 \pm 1.17$ & $ 2.79\pm 1.09$ & $2.71 \pm 1.08$ & $ 2.21\pm 1.09$ & $2.27 \pm 1.14$ & $1.98 \pm 1.11$ & $2.05 \pm 1.13$ & $2.46 \pm 1.10$ & $2.38 \pm 1.04$ & $2.22 \pm 1.16$ & $ 2.25\pm 1.19$\\
        depression & $ 2.58\pm 1.24 $  & $ 2.64\pm 1.22$ & $ 2.68\pm 1.23$ & $ 2.66\pm 1.18$ & $1.95 \pm 1.07$ & $2.00 \pm 1.12$ & $1.85 \pm 1.15$ & $1.84 \pm 1.12$ & $2.29 \pm 1.17$ & $2.31 \pm 1.16$ & $2.02 \pm 1.14$ & $2.06 \pm 1.11$\\
        gender & $ 2.77\pm 1.24 $  & $2.73 \pm 1.25$ & $2.63 \pm 1.12$ & $2.76 \pm 1.17$ & $2.32 \pm 1.11$ & $ 2.40\pm 1.13$ & $2.27 \pm 1.16$ & $2.31 \pm 1.18$ & $ 2.44\pm 1.13$ & $2.49 \pm 1.07$ & $2.40 \pm 1.18$ & $ 2.43\pm 1.19$\\
        heart condition & $2.83 \pm 1.39 $  & $2.79 \pm 1.39$ & $ 2.79\pm 1.40$ & $2.88 \pm 1.40$ & $2.24 \pm 1.28$ & $2.28 \pm 1.24$ & $2.19 \pm 1.24$ & $2.20 \pm 1.24$ & $2.49 \pm 1.34$ & $2.48 \pm 1.33$ & $2.28 \pm 1.28$ & $2.25 \pm 1.30$\\
        location & $ 2.64\pm 1.26 $  & $2.81 \pm 1.25$ & $2.67 \pm 1.21$ & $2.72 \pm 1.11$ & $ 2.16\pm 1.16$ & $2.26 \pm 1.15$ & $1.98 \pm 1.12$ & $2.07 \pm 1.06$ & $2.45 \pm 1.14$ & $2.46 \pm 1.07$ & $2.18 \pm 1.24$ & $2.24 \pm 1.23$\\
        personal identity & $ 2.44\pm 1.23 $  & $2.48 \pm 1.24$ & $2.32 \pm 1.11$ & $2.38 \pm 1.03$ & $1.81 \pm 0.94$ & $1.87 \pm 1.03$ & $1.95 \pm 1.03$ & $1.90 \pm 1.02$ & $2.02 \pm 1.01$ & $2.06 \pm 1.05$ & $1.77 \pm 0.96$ & $1.87 \pm 1.05$\\
        sexual preference & $ 2.06\pm 1.16 $  & $2.04 \pm 1.18$ & $ 2.19\pm 1.10$ & $2.22 \pm 1.12$ & $1.76 \pm 0.92$ & $1.77 \pm 0.96$ & $1.55 \pm 0.85$ & $1.66 \pm 0.92$ & $1.91 \pm 1.01$ & $ 1.83\pm 1.03$ & $ 1.85\pm 1.04$ & $ 1.79\pm 1.05$\\
        stress & $ 3.02\pm 1.23 $  & $ 3.02\pm 1.18$ & $ 3.07\pm 1.20$ & $ 3.04\pm 1.09$ & $2.33 \pm 1.14$ & $ 2.41\pm 1.21$ & $2.19 \pm 1.19$ & $ 2.22\pm 1.11$ & $ 2.62\pm 1.07$ & $ 2.73\pm 1.11$ & $ 2.44\pm 1.18$ & $2.47 \pm 1.17$\\
        \bottomrule
    \end{tabular}
    \label{tab_main_surveys_datareceivers_mean_std}
\end{sidewaystable}

\begin{sidewaystable}[htbp]
    \caption{Mean and standard deviation values of privacy concerns from our privacy concerns survey when different priming conditions are considered (Figure~\ref{fig_main_priming_results}). AR and SP indicate AR glasses and smartphone, respectively.}
    \alt{Mean and standard deviation values of privacy concerns for different priming types, namely beneficial, not-beneficial, and no priming, organized according to different devices. The first column represents the user attributes we have. The rest of the columns include the summary statistics.}
    \centering
    \begin{tabular}{|c|c|*{7}{c|}}
        Attribute & Beneficial (AR) & Beneficial (SP) & Not-beneficial (AR) & Not-beneficial (SP) & No priming (AR) & No priming (SP) \\
        \toprule
        activity & $2.99 \pm 1.25 $  & $3.0 \pm 1.21$ & $1.55 \pm 0.87$ & $1.52 \pm 0.86$ & $2.38 \pm 1.13$ & $2.36 \pm 1.07$ \\
        alertness & $ 3.37\pm 1.28 $  & $3.34 \pm 1.25$ & $2.71 \pm 1.21$ & $2.69 \pm 1.20$ & $2.55 \pm 1.19$ & $2.46 \pm 1.16$ \\
        BMI & $2.68 \pm 1.27 $  & $2.71 \pm 1.24$ & $2.31 \pm 1.18$ & $2.35 \pm 1.19 $ & $2.35 \pm 1.20$ & $2.39 \pm 1.21$ \\
        cognitive load & $ 2.77\pm 1.22 $  & $2.75 \pm 1.21 $ & $2.21 \pm 1.08$ & $2.19 \pm 1.07$ & $2.29 \pm 1.11$ & $2.29 \pm 1.11$ \\
        depression & $ 2.48\pm 1.27 $  & $2.45 \pm 1.26$ & $2.25 \pm 1.19$ & $2.29 \pm 1.16$ & $1.96 \pm 1.10$ & $2.01 \pm 1.12$ \\
        gender & $2.50 \pm 1.16 $  & $2.51 \pm 1.16$ & $2.13 \pm 1.08$ & $2.18 \pm 1.06$ & $2.78 \pm 1.17$ & $2.87 \pm 1.20$ \\
        heart condition & $3.09 \pm 1.35 $  & $3.08 \pm 1.30$ & $1.75 \pm 1.05$ & $1.80 \pm 1.10$ & $2.57 \pm 1.27$ & $2.57 \pm 1.30$ \\
        location & $2.91 \pm 1.32 $  & $2.95 \pm 1.26$ & $1.98 \pm 0.99$ & $2.06 \pm 0.97$ & $2.15 \pm 1.09$ & $2.27 \pm 1.09$ \\
        personal identity & $ 2.42\pm 1.19 $  & $2.46 \pm 1.22$ & $1.84 \pm 0.92$ & $1.91 \pm 0.98$ & $1.89 \pm 1.02$ & $1.90 \pm 0.99$ \\
        sexual preference & $1.94 \pm 1.09 $  & $1.94 \pm 1.11$ & $1.66 \pm 0.86$ & $1.67 \pm 0.92$ & $2.06 \pm 1.09$ & $2.04 \pm 1.11$ \\
        stress & $ 2.95\pm 1.24 $  & $2.93 \pm 1.22$ & $2.50 \pm 1.17$ & $ 2.64\pm 1.16$ & $2.38 \pm 1.16$ & $2.37 \pm 1.12$ \\
        \bottomrule
    \end{tabular}
    \label{tab_main_surveys_priming_mean_std}
\end{sidewaystable}

\begin{sidewaystable}[htbp]
    \caption{Mean and standard deviation values of beneficialness scores from our calibration survey when different priming conditions are considered (Figure~\ref{fig_priming_results}). AR, SP, B, NB, and GER indicate AR glasses, smartphone, beneficial, not-beneficial, and Germany,  respectively.}
    \alt{Mean and standard deviation values of beneficialness scores from our calibration surveys, including different devices, countries, and priming types. The first column represents the user attributes we have. The rest of the columns include the summary statistics.}
    \centering
    \begin{tabular}{|c|c|*{9}{c|}}
        Attribute & AR (US) - B & AR (US) - NB & SP (US) - B & SP (US) - NB & AR (GER) - B & AR (GER) - NB & SP (GER) - B & SP (GER) - NB \\
        \toprule
        activity & $ 3.87\pm 0.87 $  & $1.65 \pm 0.93$ & $3.95 \pm 0.49$ & $1.59 \pm 0.73$ & $4.29 \pm 0.62$ & $1.62 \pm 1.05$ & $4.29 \pm 0.81$ & $1.96 \pm 1.30$ \\
        alertness & $4.43 \pm 0.79 $  & $3.52 \pm 1.24 $ & $4.45 \pm 0.74$ & $3.68 \pm 1.17$ & $4.54 \pm 0.51$ & $4.12 \pm 0.74$ & $4.54 \pm 0.66$ & $3.83 \pm 1.01$ \\
        BMI & $ 3.91\pm 0.90 $  & $ 2.74 \pm 1.18$ & $ 3.91\pm 0.92$ & $ 2.82 \pm 1.26$ & $ 3.37\pm 1.21$ & $2.67 \pm 1.05$ & $3.91 \pm 0.91$ & $2.54 \pm 1.06$ \\
        cognitive load & $ 4.30\pm 0.63 $  & $ 2.48\pm 1.24$ & $ 4.23\pm 0.53$ & $2.73 \pm 1.24$ & $4.04 \pm 0.75$ & $2.67 \pm 1.31$ & $4.00 \pm 0.88$ & $ 2.67\pm 1.49$ \\
        depression & $ 3.74\pm1.14$ & $ 2.96\pm1.15 $ & $4.36 \pm 0.58$ & $3.18 \pm 0.96 $ & $3.71 \pm 0.95$ & $ 3.25 \pm 1.03 $ & $4.00 \pm 0.93$ & $3.67 \pm 1.17 $ \\
        gender & $ 2.56\pm 1.16 $  & $2.04 \pm 1.06$ & $3.09 \pm 0.87$ & $2.27 \pm 0.98$ & $2.79 \pm 0.88$ & $2.62 \pm 1.10$ & $2.87 \pm 1.23$ & $2.46 \pm 1.21$ \\
        heart condition & $ 4.43\pm 0.66 $  & $1.91 \pm 1.16$ & $4.50 \pm 0.51$ & $1.91 \pm 1.15$ & $4.37 \pm 0.87$ & $2.00 \pm 1.28$ & $4.50 \pm 0.59$ & $1.96 \pm 1.23$ \\
        location & $ 4.43\pm 0.59 $  & $2.22 \pm 1.04$ & $4.54 \pm 0.51$ & $2.73 \pm 1.03$ & $4.12 \pm 1.08$ & $1.96 \pm 0.91$ & $ 4.25\pm 0.68$ & $2.00 \pm 0.98$ \\
        personal identity & $3.48 \pm 1.08 $  & $1.96 \pm 0.98$ & $3.04 \pm 1.21$ & $1.73 \pm 0.77$ & $2.87 \pm 1.15$ & $2.29 \pm 1.27$ & $2.79 \pm 1.31$ & $2.08 \pm 1.25$ \\
        sexual preference & $2.83 \pm 1.37 $  & $1.74 \pm 0.81$ & $2.54 \pm 1.06$ & $2.14 \pm 1.04$ & $3.00 \pm 1.10$ & $2.25 \pm 0.79$ & $3.12 \pm 1.19$ & $2.29 \pm 0.95$ \\
        stress & $3.91 \pm 0.60 $  & $3.30 \pm 1.22$ & $ 4.18\pm 0.59$ & $ 3.45\pm 1.18$ & $3.96 \pm 0.81$ & $ 3.67\pm 1.01$ & $ 4.04\pm 0.81$ & $ 3.46\pm 0.93$ \\
        \bottomrule
    \end{tabular}
    \label{tab_priming_survey_mean_std}
\end{sidewaystable}

\clearpage
\newpage

\section{Appendix C: Details for qualitative data analysis}
\label{appendix_qualitative_coding}
Table~\ref{qualitative_codebook} provides the codebook that we created and used for analyzing free-text responses to the \emph{What other capabilities of your personal assistant app on your [AR glasses/smartphone] would make you uncomfortable?} question. While conducting the qualitative analysis according to our codebook, we excluded mentions of heart condition and depression, and sexual preferences from the categories of health and sexuality, as we asked about these user attributes in our surveys, and for the qualitative analysis, we explicitly sought additional concerns about other data types. 

\begin{table*}[ht]
    \centering
    \caption{Qualitative codebook for the free-text responses.}
    \alt{Qualitative codebook for the free-text responses.}
    \footnotesize
    \begin{tabular}{p{3.15cm}|p{6.8cm}|p{6.8cm}}
        \toprule
        \textbf{Code name} & \textbf{Definition} & \textbf{Example quote} \\
        \hline
        Health & References to health conditions, doctor history, prescription reminders, and disabilities. & ``If the system can tell what diseases or disorders I have.''\\
        \hline
        Social relationships/Contacts & Mentions of relationships, personal contacts, family, and friends.  & ``If the system can track my contacts with various people throughout a period of time. ''\\
        \hline
        Video/Audio & References to recording video and audio, collecting data from the front camera, and audiovisual recording. & ``If the system does audiovisual recording.''\\
        \hline
        Financial information & Mentions of banking, salary, and investment. & ``The system tracks my financial history.''\\
        \hline
        Online tracking/Messaging & References to browsing, Internet use, online messages, and conversations. & ``If the system can track what I am searching on the Internet.''\\
        \hline
        Emotions & Mentions of emotions, mood, feelings towards others or situations. & ``If the system is able to determine my feelings towards others.''\\
        \hline
        Routines & References to routines, habits, and behavioral patterns. & ``If the system can know my tendencies to do daily routines.''\\
        \hline
        Environment/Bystander & Mentions of physical environment, bystanders, or other people around the user & ``If the system is able to record my surroundings.''\\
        \hline
        Thoughts & References to thoughts and mind reading. & ``If the system could read my thoughts.''\\
        \hline
        Sexuality & Mentions of intimacy, sexual activity, desires, and sexual intent. & ``If the system can suggest anything about intimacy and sex.''\\
        \hline
        Socio-economics & References to race, ethnicity, age, address, civil record, and beliefs. & ``If the system can collect and store my home's address.''\\
        \hline
        Physiological & Mentions of appearance, height, weight, and body. & ``If the system can record and reveal anything related to my appearance.''\\
        \hline
        Gaze & References to gaze, eye movements, eye tracking, and where the user looks. & ``If the system can determine where I look when looking at advertisements or other people.''\\
        \hline
        Political views & Mentions of political views, political preferences, and voting. & ``If the system can understand my political preferences.''\\
        \hline
        Psychological & References to mental states, psychological evaluations, and psychoanalysis. & ``If the system can make psychological evaluations.''\\
        \hline
        \hline
        Overlapping concerns & Explicit mentions of same as AR glasses, same as smartphones, or same as before. & ``Ditto to my last response.''\\
        \bottomrule
    \end{tabular}
    \label{qualitative_codebook}
\end{table*}

\section*{Author contributions statement}
E.B., L.B., and L.F.C. conceptualized and conceived the experiments. E.B. conducted the experiments and analyzed the results with support from B.B. and X.W., incorporating E.K.'s, L.B.'s, and L.F.C.'s feedback. L.B. and L.F.C. supervised the project. E.B. drafted the original manuscript. All the authors reviewed and revised the manuscript. 

\section*{Acknowledgments}
Most of this research was conducted while E.B. was a visiting researcher at Carnegie Mellon University during his time at the University of Tübingen. The authors used Grammarly for proofreading. 

\section*{Funding}
This research was partly supported by the Deutsche Forschungsgemeinschaft (DFG, German Research Foundation) under Germany’s Excellence Strategy - EXC number 2064/1 - Project number 390727645 and by Meta. E.B. acknowledges support from the Visit Abroad Programme of the Cluster of Excellence – Machine Learning for Science through the University of Tübingen for his stay at Carnegie Mellon University.

\bibliographystyle{unsrt}   
\bibliography{references}   
\end{document}